\documentclass[11pt]{article}
\usepackage{times}

\usepackage[margin=1in]{geometry}
\usepackage{graphicx}
\usepackage{amsmath, amsthm, amssymb}
\usepackage{braket}
\usepackage{subcaption}
\usepackage{comment}
\usepackage{url}
\usepackage[ruled,lined,linesnumbered]{algorithm2e}
\usepackage{algpseudocode}
\usepackage{tikz}
\usepackage{diagbox}
\usepackage[colorlinks=true,citecolor=green!70!black,linkcolor=red,urlcolor=blue,bookmarks,breaklinks]{hyperref}
\usepackage{natbib}
\usepackage{bm}
\usepackage{ulem}
\setcitestyle{numbers}

\allowdisplaybreaks

\makeatletter
\newtheorem*{rep@theorem}{\rep@title}
\newcommand{\newreptheorem}[2]{%
\newenvironment{rep#1}[1]{%
 \def\rep@title{#2 \ref{##1} (restated)}%
 \begin{rep@theorem}}%
 {\end{rep@theorem}}}
\makeatother

\makeatletter
\newcommand{\subalign}[1]{%
  \vcenter{%
    \Let@ \restore@math@cr \default@tag
    \baselineskip\fontdimen10 \scriptfont\tw@
    \advance\baselineskip\fontdimen12 \scriptfont\tw@
    \lineskip\thr@@\fontdimen8 \scriptfont\thr@@
    \lineskiplimit\lineskip
    \ialign{\hfil$\m@th\scriptstyle##$&$\m@th\scriptstyle{}##$\hfil\crcr
      #1\crcr
    }%
  }%
}
\makeatother

\newtheorem{thm}{Theorem}
\newtheorem*{thm*}{Theorem}
\newtheorem{cor}[thm]{Corollary}

\newtheorem{lem}[thm]{Lemma}
\newtheorem*{lem*}{Lemma}

\newtheorem{prop}[thm]{Proposition}
\newtheorem{defn}[thm]{Definition}

\newreptheorem{thm}{Theorem}
\newreptheorem{lem}{Lemma}
\newreptheorem{prop}{Proposition}
\newreptheorem{cor}{Corollary}

\DeclareMathOperator*{\er}{ErdosRenyi}
\DeclareMathOperator{\Vect}{Vec}
\DeclareMathOperator{\Span}{span}
\DeclareMathOperator{\Bernoulli}{Bernoulli}

\definecolor{darkgreen}{rgb}{0,0.6,0}

\newcommand{\FIXME}[1]{}
\newcommand{\am}[1]{{\color{blue} AM: #1}}

\begin{document}

%\scalebox{10}{$\varphi$} \hspace{3cm} \scalebox{10}{$\Psi$}
%\end{document}

\title{Predicting parameters for the Quantum Approximate Optimization Algorithm for MAX-CUT from the infinite-size limit}
\author{Sami Boulebnane\thanks{University College London and Phasecraft Ltd; {\tt sami.boulebnane.19@ucl.ac.uk}.}\ \ and
Ashley Montanaro\thanks{Phasecraft Ltd and University of Bristol; {\tt ashley@phasecraft.io}.}
}
\maketitle

\begin{abstract}
Combinatorial optimization is regarded as a potentially promising application of near and long-term quantum computers. The best-known heuristic quantum algorithm for combinatorial optimization on gate-based devices, the Quantum Approximate Optimization Algorithm (QAOA), has been the subject of many theoretical and empirical studies. Unfortunately, its application to specific combinatorial optimization problems poses several difficulties: among these, few performance guarantees are known, and the variational nature of the algorithm makes it necessary to classically optimize a number of parameters. In this work, we partially address these issues for a specific combinatorial optimization problem: diluted spin models, with MAX-CUT as a notable special case. Specifically, generalizing the analysis of the Sherrington-Kirkpatrick model by Farhi et al., we establish an explicit algorithm to evaluate the performance of QAOA on MAX-CUT applied to random Erdos-Renyi graphs of expected degree $d$ for an arbitrary constant number of layers $p$ and as the problem size tends to infinity. This analysis yields an explicit mapping between QAOA parameters for MAX-CUT on Erdos-Renyi graphs of expected degree $d$, in the limit $d \to \infty$, and the Sherrington-Kirkpatrick model, and gives good QAOA variational parameters for MAX-CUT applied to Erdos-Renyi graphs. We then partially generalize the latter analysis to graphs with a degree distribution rather than a single degree $d$, and finally to diluted spin-models with $D$-body interactions ($D \geq 3$). We validate our results with numerical experiments suggesting they may have a larger reach than rigorously established; among other things, our algorithms provided good initial, if not nearly optimal, variational parameters for very small problem instances where the infinite-size limit assumption is clearly violated.
\end{abstract}

\tableofcontents

\newpage
% ------------------------------------------------------------------------------

\section{Introduction}

The quantum approximate optimization algorithm (QAOA) is a prominent use for near-term quantum computers. Farhi, Goldstone and Gutmann~\cite{1411.4028,1412.6062} proposed that QAOA could be used to approximately solve combinatorial optimization problems efficiently, showed that its performance for certain problems of this form could be analysed, and proved that in some cases it could outperform the best classical algorithms known at that time. (An essentially identical algorithm to QAOA was introduced 14 years previously by Hogg~\cite{hogg00} under the title of a ``quantum search heuristic'', although Hogg's algorithm assumes a particular form for the parameters in the algorithm, whereas Farhi, Goldstone and Gutmann propose optimising over these.)

Appealing features of QAOA include that it can sometimes be implemented with a low-depth quantun circuit; it can easily be applied to a variety of combinatorial optimization problems; its performance can sometimes be rigorously analysed; given certain computational complexity assumptions, it is hard to simulate classically~\cite{farhi16}; and with a sufficiently large number of layers, it encompasses the quantum adiabatic algorithm. However, QAOA also has some limitations. These include that for a given instance, its performance cannot in general be determined in advance, and one has to run the algorithm to see how well it performs; that the algorithm depends on a sequence of parameters which are generally unknown in advance and must be found using a classical optimization loop at significant cost; and that there are no cases known where it provably outperforms the best classical methods.

In this work we aim to significantly ameliorate these issues by developing theoretical and algorithmic procedures for determining provably good initial parameters, on average over a class of problems. The problem family we choose is MAX-CUT on random graphs. MAX-CUT is a combinatorial optimization problem of practical importance and a well-studied test case for QAOA, being one of the easiest optimization problems for which QAOA can be implemented.
We build on recent work by Farhi et al.~\cite{1910.08187}, who performed an analysis of this form for QAOA applied to the Sherrington-Kirkpatrick (SK) model. The SK model is a Hamiltonian with randomly chosen interactions between all pairs of sites. Farhi et al.\ showed that the performance of QAOA, and its optimal parameters, could be analysed in the infinite-size limit of the SK model using an algorithm whose runtime is exponential in the number of algorithm layers (not the number of qubits). The same work argued that it could be possible to generalize these results to the infinite-size limit for MAX-CUT on Erdos-Renyi graphs; however, the argument is restricted to \textit{dense} Erdos-Renyi graphs, i.e.\ with constant edge probability, contrary to the sparse graphs considered in this paper.

Here, we perform a similar analysis for the practically relevant setting of MAX-CUT on varying random graph families. First, we establish an exact expression for the QAOA energy on sparse Erdos-Renyi random graphs as the number of vertices $n \to \infty$. This exact expression can be optimised over QAOA parameters to allow the optimal parameters (in this limiting case) to be determined. We then show that this expression is closely connected to an analogous expression for the SK model, allowing parameters obtained for one model to be mapped to the other. We generalise these results to a broader family of nonuniform random graphs closely related to the Chung-Lu model, and also to a variant of MAX-CUT where each constraint acts on $D$ variables (in the case where the number of QAOA layers $p=1$).
These theoretical results all hold in the limit, and not for finite-size instances. However, we also give an efficient Monte Carlo algorithm for estimating the QAOA energy for finite instances of the SK model when $p \le 3$.

Next, we carry out numerical experiments to validate our results. First we compute the optimal limiting parameters for several families of random graphs. Next we evaluate how well these parameters (angles) perform on small finite-size instances (16 vertices). We test this in several ways: how the energy evaluated at these ``guessed'' angles compares with that obtained from optimization; how many optimization attempts starting from random angles are needed to outperform evaluation/optimization from guessed angles; and how far the angles are from the optimal angles. We find that just evaluating the QAOA energy with guessed angles outperforms random angles substantially, but usually does not achieve results competitive with optimizing the angles. However, using the guessed angles as a starting point for optimization achieves a reduction in optimization cost by a factor of 40 on average, as measured by the number of attempts required to achieve a better result, starting with random angles. We attribute this to the guessed parameters being close enough to the optimal parameters to enable the optimization algorithm to find the global optimum, rather than a local optimum.

Following a discussion of related work in the literature, we give a more detailed technical summary of our results below.

\subsection{Related work}

The Quantum Approximate Optimization Algorithm and variational quantum algorithms in general have been the subject of numerous theoretical and empirical studies; we refer e.g.\ to \cite{McClean2016,Cerezo2021} for an extensive review. In the following, we focus on the specific task of finding good variational parameters for QAOA.

Optimizing QAOA parameters was shown to be NP-hard including for some classically ``easy" optimization problems \cite{2101.07267}. It also proves challenging in practice at large depth owing, among other things, to the non-convexity of the optimization landscape \cite{Zhou2020}. As a result, optimizing variational parameters based on a random initialization typically requires a number of trials growing exponentially with the ansatz depth \cite{Zhou2020}. Several works have considered more efficient strategies to train QAOA with a limited number of executions of the quantum circuit. 

A first approach is based on analytically predicting good variational parameters. Hogg~\cite{hogg00} developed a method for finding good angles for QAOA when applied to 3-SAT, based on a mean-field approximation. Wang et al.~\cite{Wang2018} report optimal QAOA angles for MAX-CUT on large-girth $d$-regular graphs at depth $p = 1$. More recently, \cite{2103.11976} applied QAOA to unstructured search and studied the asymptotic behaviour of optimal variational parameters as the problem size went to infinity for small constant depth. Farhi et al.'s work on the SK model identified the correct scaling of QAOA angles with the problem size at arbitrary constant depth~\cite{1910.08187}. Unfortunately, such theoretical results are rather scarce.

An alternative is to fully optimize QAOA on a problem instance and reuse the determined optimal angles on a different instance of the same problem. \cite{1812.04170} shows empirical success of this method on MAX-CUT on small 3-regular graphs: optimal angles depend little on the instance but \textit{concentrate} around a typical value. For large regular graphs, this concentration property can be theoretically justified \cite{Wang2018} from the locality of QAOA \cite{1411.4028}. However, for other problems such as the SK model \cite{1910.08187,2102.12043}, concentration of optimal parameters over instances was shown to hold even in the absence of locality.

Other methods eschew the question of predicting good variational parameters while still reducing the number of optimization iterations or increasing the success probability. \cite{Zhou2020} proposes to optimize QAOA layer after layer, initializing each layer with QAOA angles extrapolated from the optimal parameters of the previous one; empirically, this deterministic initialization strategy strictly outperforms random initialization, including for moderate, single-digit ansatz depth. More recently, \cite{Sack2021} proposed to initialize QAOA angles to a linear schedule, corresponding to a first-order Trotter approximation to quantum annealing, and successfully illustrated the approach on MAX-CUT. Finally, several works \cite{pmlr-v107-yao20a,Khairy2020,1911.04574} demonstrated the use of reinforcement learning to optimize the expected energy of QAOA, seen as a black-box function of the variational parameters.

\section{Technical preliminaries}

\subsection{Random graphs ensembles}
\label{sec:random_graphs_ensembles}

In this work, QAOA is applied to the MaxCut problem on random graphs. In this section, we define the random graph ensembles we use.

\begin{defn}[Random Erdos-Renyi graph {\cite{Bollobs2001}}]
\label{def:erdos_renyi_graph}
Let $n \geq 2$ be an integer (number of vertices) and $q \in [0, 1]$ (edge probability). The \textnormal{Erdos-Renyi} random unweighted graph ensemble on $n$ vertices with edge probability $q$ is defined by including each of the possible $\binom{n}{2}$ edges in the graph with probability $q$. We will occasionally denote by $G \sim \er\left(n, q\right)$ graph $G$ sampled from the Erdos-Renyi ensemble on $n$ vertices with edge probability $q$.
\end{defn}

It follows from this definition that each vertex of a random graph $G \sim \er\left(n, q\right)$ has an expected degree $(n - 1)q$. Graphs from $\er\left(n, \frac{d}{n - 1}\right)$ may then provide a convenient proxy to understand $d$-regular graphs; as an example, both ensembles have asymptotically equivalent maximum cuts \cite{Dembo2017}. Random Erdos-Renyi graphs are a particular case of the following ensemble, inspired from the Chung-Lu model~\cite{Chung2002}, which allows a notion of non-uniformity across the graph. The difference from the original Chung-Lu model is that loops are excluded.
\begin{defn}[Pseudo Chung-Lu model {\cite{Chung2002}}]
\label{def:pseudo_chung_lu_model}
Let $n \geq 2$ (number of vertices) and $N_L \geq 1$ (number of distinct labels) be integers. Let $\left(d_l\right)_{1 \leq l \leq N_L}$ be positive reals ($d_l$ corresponds to the expected degree of a vertex labelled $l$) and $\left(q_l\right)_{1 \leq l \leq N_L}$ be a probability distribution on the labels. A random graph from the \textnormal{pseudo Chung-Lu} ensemble is constructed as follows.
\begin{itemize}
    \item For each vertex $k \in [n]$, label $k$ by $l \in \{1, \ldots, N_L\}$ with probability $q_l$.
    \item Given this labelling, add an edge between a pair of vertices labelled $l, l'$ with probability $\frac{1}{n - 1}\frac{d_ld_{l'}}{\sum_{1 \leq l'' \leq N_L}q_{l''}d_{l''}}$.
\end{itemize}
(This model is only well-defined for $n$ sufficiently large, e.g. $n \geq 1 + \frac{\left(\max_{1 \leq l \leq N_L}d_l\right)^2}{\min_{1 \leq l \leq N_L}d_l}$)
\end{defn}

\subsection{Optimization problems}
\label{sec:optimization_problems}
In this section, we precisely define the optimization problems considered in this work: dense and diluted spin models, with the Sherrington-Kirkpatrick model and MaxCut on Erdos-Renyi graphs as particular cases. All these problems consist of minimizing a Hamiltonian of the form:
\begin{align}
\label{eq:general_optimization_problem}
    H[J]\left((\sigma_k)_{k \in [n]}\right) & := \sum_{\{k_1, \ldots, k_D\} \subset [n]}J_{k_D \ldots k_1}\sigma_{k_D} \ldots \sigma_{k_1}
\end{align}
over $\left(\sigma_k\right)_{k \in [n]} \in \{-1, 1\}^n$. $J$ in equation \ref{eq:general_optimization_problem} is a $D$-indices tensor, where each index has dimension $n$. In the following, we will be interested in \textit{random} optimization problems, where $J$ is drawn from a certain probability distribution. We now describe the particular distributions considered in this work.

The dense $D$-spin model and the performance of QAOA on its optimization was already considered in \cite{2102.12043}:
\begin{defn}[Dense $D$-spin model]
The dense $D$-spin model on $n$ vertices corresponds to the random Hamiltonian in equation \ref{eq:general_optimization_problem}, with
\begin{align}
    J_{k_D \ldots k_1} & \stackrel{\textrm{i.i.d}}{\sim} n^{\frac{1 - D}{2}}\mathcal{N}(0, 1).
\end{align}
In the following, we denote $J \sim \textnormal{Dense}(D)$ for a tensor $J$ sampled according to this distribution.
\end{defn}

We now introduce the diluted $D$-spin model, which generalizes MaxCut on Erdos-Renyi graphs and will be considered in this work:
\begin{defn}[Diluted $D$-spin model]
\label{def:diluted_d_spin}
The diluted $D$-spin model on $n$ vertices with degree parameter $d$ corresponds to the random Hamiltonian in equation \ref{eq:general_optimization_problem}, with
\begin{align}
    J_{k_D \ldots k_1} & \stackrel{\textrm{i.i.d}}{\sim} \textnormal{Bernoulli}\left(\frac{nd}{D\binom{n}{D}}\right).
\end{align}
In the following, we denote $J \sim \textnormal{Diluted}(D, n, d)$ for a tensor $J$ sampled according to this distribution.
\end{defn}

The performance of QAOA on the Sherrington-Kirkpatrick model, a particular case of the dense $D$-spin model, was considered in \cite{1910.08187}:
\begin{defn}[Sherrington-Kirkpatrick model {\cite{panchenko_2013}}]
\label{def:sk}
The Sherrington-Kirkpatrick (SK) model is the dense 2-spin model. Explicitly, it is characterized by Hamiltonian \ref{eq:general_optimization_problem} with
\begin{align}
    J & \stackrel{\textrm{i.i.d}}{\sim} n^{-1/2}\mathcal{N}(0, 1).
\end{align}
In the following, we denote $J \sim \textnormal{SK}(n)$ for a tensor $J$ sampled according to this distribution.
\end{defn}

Finally, MaxCut on random Erdos-Renyi graphs is a particular case of the diluted $D$-spin model, for which we will prove stronger results:
\begin{defn}[MaxCut on Erdos-Renyi graphs]
\label{def:maxcut_erdos_renyi}
The MaxCut problem on random Erdos-Renyi graphs with degree parameter $d$ corresponds to the diluted $2$-spin model with degree parameter $d$. Explicitly, it is characterized by Hamiltonian \ref{eq:general_optimization_problem} with
\begin{align}
    J_{kl} & \stackrel{\textrm{i.i.d}}{\sim} \textnormal{Bernoulli}\left(\frac{d}{n - 1}\right).
\end{align}
In the following, we denote $J \sim \er\left(n, \frac{d}{n - 1}\right)$ for a tensor $J$ sampled according to this distribution.
\end{defn}
The connection between the distribution of $J$ in the latter case and MaxCut on Erdos-Renyi graphs can be understood by interpreting $J$ as the adjacency matrix of an $n$-vertices graph ($J_{kl} = 1$ if $\{k, l\}$ is an edge of the graph and $0$ otherwise). Then $J_{kl} \stackrel{\textrm{i.i.d}}{\sim} \textnormal{Bernoulli}\left(\frac{d}{n - 1}\right)$ means that the corresponding graph is sampled from the Erdos-Renyi ensemble with edge probability $\frac{d}{n - 1}$. In such a graph, each vertex has expected degree $d$, justifying calling $d$ the \textit{degree parameter}. Given a cut $\left(\sigma_k\right)_{k \in [n]} \in \{-1, 1\}^n$, $\sum_{\{k, l\} \subset [n]}J_{kl}\frac{1 - \sigma_k\sigma_l}{2}$ is the number of satisfied edges in the cut, which is easily related to Hamiltonian \ref{eq:general_optimization_problem}.

\subsection{The Quantum Approximate Optimization Algorithm}
\label{sec:qaoa}
The Quantum Approximate Optimization Algorithm (QAOA), as introduced in \cite{1411.4028}, aims to find an approximate ground state of a Hamiltonian $H_C$ acting on $n$ qubits. It does so by preparing a variational trial state depending on $2p$ variables $\bm\beta = (\beta_0, \ldots, \beta_{p - 1}) \in \mathbf{R}^p, \bm\gamma = (\gamma_0, \ldots, \gamma_{p - 1}) \in \mathbf{R}^p$:
\begin{align}
\label{eq:qaoa_variational_state}
    \ket{\Psi_{\textnormal{QAOA}}(H_C, \bm{\beta}, \bm{\gamma})} & = \prod_{0 \leq j < p}e^{-i\frac{\beta_j}{2}H_B}e^{-i\frac{\gamma_j}{2}H_C}\ket{+},
\end{align}
where the ordering of the product is such that $e^{-i\frac{\gamma_0}{2}H_C}$ is closest to $\ket{+}$,
\begin{align}
    H_B & := \sum_{0 \leq k < n}X_k
\end{align}
and we have slightly abused notation by writing $\ket{+}$ for $\ket{+}^{\otimes n}$. The parameters $\bm\beta, \bm\gamma$ are sometimes referred to as \textit{angles}. In the following, $H_B$ will be called the \textit{mixer Hamiltonian} and $H_C$ the \textit{problem Hamiltonian}. The most common cost function for optimizing $\bm\beta, \bm\gamma$ is the expected energy of the variational state:
\begin{align}
    & \braket{\Psi_{\textnormal{QAOA}}(H_C, \bm\beta, \bm\gamma)|H_C|\Psi_{\textnormal{QAOA}}(H_C, \bm\beta, \bm\gamma)}
\end{align}
but other cost functions have been proposed, such as the CVar \cite{Barkoutsos2020} or the Gibbs function \cite{Li2020}. A particular case is when the problem Hamiltonian $H_C$ is diagonal in the computational basis, corresponding to the classical Hamiltonian of an unconstrained binary optimization problem (see e.g. \cite{1411.4028,2004.09002,2005.08747,1910.08187}). In this case, measuring the variational state $\ket{\Psi_{\textnormal{QAOA}}(H_C, \bm\beta, \bm\gamma)}$ in the computational basis yields bitstrings that are approximate optima of the binary optimization problem.

In this work, we apply the QAOA to the classical Hamiltonian defined in section \ref{sec:optimization_problems}. This Hamiltonian is naturally converted to a diagonal quantum Hamiltonian
\begin{align}
    \hat{H}[J] & := \sum_{\{k_1, \ldots, k_D\} \subset [n]}J_{k_1\ldots k_D}Z_{k_1}\ldots Z_{k_D}.
\end{align}
We shall omit the hat from now; besides, we will use the shorter notation $\ket{\Psi_{\textrm{QAOA}}(J, \bm\beta, \bm\gamma)}$ for the associated variational QAOA state $\ket{\Psi_{\textnormal{QAOA}}\left(H[J], \bm\beta, \bm\gamma\right)}$ defined in equation \ref{eq:qaoa_variational_state}.

\subsection{Bitstrings}
\label{sec:bitstrings}

In the statement of the algorithms and the technical derivations, we will frequently refer to the \textit{level of symmetry} of a bitstring. This is equivalent to the partition of the set of tuples from $\{-1, 1\}$ into subsets $A_l$ defined in \cite{1910.08187}.
\begin{defn}[Level of symmetry of a bitstring]
Let $p \geq 0$ an integer. For all bitstrings $s \in \{0, 1\}^{2p + 1}$, we define the level of symmetry of $s$, denoted by $L(s)$, as:
\begin{align}
    L(s) & := \max\left\{0 \leq j \leq p\,:\,\forall k \in [0, j], s_{p + k} = s_{p - k}\right\}
\end{align}

\end{defn}
For instance, for $p = 2$, $L(01001) = 0, L(10100) = 1, L(01010) = 2$. The bitstrings can be grouped by symmetry levels and also according to whether they are odd or even. We defined the latter in the following:
\begin{defn}[Odd and even bitstrings]
\label{def:odd_bitstrings}
Let $p \geq 1$ an integer and $s \in \{0, 1\}^{2p + 1}$ a bitstring. $s$ is said to be \textit{odd} if $s_0 \neq s_p$; otherwise, it is said to be \textit{even}. For a fixed $p$, the set of odd bitstrings is denoted by $\mathcal{L}$.
\end{defn}

One can then classify bitstrings according to their levels of symmetry and parities, for which the following notations help:
\begin{defn}[Bitstrings by level and parity]
\label{def:mathcal_l}
Let $p \geq 1$ an integer. For all level of symmetry $l \in [0, p]$, one defines:
\begin{align}
    \mathcal{L}_l & := \{s \in \{0, 1\}^{2p + 1}\,:\,L(s) = l, s\,\textrm{odd}\}\\
    \mathcal{L}'_l & := \{s \in \{0, 1\}^{2p + 1}\,:\,L(s) = l, s\,\textrm{even}\}
\end{align}
\end{defn}

It will also be helpful to define a partial flip operation on bitstrings, whereby some bits of the string are flipped, starting from the center:

\begin{defn}[Partial bitstring flip]
Let $p \geq 1$ an integer. Given a bitstring $s \in \{0, 1\}^{2p + 1}$, one defines the \textnormal{partial flip} of $s$, denoted by $F(s)$, as the following length $(2p + 1)$ bitstring:
\begin{align}
    F(s)_{p + j} & = \left\{\begin{array}{lr}
        \textnormal{NOT}(s_{p + j}) & \textrm{if } -L(s) \leq j \leq L(s)\\
        s_{p + j} & \textrm{otherwise}
    \end{array}\right. \qquad \forall -p \leq j \leq p.
\end{align}
\end{defn}
For example, for $p = 2$, $F(01001) = 01101, F(10100) = 11010, F(01010) = 10101$. The following useful fact is easily proven:
\begin{prop}[Partial bitstring flip changes parity]
\label{prop:partial_flip_parity}
Let $p \geq 1$ an integer and let $\mathcal{L}_l, \mathcal{L}'_l, 0 \leq l \leq p$ be as in definition \ref{def:mathcal_l}. For $0 \leq l < p$, $F$ maps $\mathcal{L}_l$ onto $\mathcal{L}'_l$ and $\mathcal{L}'_l$ onto $\mathcal{L}_l$.
\end{prop}
In other words, the partial flip changes the parity of a bitstring, unless it is completely symmetric ($L(s) = p$). In the calculations and algorithms to follow, we will often exploit cancellation bitstrings $s$ and $F(s)$; besides, we will be led to consider bitstrings by decreasing level of symmetry. This motivates us to define an ordering of the bitstrings:
\begin{defn}[Ordering of bitstrings]
\label{def:bitstrings_ordering}
Given an integer $p \geq 1$, bitstrings $s \in \{0, 1\}^{2p + 1}$ are totally ordered with non-decreasing level of symmetry, such that for all $s \in \mathcal{L}$, $F(s)$ is the successor of $s$.
\end{defn}
This ordering is non-unique; the ordering of pairs $\left(s, F(s)\right)$ among strings with a fixed level of symmetry is irrelevant. We then define functions $B_{\bm{\beta}, s}$, $\varphi(\bm{\gamma}, s)$ depending on a bitstring $s$ and the QAOA angles $\bm{\beta}, \bm{\gamma}$.
\begin{defn}[$B_{\bm{\beta}, s}$ and $\varphi(\bm{\gamma}, s)$]
\label{def:B_phi}
Let $p \geq 0$ an integer, $s \in \{0, 1\}^{2p + 1}$ a bitstring and $\bm{\beta}, \bm{\gamma} \in \mathbf{R}^p$ a set of level-$p$ QAOA angles. One defines:
\begin{align}
    B_{\bm{\beta}, s} & := \prod_{0 \leq j < p}\left(\cos\beta_j\right)^{(s_j = s_{j + 1}) + (s_{2p - j} = s_{2p - j - 1})}\left(i\sin\beta_j\right)^{(s_j \neq s_{j + 1}) + (s_{2p - j} \neq s_{2p - j - 1})}\\
    \varphi(\bm{\gamma}, s) & = \sum_{0 \leq j < p}\frac{\gamma_j}{2}\left((-1)^{s_{2p - j}} - (-1)^{s_j}\right)
\end{align}
\end{defn}
We will repeatedly use the following fact concerning $B_{\bm{\beta}, s}$ and $\varphi$ defined above:
\begin{prop}
\label{prop:beta_phi_invariance_partial_flip}
Let $p \geq 1$ an integer and $\bm{\gamma} \in \mathbf{R}^p$. Then for all $s \in \{0, 1\}^{2p + 1}$,
\begin{align}
    B_{\bm{\beta}, F(s)} & = B_{\bm{\beta}, s}.
\end{align}
Besides, for $s, t \in \{0, 1\}^{2p + 1}$,
\begin{align}
    s < t & \implies \varphi(\bm{\gamma}, F(s) \oplus t) = \varphi(\bm{\gamma}, s \oplus t).
\end{align}
\end{prop}

Finally, we will occasionally use the following convention to denote the XOR operation between a subset of bitstrings:

\begin{defn}{XOR of a subset of bitstrings.}
Let $p \geq 1$ an integer and $\left(D_s\right)_{s \in \{0, 1\}^{2p + 1}}$ a family of non-negative integers. We define:
\begin{align}
    \bigoplus_{s \in \{0, 1\}^{2p + 1}}s^{\oplus D_s} & := \bigoplus_{\substack{s \in \{0, 1\}^{2p + 1}\,:\,D_s\,\textrm{odd}}}s,
\end{align}
where $\oplus$ denotes the XOR addition. In other words, $\bigoplus_{s \in \{0, 1\}^{2p + 1}}s^{\oplus D_s}$ is the XOR addition between bitstrings $s$ such that $D_s$ is odd. In particular, if $\left(D'_s\right)_{s \in \{0, 1\}^{2p + 1}}$ is another family of nonnegative integers,
\begin{align}
    \bigoplus_{s \in \{0, 1\}^{2p + 1}}s^{\oplus D_s} \oplus \bigoplus_{s \in \{0, 1\}^{2p + 1}}s^{\oplus D'_s} & = \bigoplus_{s \in \{0, 1\}^{2p + 1}}s^{\oplus(D_s + D'_s)}.
\end{align}
\end{defn}

\section{Results}
In this section, we discuss the main results of the paper. This includes an estimate for the performance of MaxCut-QAOA on random Erdos-Renyi graphs of constant average degree in the infinite size limit for any constant $p$; a similar result for QAOA on the diluted and dense $D$-spin model but proven at $p = 1$ only; and finally, a classical algorithm to evaluate the energy of SK-QAOA on finite size instances for $p > 1$. We start by introducing our proven results (section \ref{subsec:theoretical_results}), before illustrating and extrapolating them numerically (section \ref{subsec:numerical_experiments}).

\subsection{Theoretical results}
\label{subsec:theoretical_results}

Our first result concerns the performance of MaxCut-QAOA on an average Erdos-Renyi large random graph of constant average degree $d$ and applies to arbitrary $p$ (however, $p$ has to remain constant as the graph size goes to infinity). It can be phrased as an explicit algorithm to evaluate the energy of MaxCut-QAOA for fixed angles on such a random instance, similar to the one given in \cite{1910.08187} for the SK model.

\begin{thm}
\label{th:qaoa_erdos_renyi_energy}
Let $d \geq 3$ be the average degree of an Erdos-Renyi graph. Let $p \geq 1$ an integer and $\bm\beta, \bm\gamma \in \mathbf{R}^p$. The energy of level-$p$ MaxCut-QAOA on an average Erdos-Renyi graph $G \sim \er\left(n, \frac{d}{n - 1}\right)$ is given, in the infinite size limit $n \to \infty$, by:
\begin{align}
    & \lim_{n \to \infty}\frac{\mathbf{E}_{J \sim \er\left(n, \frac{d}{n - 1}\right)}\left[\left\langle\Psi_{\textnormal{QAOA}}(J, \bm\beta, \bm\gamma)|H[J]|\Psi_{\textnormal{QAOA}}(J, \bm\beta, \bm\gamma)\right\rangle\right]}{nd/2}\nonumber\\
    & = 2\sum_{u \in \{0, 1\}^{2p + 1}}S_u^{\infty} + 2\sum_{\substack{u, v \in \{0, 1\}^{2p + 1}\\u < v}}S_{u, v}^{\infty},
\end{align}
where
\begin{align}
    S^{\infty}_{u, v} & = \frac{(-1)^{u \in \mathcal{L}}B_{\bm{\beta}, u}R_u(-1)^{v \in \mathcal{L}}B_{\bm{\beta}, v}R_v}{4}\\
    S^{\infty}_u & = \frac{B_{\bm{\beta}, u}^2R_u^2}{4}
\end{align}
for bitstrings $u, v \in \{0, 1\}^{2p + 1}$. The notations $B_{\bm\beta, u}$ and $\mathcal{L}$ are defined in section \ref{sec:bitstrings}. Finally, the quantities $R_u$, similar to the ones defined in \cite{1910.08187} for the SK model, can be evaluated using algorithm \ref{alg:compute_r}.
\end{thm}

\begin{algorithm}[!htbp]
	\caption{Compute $\left(R_u\right)_{u \in \{0, 1\}^{2p + 1}}$}
	\label{alg:compute_r}
	\KwData{QAOA level $p$, QAOA angles $\bm{\beta}$, $\bm{\gamma}$}
	\KwResult{$\left(R_u\right)_{u \in \{0, 1\}^{2p + 1}}$}
	$\mathcal{L}_{\textrm{list}} \leftarrow \textrm{bitstrings  } s \in \mathcal{L} - \mathcal{L}_p \textrm{ ordered by increasing level of symmetry}$\\
	\For{$s \in \{0, 1\}^{2p + 1}\,:\,L(s) \in \{p - 1, p\}$}{
	    $R_s \leftarrow  e^{-d\left(1 - \frac{1}{2}\sum_{t \in \mathcal{L}_p \sqcup \mathcal{L}'_p}(-1)^{t \in \mathcal{L}}B_{\bm{\beta}, t}e^{i\varphi(\bm{\gamma}, s \oplus t)}\right)}$
	}
	\For{$s \in \mathcal{L}_{\textnormal{list}}$ (backwards)}{
	    $R_s, R_{F(s)} \leftarrow e^{-d\left(1 - \frac{1}{2}\sum_{t \in \mathcal{L}_p \sqcup \mathcal{L}'_p}(-1)^{t \in \mathcal{L}}B_{\bm{\beta}, t}e^{i\varphi(\bm{\gamma}, s \oplus t)}\right)}$\\
	    \For{$t \in \{0, 1\}^{2p + 1}\,:\,L(s) < L(t) < p$}{
	        $R_s, R_{F(s)} \leftarrow R_se^{\frac{d}{2}\sum_{t, L(s) < L(t) < p}(-1)^{t \in \mathcal{L}}B_{\bm{\beta}, t}R_te^{i\varphi(\bm{\gamma}, s \oplus t)}}$
	    }
	}
	\Return $\left(R_u\right)_{u \in \{0, 1\}^{2p + 1}}$
\end{algorithm}

Using theorem \ref{th:qaoa_erdos_renyi_energy}, one can relate the performance of MaxCut-QAOA on Erdos-Renyi graphs to the performance achieved on the SK model in the infinite size limit as analyzed in \cite{1910.08187}. This is expressed in the following result:

\begin{thm}[QAOA energy on Erdos-Renyi MAX-CUT]
\label{th:qaoa_sk_qaoa_maxcut}
Let $d \geq 3$ and $p \geq 1$ integers, $\bm{\beta}^{SK}, \bm{\gamma}^{SK} \in \mathbf{R}^{p}$. Consider a random $n$-vertices Erdos-Renyi graph of average degree $d$: $G \sim \er\left(n, \frac{d}{n - 1}\right)$. Then the expected energy of MaxCut-QAOA on $G$ is given as follows as $n \to \infty$:
{
\small
\begin{align}
    \label{eq:qaoa_sk_qaoa_maxcut_relation}
    & \lim_{n \to \infty}\mathbf{E}_{J \sim \er\left(n, \frac{d}{n - 1}\right)}\left\langle\Psi_{\textnormal{QAOA}}\left(J, \bm{\beta}^{SK}, \frac{\bm{\gamma}^{SK}}{\sqrt{d}}\right)\bigg|\frac{H[J]}{n}\bigg|\Psi_{\textnormal{QAOA}}\left(J, \bm{\beta}^{SK}, \frac{\bm{\gamma}^{SK}}{\sqrt{d}}\right)\right\rangle\nonumber\\
    & = \frac{1}{\sqrt{d}}\lim_{n \to \infty}\mathbf{E}_{J \sim \textnormal{SK}(n)}\left\langle\Psi_{\textnormal{QAOA}}\left(J, \bm{\beta}^{SK}, \bm{\gamma}^{SK}\right)\bigg|\frac{H[J]}{n}\bigg|\Psi_{\textnormal{QAOA}}\left(J, \bm{\beta}^{SK}, \bm{\gamma}^{SK}\right)\right\rangle + \mathcal{O}\left(\frac{1}{d}\right)
\end{align}
}
\end{thm}
This relation between the energy of MaxCut-QAOA and SK-QAOA is essentially identical to the relation established in \cite{Dembo2017} between the ground state energy of the MaxCut Hamiltonian on a typical Erdos-Renyi graph of average degree $d$ and the ground state energy of the SK Hamiltonian on an average instance. It also establishes a parallel between QAOA and the Approximate Message Passing Algorithm developed in \cite{Montanari2019}, whose performances on both problems are similarly related. Note that equation \ref{eq:qaoa_sk_qaoa_maxcut_relation} is sharp only in the $d \to \infty$ limit; this limitation is also common in the random graph theory literature and our proof techniques unfortunately do not allow to quantify the error more sharply. Yet, as illustrated in numerical experiments (section \ref{subsec:numerical_experiments}), the result remains remarkably robust for finite $d$.

Another open question is whether the result generalizes to random graph ensembles closely related to $\er\left(n, \frac{d}{n - 1}\right)$, such as random $d$-regular (multi)graphs sampled from the configuration model \cite{Dembo2017}. In the last reference, the authors construct a coupling between (essentially) the Erdos-Renyi ensemble and the configuration model, so that all cuts $\bm{\sigma} \in \left\{\bm{\tau} \in \{-1, 1\}^n\,:\,\sum_{0 \leq j < n}\tau_j = 0\right\}$ evaluate to the same value up to a negligible error on almost all pairs $\left(G_{\er}, G_{\textnormal{conf}}\right)$ sampled from the coupling. The main roadblock to exploiting this result for QAOA is to show that the distributions sampled from QAOA applied to $G_{\er}$, $G_{\textnormal{conf}}$ are close enough with high probability on a pair $\left(G_{\er}, G_{\textnormal{conf}}\right)$ sampled from the coupling. Although we were not able to prove that, our numerical results suggest that the result from proposition \ref{th:qaoa_sk_qaoa_maxcut} can be successfully applied to low-degree random regular graphs beyond $p = 1$. Theorem \ref{th:qaoa_sk_qaoa_maxcut} is proven in section \ref{sec:analysis_maxcut_constant_degree} of the appendix. Essentially, it relies on an algorithm to compute the energy of QAOA on an average instance of Erdos-Renyi MaxCut, similar to the algorithm developed for the Sherrington-Kirkpatrick model in \cite{1910.08187}. In appendix \ref{sec:qaoa_maxcut_degree_distribution}, we further generalize the algorithm to random graphs with an expected degree distribution, via the pseudo Chung-Lu model (for Erdos-Renyi graphs, all vertices have the same expected degree). The details are in propositions \ref{prop:qaoa_energy_maxcut_degree_distribution} and \ref{prop:qaoa_energy_maxcut_degree_distribution_infinite_size} from the appendix. Unfortunately, in this case, we do not know of a simple relation like equation \ref{eq:qaoa_sk_qaoa_maxcut_relation} predicting the QAOA energy from the degree distribution in terms of a universal model.

Our second main result is a Monte-Carlo algorithm to estimate the QAOA energy on the Sherrington-Kirkpatrick model for a finite-size instance:
\begin{thm}[Estimating finite-size SK QAOA energy by Monte-Carlo]
\label{th:sk_finite_size_monte_carlo}
Let $n \geq 2$, $p \geq 1$ integers and $\bm\beta, \bm\gamma \in \mathbf{R}^p$ QAOA angles. Given these parameters, the randomized algorithm \ref{alg:sample_sk_energy} outputs a sample whose expectation equals the expectation QAOA energy on an average size-$n$ instance of the SK model:
\begin{align}
    & \mathbf{E}_{J \sim \textnormal{SK}(n)}\left[\braket{\Psi_{\textnormal{QAOA}}(J, \bm\beta, \bm\gamma)|H[J]|\Psi_{\textnormal{QAOA}}(J, \bm\beta, \bm\gamma)}\right]
\end{align}
and whose variance is upper-bounded by the output of algorithm \ref{alg:upper_bound_variance_sk_energy}.
\end{thm}
While the energy of QAOA on finite-size instances of the SK model can be exactly calculated for $p = 1$ (see \cite{1910.08187}), no analytic formulae are available for $p > 1$ to the best of our knowledge. Theorem \ref{th:sk_finite_size_monte_carlo} then provides a method of estimating the QAOA energy in this case. The proof is based on an analysis of the algorithm developed in \cite{1910.08187} before taking the infinite size limit. In this case, the QAOA energy takes the form of a $2^{2p + 1}$-variables sum which is too costly to evaluate exactly but can be estimated by importance sampling. Unfortunately, algorithms \ref{alg:sample_sk_energy} and \ref{alg:upper_bound_variance_sk_energy} are still limited: we numerically conjecture they break down for $p > 3$, where the upper bound on the variance output by algorithm \ref{alg:upper_bound_variance_sk_energy} blows up (see section \ref{sec:sk_energy_numerics} for details), as does the empirical variance of the samples output by algorithm \ref{alg:sample_sk_energy}. Besides, we do not expect algorithms \ref{alg:sample_sk_energy} and \ref{alg:upper_bound_variance_sk_energy} to generalize to the finite-size Erdos-Renyi MaxCut problem; the reason is that the exact expression for the energy of the SK model possesses symmetries that are crucial to the efficiency of the algorithms, but are only recovered in the infinite size limit for MaxCut.

\begin{algorithm}[!htbp]
	\caption{Sample finite-size SK-QAOA energy}
	\label{alg:sample_sk_energy}
	\KwData{Number of spins $n$, QAOA level $p$, QAOA angles $\bm{\beta}$, $\bm{\gamma}$}
	\KwResult{Energy sample $E$}
	$\mathcal{L}_{\textrm{list}} \leftarrow \textrm{bitstrings  } s \in \mathcal{L} - \mathcal{L}_p \textrm{ ordered by increasing level of symmetry}$\\
	\For{$s, t \in \{0, 1\}^{2p + 1}$}{
	    $M_{s, t} \leftarrow 1$\\
	    $\left(m_u\right)_{u \in \mathcal{L}_{\textrm{list}}} \leftarrow 1$\\
    	\For{$u \in \mathcal{L}_{\textrm{list}}$}{
    	    %\vspace*{-20px}
            \begin{align*}
                \lambda & \leftarrow nB_{\bm{\beta}, u}\left(-e^{-\frac{\varphi(\bm{\gamma}, s \oplus u)^2}{2n}}e^{-\frac{\varphi(\bm{\gamma}, t \oplus u)^2}{2n}}\prod_{\substack{v \in \mathcal{L}_{\textrm{list}}\\v < u}}e^{-\frac{\varphi(\bm{\gamma}, v \oplus u)^2}{2n}m_v}\right.\\
        	    & \left.\hspace*{170px}e^{-\frac{\varphi(\bm{\gamma}, s \oplus F(u))^2}{2n}}e^{-\frac{\varphi(\bm{\gamma}, F(u))^2}{2n}}\prod_{\substack{v \in \mathcal{L}_{\textrm{list}}\\v < u}}e^{-\frac{\varphi(\bm{\gamma}, v \oplus F(u))^2}{2n}m_v}\right)
        	\end{align*}
        	\If{$\lambda \neq 0$}{
        	    $m_u \leftarrow \textrm{Poisson}(|\lambda|)$\\
        	    $M_{s, t} \leftarrow M_{s, t}e^{|\lambda|}\left(\frac{\lambda}{|\lambda|}\right)^{m_u}$
        	}
    	}
    	%<\vspace*{-30px}
    	\begin{align*}
    	    \hspace*{-5px} M_{s, t} & \leftarrow M_{s, t}\frac{n!}{\left(n - \sum_{v \in \mathcal{L}_{\textrm{list}}}m_v\right)!n^{\sum_{v \in \mathcal{L}_{\textrm{list}}}m_v}}e^{-\frac{\varphi(\bm{\gamma}, s \oplus t)^2}{2n}}\\
    	    & \hspace*{20px} \times \left(\sum_{\substack{u \in \{0, 1\}^{2p + 1}\\L(u) = p}}(-1)^{u \in \mathcal{L}}B_{\bm{\beta}, u}e^{-\frac{\varphi(\bm{\gamma}, s \oplus u)^2}{2n}}e^{-\frac{\varphi(\bm{\gamma}, t \oplus u)^2}{2n}}\prod_{v \in \mathcal{L}_{\textrm{list}}}e^{-\frac{\varphi(\bm{\gamma}, u \oplus v)^2}{2n}m_v}\right)^{n - \sum_{v \in \mathcal{L}_{\textrm{list}}}m_v - 2}
    	\end{align*}
	}
	$E \leftarrow \Re\left\{\frac{i}{8}\sum_{s, t \in \{0, 1\}^{2p + 1}}(-1)^{s_p + t_p}B_{\bm{\beta}, s}B_{\bm{\beta}, t}(-1)^{s \in \mathcal{L}}(-1)^{t \in \mathcal{L}}\varphi(\bm{\gamma}, s \oplus t)M_{s, t}\right\}$\\
	\Return $E$
\end{algorithm}

\begin{algorithm}[!htbp]
    \caption{Upper-bound variance of sampled finite-size SK-QAOA energy}
    \label{alg:upper_bound_variance_sk_energy}
    \KwData{QAOA level $p$, QAOA angles $\bm{\beta}, \bm{\gamma}$}
    \KwResult{Upper bound $\Delta E_u^2$ on the variance of a sample generated by algorithm \ref{alg:sample_sk_energy}}
    \For{$s \in \mathcal{L}_p \sqcup \mathcal{L}_p'$}{
        $\hat{R}_s \leftarrow 1$
    }
    $\mathcal{L}_{\textrm{list}} \leftarrow \textrm{bitstrings } s \in \mathcal{L} - \mathcal{L}_p \textrm{ ordered by increasing level of symmetry}$\\
    \For{$s \in \mathcal{L}_{\textrm{list}}$}{
        $\hat{R}_s, \hat{R}_{F(s)} \leftarrow \exp\left(\frac{1}{4}\sum_{t \in \mathcal{L} - \mathcal{L}_p}\left|\varphi(\bm\gamma, s \oplus t)^2 - \varphi(\bm\gamma, s \oplus F(t))^2\right|\right)$
    }
    \For{$s \in \mathcal{L}_{\textrm{list}}$ (backwards)}{
        \For{$t \in \mathcal{L}_{\textrm{list}}, t < s$}{
            $\hat{R}_t, \hat{R}_{F(t)} \leftarrow \hat{R}_t\exp\left|\frac{1}{4}\hat{R}_sB_{\bm{\beta}, s}\left(\varphi(\bm{\gamma}, F(s) \oplus t)^2 - \varphi(\bm{\gamma}, t \oplus s)^2\right)\right|$ 
        }
    }
    $\Delta E_u^2 \leftarrow \frac{1}{8}\sum_{s, t \in \{0, 1\}^{2p + 1}}\left|\varphi(\bm{\gamma}, s \oplus t)\right|^2\left|B_{\bm{\beta}, s}\right|\hat{R}_s\left|B_{\bm{\beta}, t}\right|\hat{R}_t$\\
    \Return $\Delta E_u^2$
\end{algorithm}

Our last result is a generalization of theorems \ref{th:qaoa_erdos_renyi_energy} and \ref{th:sk_finite_size_monte_carlo} to $D$-spin models ($D \geq 3$), which is unfortunately restricted to $p = 1$.

\begin{prop}
\label{prop:qaoa_dense_qaoa_diluted}
Let $D \geq 3$. Let $d \geq 3$ denote the degree parameter of the diluted spin model. For large $d$, the QAOA energy of the diluted $D$-spin model in the infinite size limit is approximated as follows:
\begin{align}
    & \lim_{n \to \infty}\mathbf{E}_{J \sim \textnormal{Diluted}(D, n, d)}\left[\left\langle\Psi_{\textnormal{QAOA}}\left(J, \beta^{\textnormal{Dense}}, \frac{\gamma^{\textnormal{Dense}}}{\sqrt{(D - 1)!d}}\right)\bigg|\frac{H[J]}{nd/D}\bigg|\Psi_{\textnormal{QAOA}}\left(J, \beta^{\textnormal{Dense}}, \frac{\gamma^{\textnormal{Dense}}}{\sqrt{(D - 1)!d}}\right)\right\rangle\right]\nonumber\\
    & = -\frac{i\gamma^{\textnormal{Dense}}}{2\sqrt{(D - 1)!d}}\left[\left(\cos\beta^{\textnormal{Dense}} - i\sin\beta^{\textnormal{Dense}} e^{-\frac{\left(\gamma^{\textnormal{Dense}}\right)^2}{2(D - 1)!}}\right)^D - \left(\cos\beta^{\textnormal{Dense}} + i\sin\beta^{\textnormal{Dense}} e^{-\frac{\left(\gamma^{\textnormal{Dense}}\right)^2}{2(D - 1)!}}\right)^D\right] + \mathcal{O}\left(\frac{1}{d}\right)\nonumber\\
    & = \frac{1}{\sqrt{(D - 1)!d}}\lim_{n \to \infty}\mathbf{E}_{J \sim \textnormal{Dense}(D, n)}\left[\left\langle\Psi_{\textnormal{QAOA}}\left(J, \beta^{\textnormal{Dense}}, \gamma^{\textnormal{Dense}}\right)\bigg|\frac{H[J]}{n}\bigg|\Psi_{\textnormal{QAOA}}\left(J, \beta^{\textnormal{Dense}}, \gamma^{\textnormal{Dense}}\right)\right\rangle\right] + \mathcal{O}\left(\frac{1}{d}\right)
\end{align}
where the leading term is, up to the factor $\frac{1}{\sqrt{(D - 1)!d}}$, the energy achieved on the dense $D$-spin model by level-1 QAOA evaluated at angles $\beta^{\textnormal{Dense}}, \gamma^{\textnormal{Dense}} = O(1)$.
\end{prop}
This result relates the performance of level-1 QAOA on the diluted $D$-spin model of degree parameter $d$ to the performance on the dense $D$-spin model, provided the $\beta, \gamma$ angles follow an explicit scaling with respect to $d, D$. We expect the extension of this result to QAOA level $p > 1$ to be challenging and to require significantly different techniques from the ones used in this work.

\subsection{Numerical experiments}
\label{subsec:numerical_experiments}
We now validate the results stated in section \ref{subsec:theoretical_results}, and further detailed in the appendix, by numerical experiments. In section \ref{sec:maxcut_erdos_renyi_numerics}, we consider the infinite size energy of QAOA applied to MaxCut on random Erdos-Renyi and Chung-Lu graphs, illustrating proposition \ref{th:qaoa_sk_qaoa_maxcut} and propositions \ref{prop:qaoa_energy_maxcut_degree_distribution}, \ref{prop:qaoa_energy_maxcut_degree_distribution_infinite_size} from the appendix. Section \ref{sec:sk_energy_numerics} benchmarks our Monte-Carlo algorithm \ref{alg:sample_sk_energy} for estimating the finite-size energy of QAOA applied to the SK model. Finally, in section \ref{sec:finite_size_instances}, we explore using the optimal parameters obtained for infinite-size MaxCut-QAOA as starting points to optimize small instances.

\subsubsection{MaxCut-QAOA}
\label{sec:maxcut_erdos_renyi_numerics}
Theorem \ref{th:qaoa_sk_qaoa_maxcut} relates the energy of QAOA applied to MaxCut on a random Erdos-Renyi graph of size $n$ and average degree $d$ to the energy of QAOA applied to the size-$n$ SK model, in the limit $n \to \infty$, $d \to \infty$ (the limits being taken in this order). Unfortunately, our methods do not explicitly quantify errors when taking these limits. In this section, we therefore evaluate the validity of approximation \ref{eq:qaoa_sk_qaoa_maxcut_relation} for finite (and small) $d$.

The optimal angles of MaxCut-QAOA on Erdos-Renyi graphs of average degree $d$ were determined for various $d \in [4, 19]$ by optimizing the exact infinite-size energy stated in proposition \ref{prop:qaoa_infinite_size_limit_energy} from the appendix. In each instance, the variational parameters were optimized with the COBYLA algorithm, starting from random initial values $\left(\bm\beta_i, \bm\gamma_i\right)$ drawn uniformly from $\left[-\pi, \pi\right]^{2p}$. Results are shown for QAOA with $p \in \{1, 3, 5\}$ levels in figure \ref{fig:maxcut_erdos_renyi_parameters_scaling}; more examples are available in appendix \ref{sec:extra_figures}. The optimal $\bm\gamma$ angles $\gamma^*_j$ are rescaled by $\sqrt{d}$ so that the $\gamma^*_j\sqrt{d}$ is expected to converge in the $d \to \infty$ limit. These results suggest that the approximation in theorem \ref{th:qaoa_sk_qaoa_maxcut} is robust even for small $d$. The results obtained on random Chung-Lu graphs are also detailed in appendix \ref{sec:extra_figures}.

\begin{figure}[!htbp]
    \centering
    \begin{subfigure}{0.48\textwidth}
        \includegraphics[width=\textwidth]{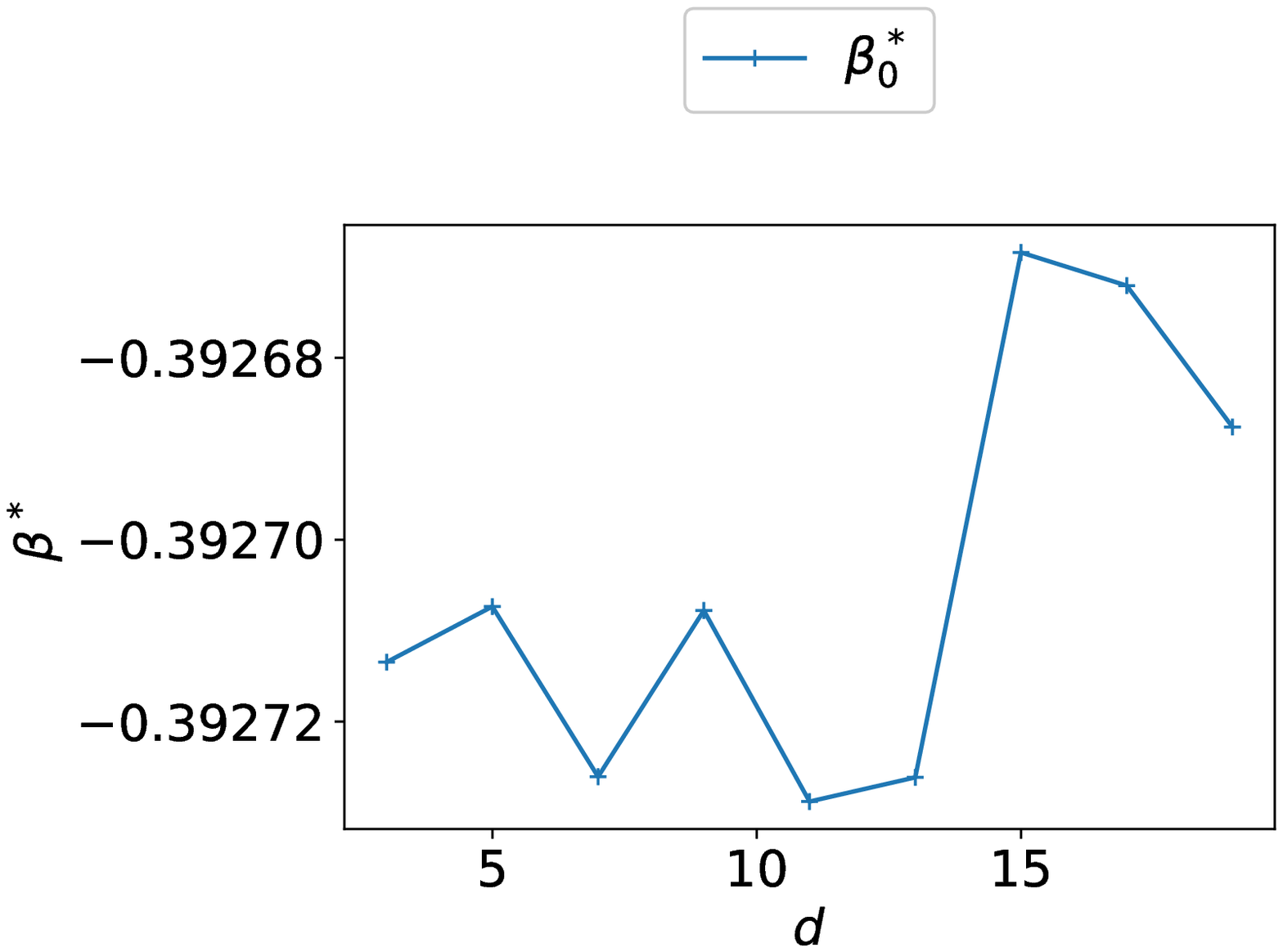}
        \caption{Optimal $\bm\beta$ angles at level $p = 1$.}
    \end{subfigure}
    \hspace*{0.03\textwidth}
    \begin{subfigure}{0.47\textwidth}
        \includegraphics[width=\textwidth]{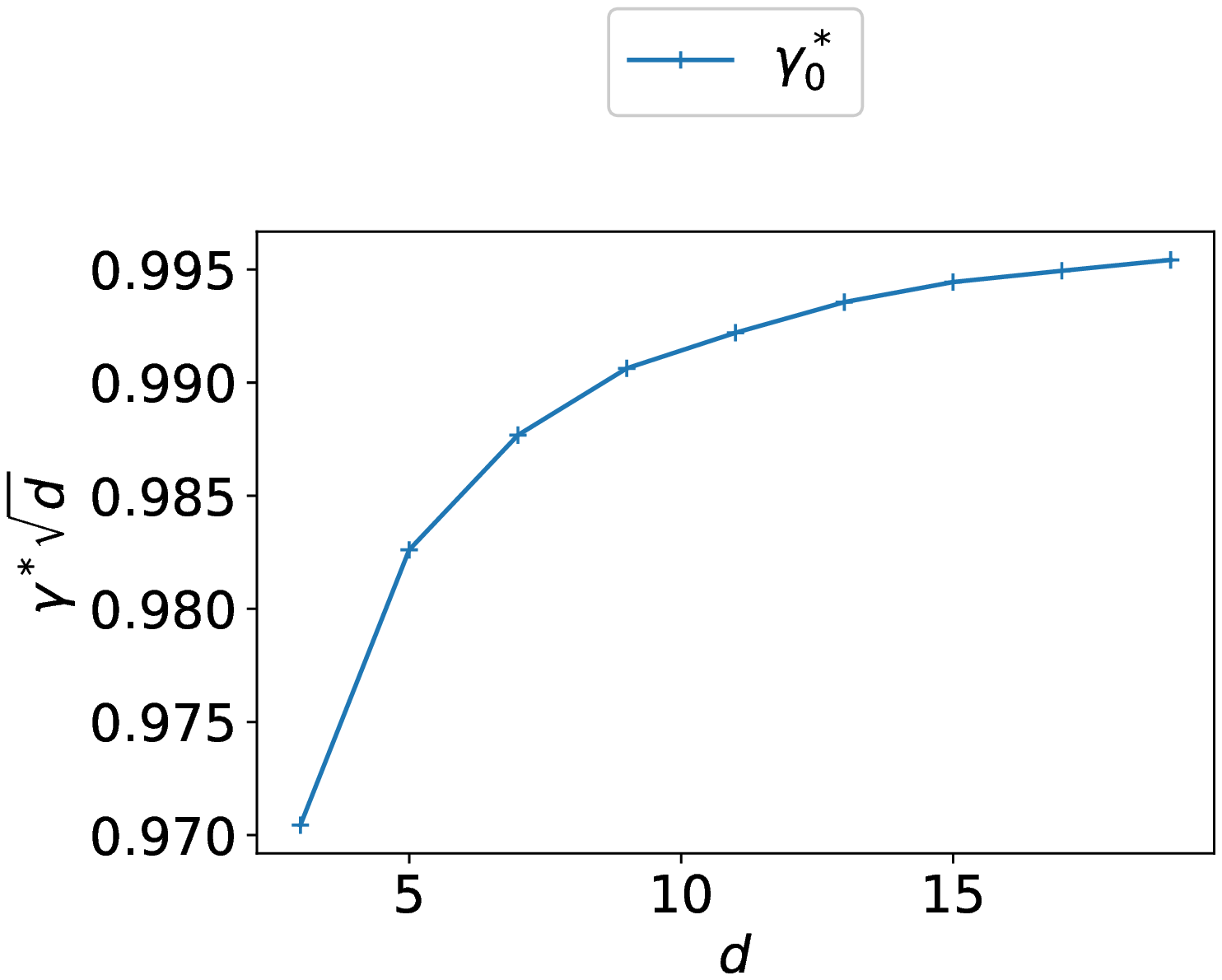}
        \caption{Optimal $\bm\gamma$ angles at level $p = 1$.}
    \end{subfigure}\\
    \vspace*{10px}
    \begin{subfigure}{0.48\textwidth}
        \includegraphics[width=\textwidth]{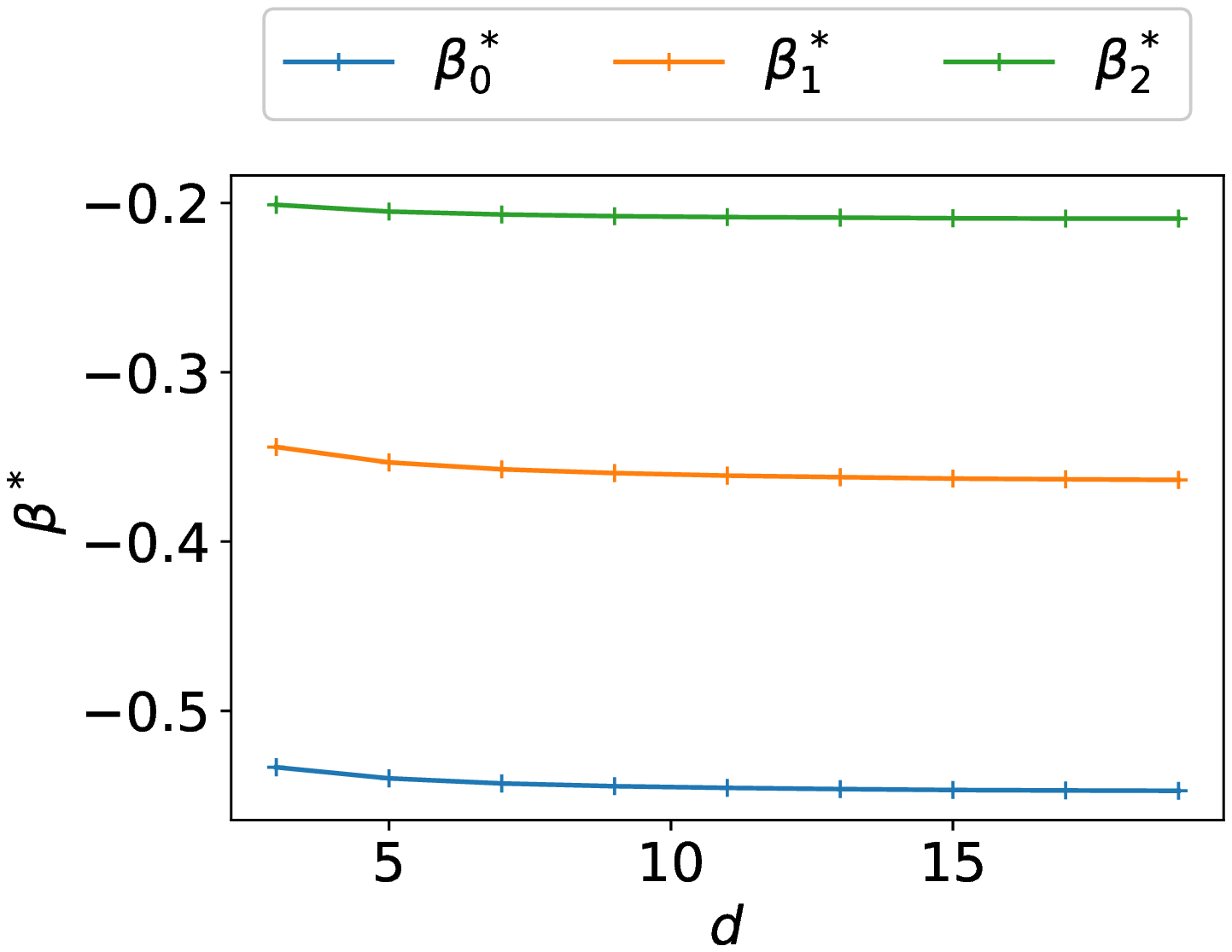}
        \caption{Optimal $\bm\beta$ angles at level $p = 3$.}
    \end{subfigure}
    \hspace*{0.03\textwidth}
    \begin{subfigure}{0.47\textwidth}
        \includegraphics[width=\textwidth]{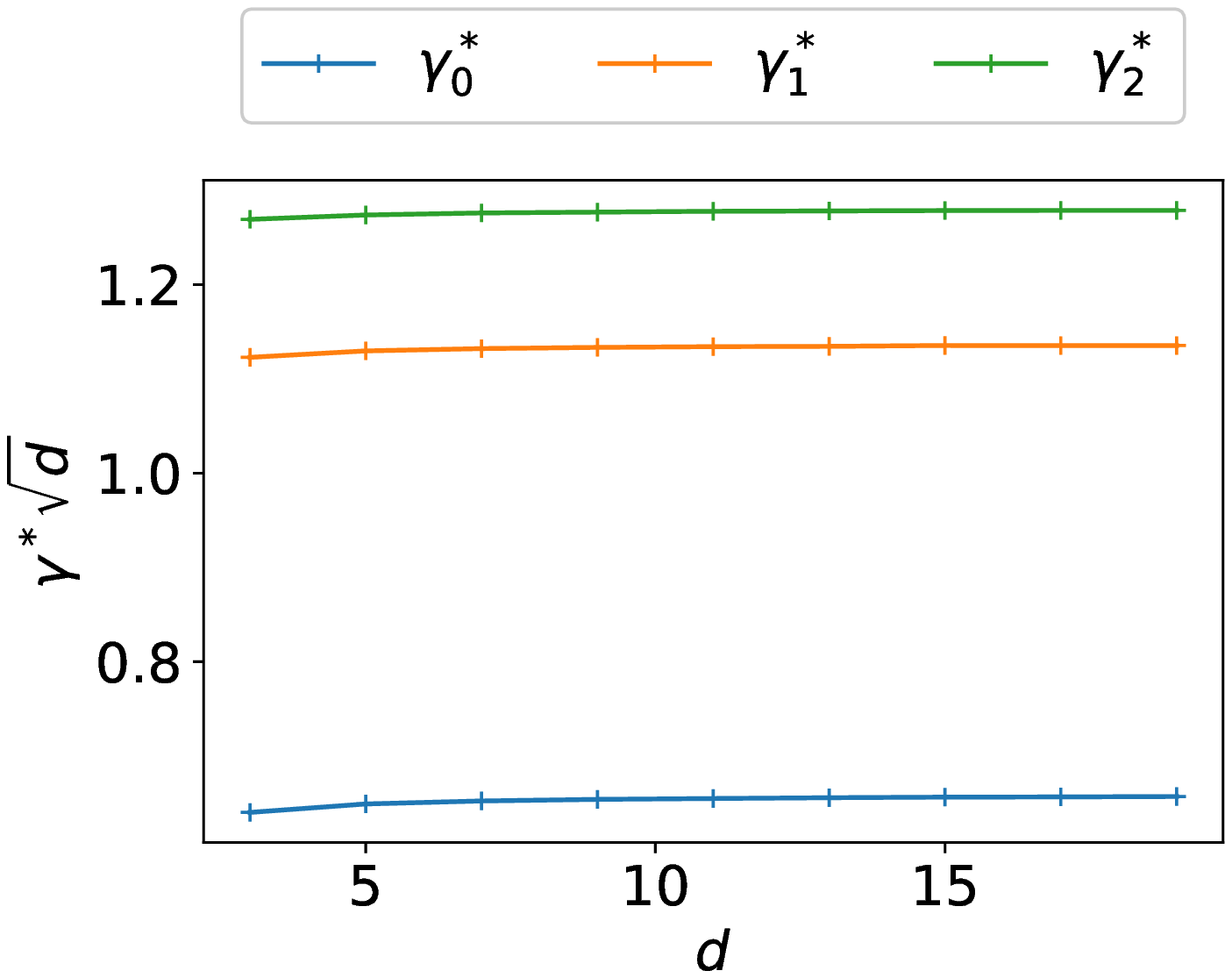}
        \caption{Optimal $\bm\gamma$ angles at level $p = 3$.}
    \end{subfigure}
    \vspace*{10px}
    \begin{subfigure}{0.48\textwidth}
        \includegraphics[width=\textwidth]{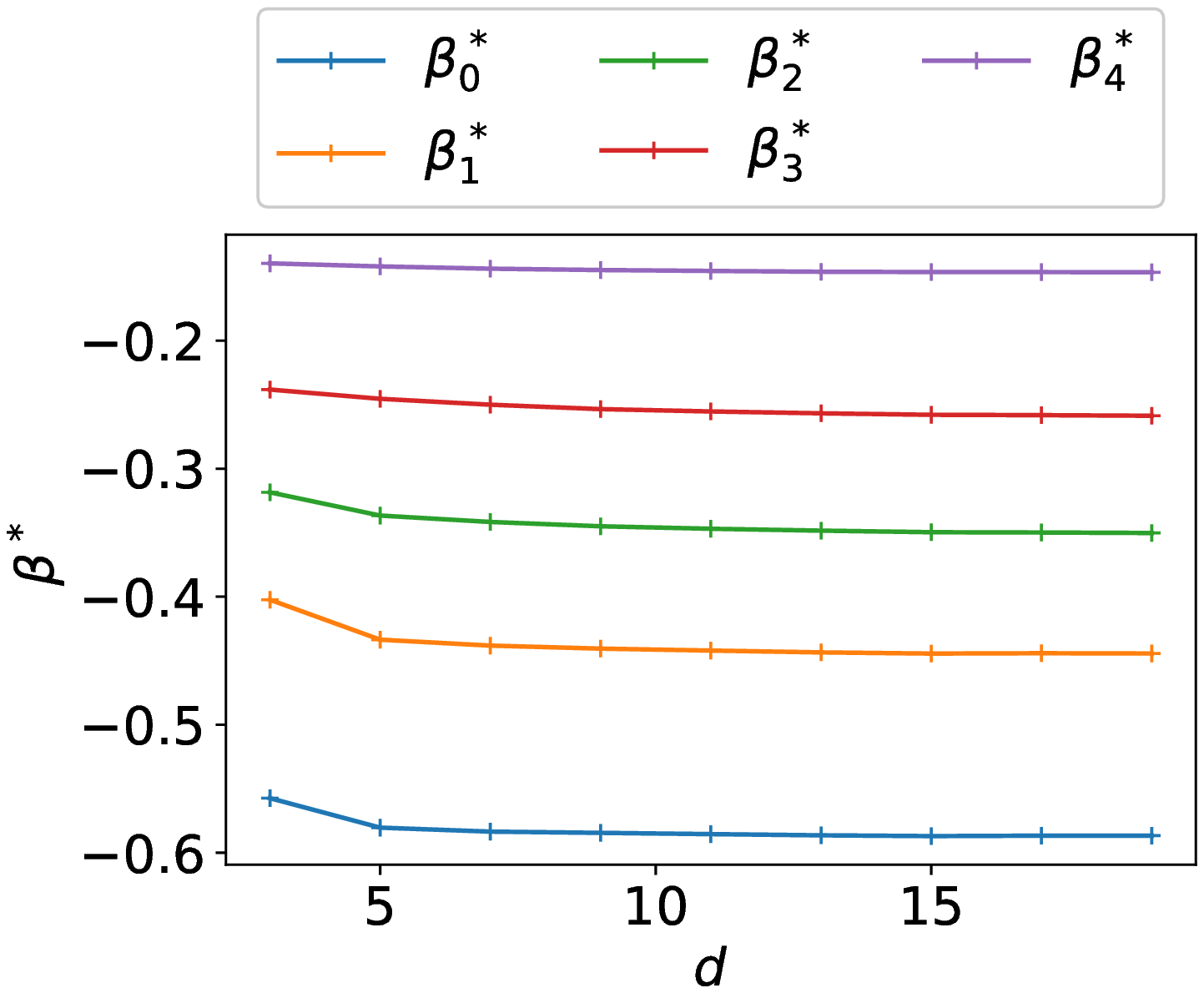}
        \caption{Optimal $\bm\beta$ angles at level $p = 5$. $\beta^*_0, \beta^*_1, \beta^*_2, \beta^*_3, \beta^*_4$ vary by 5.3\%, 10\%, 10\%, 8.6\%, 5.0\% on the $d$ range.}
    \end{subfigure}
    \hspace*{0.03\textwidth}
    \begin{subfigure}{0.47\textwidth}
        \includegraphics[width=\textwidth]{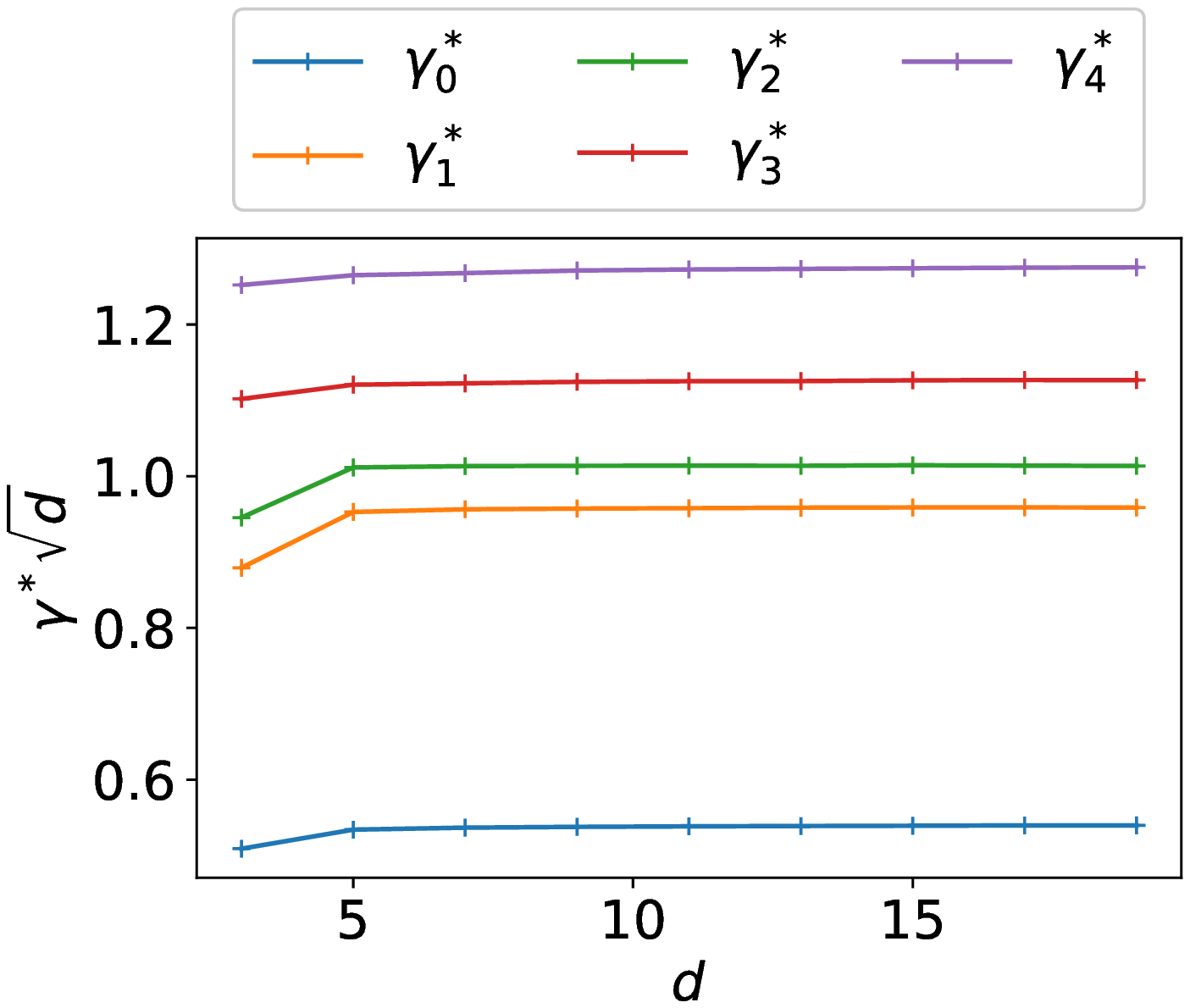}
        \caption{Optimal $\bm\gamma$ angles at level $p = 5$. $\gamma^*_0, \gamma^*_1, \gamma^*_2, \gamma^*_3, \gamma^*_4$ vary by 6.0\%, 9.0\%, 7.2\%, 2.3\%, 1.9\% on the $d$ range.}
    \end{subfigure}
    \caption{Scaling of optimal MaxCut-QAOA parameters for $p \in \{1, 3, 5\}$ and $d \in [3, 19]$. Each point was generated from 1000 optimization trials.}
    \label{fig:maxcut_erdos_renyi_parameters_scaling}
\end{figure}

\subsubsection{Finite size Sherrington-Kirkpatrick model energy}
\label{sec:sk_energy_numerics}
We now evaluate algorithms \ref{alg:sample_sk_energy}, \ref{alg:upper_bound_variance_sk_energy} introduced in section \ref{subsec:theoretical_results} to estimate the energy of QAOA at level $p > 1$ on an average finite-size instance of the Sherrington-Kirkpatrick model.

\begin{figure}[!htbp]
    \centering
    \begin{subfigure}{0.48\textwidth}
    \centering
    \includegraphics[width=\textwidth]{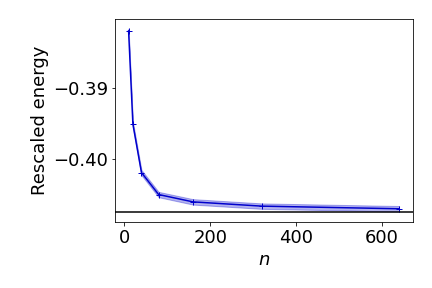}
    \caption{$p = 2$}
    \end{subfigure}
    \hspace*{0.03\textwidth}
    \begin{subfigure}{0.47\textwidth}
    \centering
    \includegraphics[width=\textwidth]{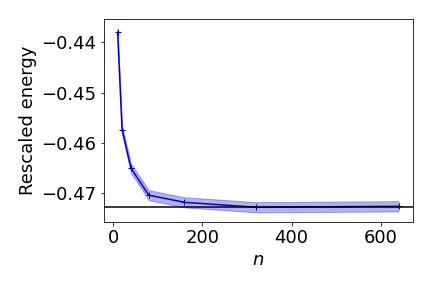}
    \caption{$p = 3$}
    \end{subfigure}
    \caption{Finite-size energy of QAOA on the Sherrington-Kirkpatrick model (black horizontal line: infinite-size limit)}
    \label{fig:finite_size_sk_qaoa}
\end{figure}

In figure \ref{fig:finite_size_sk_qaoa}, we represent the rescaled energy $\mathbf{E}_{J \sim \textrm{SK}(n)}\braket{\Psi_{\textrm{QAOA}}(J, \bm\beta, \bm\gamma)|\frac{H[J]}{n}|\Psi_{\textrm{QAOA}}(J, \bm\beta, \bm\gamma)}$ as a function of $n$, evaluated at the infinite-size optimal $\bm\beta, \bm\gamma$ angles reported in \cite{1910.08187}. Each point was obtained by averaging $10000$ samples generated by algorithm \ref{alg:sample_sk_energy}. The empirical standard deviation is represented by a shaded area around the curve. For comparison, the upper bound on the standard deviation following from algorithm \ref{alg:upper_bound_variance_sk_energy} is $\approx 0.03$ for $p = 2$ and $\approx 3$ for $p = 3$. Although these upper bounds are somehow pessimistic, they establish that estimating the QAOA energy by algorithm \ref{alg:sample_sk_energy} is feasible up to $p = 3$ (for the particular $\bm\beta, \bm\gamma$ chosen here). In contrast, at level $p = 4$, algorithm \ref{alg:upper_bound_variance_sk_energy} outputs an upper bound $\approx 10^{203}$ on the standard deviation of the sampled energy, suggesting that the method breaks down at this level. However, this upper bound, hence the performance of the Monte-Carlo algorithm \ref{alg:sample_sk_energy}, does depend on the QAOA angles $\bm\beta, \bm\gamma$. Indeed, we observed that for parameters $\bm\beta = \left(\beta_0, \beta_1, \beta_2, \beta_3\right), \bm\gamma = \left(\gamma_0, \gamma_1, \gamma_2, \gamma_3\right)$ chosen uniformly at random from $\left[-\pi, \pi\right]^8$, the upper bound from algorithm \ref{alg:upper_bound_variance_sk_energy} is frequently as low as $10$-$100$ instead of $10^{203}$ for optimal angles.

\subsubsection{Finite size MaxCut instances}
\label{sec:finite_size_instances}
Algorithms \ref{alg:sample_sk_energy} and \ref{alg:upper_bound_variance_sk_energy} that we just illustrated unfortunately apply only to the Sherrington-Kirkpatrick model and not to unweighted MaxCut. Hence, for the latter problem, it is not immediate to evaluate the performance of the optimal angles derived in section \ref{sec:maxcut_erdos_renyi_numerics} for the infinite limit on finite-size instances. Therefore, we validate the performance of these limiting angles in practice using small problem instances where QAOA is fully simulatable classically. More precisely, we considered the performance of level-$3$ QAOA on MaxCut applied to randomly generated graphs with $16$ vertices. The latters were generated either from the Erdos-Renyi or pseudo Chung-Lu model described in section \ref{sec:random_graphs_ensembles}, with expected degrees $\in \{4, 9\}$. Note that for this choice of values, QAOA ``sees the whole graph" \cite{2004.09002,2005.08747}, i.e. the QAOA ansatz unitary $U$ spreads every local operator $Z_j, j \in [n]$ over all vertices by conjugation $UZ_jU^{\dagger}$. Informally, this implies that QAOA is sensitive to the global structure of the graph; this stands in sharp contrast with the constant $p$, infinite size limit, where QAOA only probes the local structure of the graph \cite{1411.4028}. Despite this potential limitation, we numerically show that the optimal QAOA parameters for an average infinite-size instance still provide a good guess for the optimal angles on specific finite-size instances.

\begin{figure}[!t]
    \centering
    \begin{subfigure}{0.3\textwidth}
        \centering
        \includegraphics[width=\textwidth]{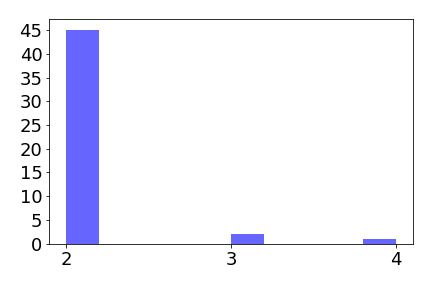}
        \caption{Number of attempts to outperform evaluation at guessed angles}
        \label{subfig:erdos_renyi_instances_num_eval_attempts}
    \end{subfigure}
    \hspace*{0.03\textwidth}
    \begin{subfigure}{0.3\textwidth}
        \centering
        \includegraphics[width=\textwidth]{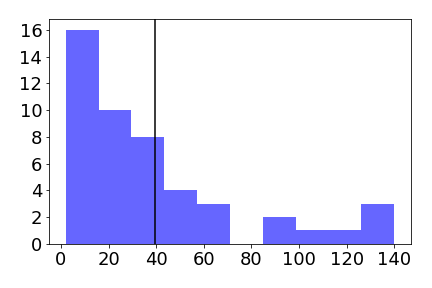}
        \caption{Number of attempts to outperform optimization from guessed angles (black vertical line: average $\approx 40$)}
        \label{subfig:erdos_renyi_instances_num_opt_attempts}
    \end{subfigure}
    \hspace*{0.03\textwidth}
    \begin{subfigure}{0.3\textwidth}
        \centering
        \includegraphics[width=\textwidth]{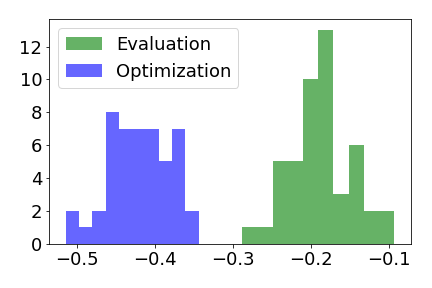}
        \caption{Energies evaluated at guessed angles vs. optimized}
        \label{subfig:erdos_renyi_instances_eval_vs_opt}
    \end{subfigure}
    \caption{Comparison of evaluation and optimization for Erdos-Renyi graphs of expected degree 4}
    \label{fig:erdos_renyi_evaluation_vs_optimization}
\end{figure}

\begin{figure}[!t]
    \centering
    \begin{subfigure}{0.3\textwidth}
        \centering
        \includegraphics[width=\textwidth]{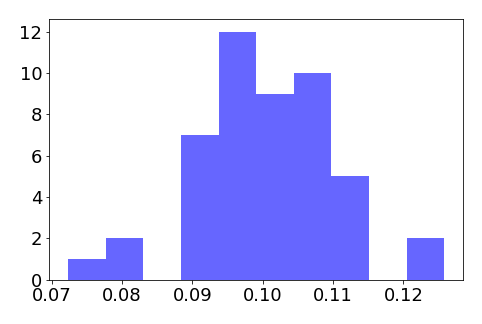}
        \caption{All angles (expected value for uniformly random angles: 0.57)}
    \end{subfigure}
    \hspace*{0.03\textwidth}
    \begin{subfigure}{0.3\textwidth}
        \centering
        \includegraphics[width=\textwidth]{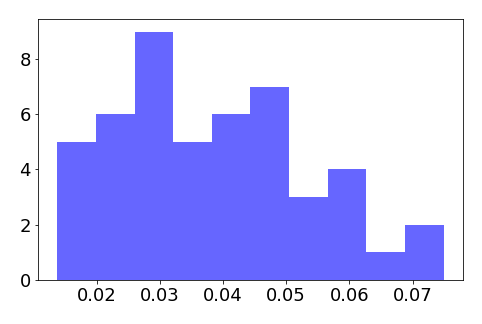}
        \caption{$\bm\beta$ angles (expected value for uniformly random angles: 0.55)}
    \end{subfigure}
    \hspace*{0.03\textwidth}
    \begin{subfigure}{0.3\textwidth}
        \centering
        \includegraphics[width=\textwidth]{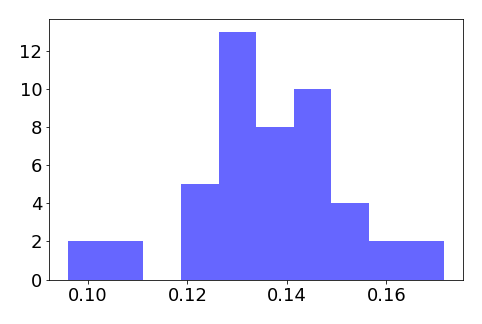}
        \caption{$\bm\gamma$ angles (expected value for uniformly random angles: 0.55)}
    \end{subfigure}
    \caption{Distance between guessed and optimal angles for Erdos-Renyi graphs of expected degree 4}
    \label{fig:erdos_renyi_instances_angle_distances}
\end{figure}

A first series of experiments considered 1000 randomly generated Erdos-Renyi graphs $G \sim \er\left(16, \frac{4}{15}\right)$ with $n = 16$ vertices and expected degree $d = 4$. The optimal angles $\bm{\beta^*}, \bm{\gamma^*} \in \mathbf{R}^3$ were guessed from the approximation given in theorem \ref{th:qaoa_sk_qaoa_maxcut} (i.e. rescaling by $\sqrt{d} = 2$ the optimal angles for the SK model). For each graph instance, the expected energy was evaluated at these angles. Then, these angles were used as an initial value to find the optimize the energy. Finally, the expected energy was repeatedly optimized again, starting from uniformly random initial angles $\in [-\pi, \pi]^6$. In all experiments, we used the L-BFGS optimizer \cite{Liu1989} provided by \verb|NLopt| \cite{Johnson2011} with at most 1000 iterations and attempted 1000 optimizations starting from random initial angles. For each of the 50 graphs, starting from guessed angles provided an optimum at least as good as starting from 1000 uniformly random angles. In addition, simply evaluating the ansatz at the guessed parameters (without any optimization) frequently led to better results than a few optimization attempts starting from uniformly random parameters. More precisely, in histograms \ref{subfig:erdos_renyi_instances_num_eval_attempts} and \ref{subfig:erdos_renyi_instances_num_opt_attempts}, we represent for all problem instances the number of optimization attempts required to outperform evaluation at the guessed angles and optimization starting from the guessed angles; the energies obtained by optimization and evaluation at the guessed parameters are compared in figure \ref{subfig:erdos_renyi_instances_eval_vs_opt}. On average, 40 attempts (140 in the worst case) are required to match optimization starting from the guessed angles. Finally, the distances between optimal angles and guessed angles are given in figure \ref{fig:erdos_renyi_instances_angle_distances}, both for the full vector of angles $\left(\bm\beta, \bm\gamma\right)$ and for $\bm\beta$, $\bm\gamma$ separately. In every case, angle vectors were standardized accounting for the symmetries of the QAOA cost function and distances were normalized to be at most $1$. Figure \ref{fig:erdos_renyi_instances_angle_distances} shows that the guessed parameters are appreciably close to the optimum, but the guess is more accurate for the $\bm\beta$ than for the $\bm\gamma$ angles.


\begin{figure}[!t]
    \centering
    \begin{subfigure}{0.3\textwidth}
        \centering
        \includegraphics[width=\textwidth]{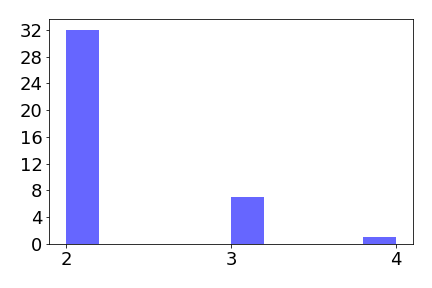}
        \caption{Number of attempts to outperform evaluation at guessed angles}
        \label{subfig:chung_lu_instances_num_attempts}
    \end{subfigure}
    \hspace*{0.03\textwidth}
    \begin{subfigure}{0.3\textwidth}
        \centering
        \includegraphics[width=\textwidth]{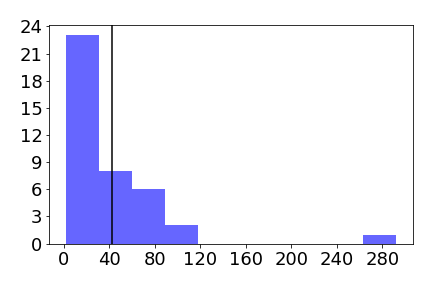}
        \caption{Number of attempts to outperform optimization from guessed angles (black vertical line: average $\approx 42$)}
        \label{subfig:chung_lu_instances_num_attempts}
    \end{subfigure}
    \hspace*{0.03\textwidth}
    \begin{subfigure}{0.3\textwidth}
        \centering
        \includegraphics[width=\textwidth]{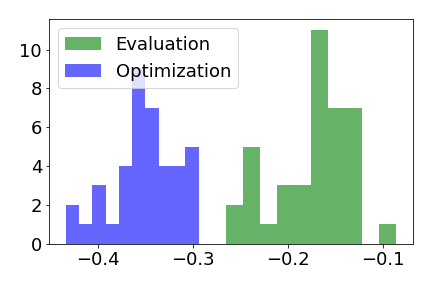}
        \caption{Energies evaluated at guessed angles vs. optimized}
        \label{subfig:chung_lu_instances_eval_vs_opt}
    \end{subfigure}
    \caption{Comparison of evaluation and optimization for Chung-Lu graphs of expected degrees 4 (probability $\frac{2}{3}$) and 9 (probability $\frac{1}{3}$)}
    \label{fig:chung_lu_evaluation_vs_optimization}
\end{figure}

\begin{figure}[!t]
    \centering
    \begin{subfigure}{0.3\textwidth}
        \centering
        \includegraphics[width=\textwidth]{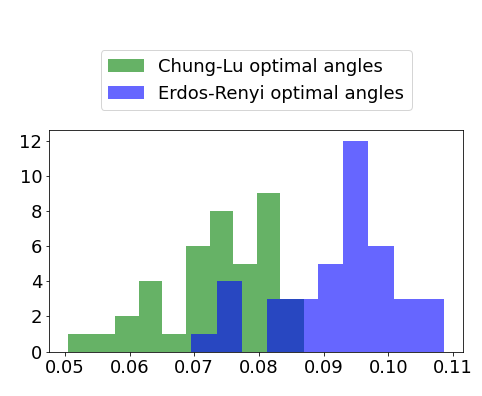}
        \caption{All angles (expected value for uniformly random angles: 0.57)}
        \label{subfig:chung_lu_instances_angle_distances_all}
    \end{subfigure}
    \hspace*{0.03\textwidth}
    \begin{subfigure}{0.3\textwidth}
        \centering
        \includegraphics[width=\textwidth]{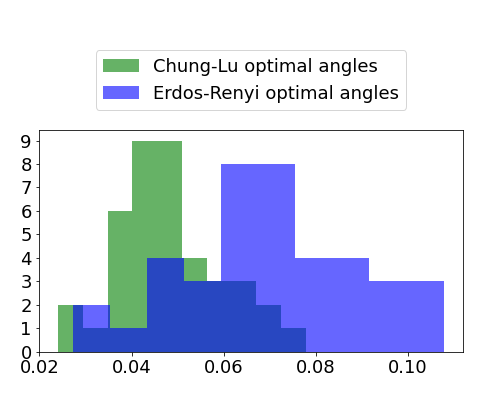}
        \caption{$\bm\beta$ angles (expected value for uniformly random angles: 0.55)}
        \label{subfig:chung_lu_instances_angle_distances_betas}
    \end{subfigure}
    \hspace*{0.03\textwidth}
    \begin{subfigure}{0.3\textwidth}
        \centering
        \includegraphics[width=\textwidth]{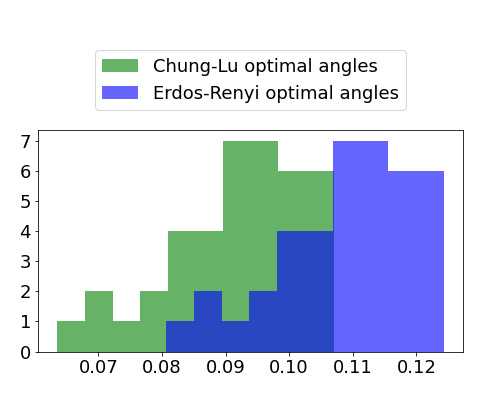}
        \caption{$\bm\gamma$ angles (expected value for uniformly random angles: 0.55)}
        \label{subfig:chung_lu_instances_angle_distances_gammas}
    \end{subfigure}
    \caption{Distance between guessed and optimal angles for Chung-Lu graphs of expected degrees 4 (probability $\frac{2}{3}$) and 9 (probability $\frac{1}{3}$)}
    \label{fig:chung_lu_instances_angle_distances}
\end{figure}

A similar analysis was performed for an ensemble of 50 random Chung-Lu graphs with $n = 16$ vertices, $\frac{2}{3}$ of which having expected degree $4$ and $\frac{1}{3}$ expected degree $9$ (i.e., following the notation of definition \ref{def:pseudo_chung_lu_model}: $N_L = 2$, $(d_1, d_2) = (4, 9)$ and $(q_1, q_2) = \left(\frac{2}{3}, \frac{1}{3}\right)$). A difference with the Erdos-Renyi case is that two sets of guessed angles are considered: the first one corresponds to the optimal variational parameters for Chung-Lu graphs in the infinite size limit (see section \ref{sec:maxcut_erdos_renyi_numerics}); the second one is given by the optimal angles for an Erdos-Renyi graph of expected degree $\sum_{1 \leq l \leq N_L}q_ld_l = \frac{2}{3} \times 4 + \frac{1}{3} \times 9 = \frac{17}{3}$. The results are reported in figures \ref{fig:chung_lu_evaluation_vs_optimization} and \ref{fig:chung_lu_instances_angle_distances}. Figure \ref{fig:chung_lu_instances_angle_distances} shows that using the knowledge of the degree distribution leads to guessed parameters closer to the optimum, both overall (figure \ref{subfig:chung_lu_instances_angle_distances_all}) and when looking at $\bm\beta$ (figure \ref{subfig:chung_lu_instances_angle_distances_betas}), $\bm\gamma$ (figure \ref{subfig:chung_lu_instances_angle_distances_gammas}) individually.
\iffalse
In particular, figure \ref{subfig:chung_lu_instances_angle_distances_all} shows that using the knowledge of the degree distribution overall leads to guessed parameters closer to the optimum; however, distinguishing between mixing angles $\bm\beta$ (figure \ref{subfig:chung_lu_instances_angle_distances_betas}) and dephasing angles $\bm\gamma$ (figure \ref{subfig:chung_lu_instances_angle_distances_gammas}), using the degree distribution yields a better guess for the former but a worse one for the latter.
\fi


\subsection{Summary of techniques}

We now give an overview of the techniques used to derive the theoretical results introduced in section \ref{subsec:theoretical_results}.

In this work, similar to \cite{1910.08187}, the energy achieved by QAOA for fixed variational angles averaged over problem instances is considered. Therefore, the problem Hamiltonian $H_C$ (see section \ref{sec:qaoa}) is regarded as a random variable. The analysis starts (section \ref{sec:preliminaries}) by establishing a general expression (proposition \ref{prop:expectation_permutation_invariant_average_hamiltonian}) for the QAOA energy given $H_C$ is invariant under qubit permutations in the following sense: for all permutations $\sigma \in \mathfrak{S}_{2^n}$ induced by a permutation of $n$ qubits, and with associated permutation matrix $P_{\sigma}$ acting on $\mathbf{C}^{2^n}$, $P_{\sigma}H_CP_{\sigma^{-1}}$ has the same distribution as $H_C$. The proof of this result regards the QAOA energy as a contracted tensor network; choosing an appropriate contraction order allows to single out the contribution of the random Hamiltonians $H_C$, separating it from mixer Hamiltonians $H_B$. The problem Hamiltonian and mixer Hamiltonian contributions are then formally expressed in a convenient basis (equivalent to the \textit{configuration basis} introduced in \cite{1910.08187}) adapted to the symmetry of the random Hamiltonian. This results in a general expression which is then specialized to several examples where $H_C$ satisfies the required symmetry assumption: from general to particular, dense or diluted $D$-spin models for arbitrary $D$ (section \ref{sec:qaoa_moments_dense_diluted}), MAX-CUT on sparse random Erdos-Renyi graphs (section \ref{sec:analysis_maxcut_constant_degree}) and MAX-CUT on random Chung-Lu graphs (section \ref{sec:qaoa_maxcut_degree_distribution}).

Specializing proposition \ref{prop:expectation_permutation_invariant_average_hamiltonian} to the latter random optimization problems leads to a formula for the energy achieved by QAOA that can be (classically) evaluated in time $\mathcal{O}\left(n^{2^{2p + 1}}\right)$ for problem size $n$ and QAOA level $p$, similar to the results obtained in \cite{1910.08187} for the Sherrington-Kirkpatrick model. We emphasize that despite this similarity, our method is more systematic and expected to apply to random problems beyond the spin models considered in this work. Although the complexity of evaluating the formula is polynomial and not exponential in $n$, the task remains intractable even for moderate $n$ and $p$. We then adapt techniques from \cite{1910.08187} to obtain a more efficiently computable expression (time $\mathcal{O}\left(16^p\right)$ and space $\mathcal{O}\left(4^p\right)$) in the infinite-size limit $n \longrightarrow \infty$. This formula requires to evaluate the quantities $R_s$ according to algorithm \ref{alg:compute_r}. In the case of MAX-CUT on a sparse Erdos-Renyi graph of expected degree $d$ (section \ref{sec:analysis_maxcut_constant_degree}), a careful analysis of the $R_s$ for the variational parameters stated in theorem \ref{th:qaoa_sk_qaoa_maxcut}: $\bm\beta := \bm\beta^{SK}, \bm\gamma := \frac{\bm\gamma^{SK}}{\sqrt{d}}$ shows they coincide with the analogous numbers (here denoted $\widetilde{R}_s$) defined in \cite{1910.08187} for the Sherrington-Kirkpatrick model in the limit $d \longrightarrow \infty$. Our methods hold for arbitrary $p$ and large $d$, but presumably fail to precisely quantify the quality of the convergence for finite $d$; more precisely, they imply one may need $d = \Omega\left(16^p\right)$, which is clearly pessimistic given the numerical results from section \ref{sec:maxcut_erdos_renyi_numerics}. A similar analysis can be carried out to relate the energy of the diluted $D$-spin model to that of the dense $D$-spin model (section \ref{sec:analysis_diluted_d_spin_model}), but is restricted to $p = 1$.

Finally, the Monte-Carlo algorithm \ref{alg:sample_sk_energy} is analyzed in section \ref{sec:analysis_sk_monte_carlo} by referring to the exact formula for the energy of QAOA on the Sherrington-Kirkpatrick model at arbitrary constant $p$, averaged over instances, established in \cite{1910.08187}. Informally, we interpret this expression as an expectation over $4^p$ (non-independent) random Poisson variables of \textit{complex} parameters; this is recast to an expectation over standard Poisson variables with positive real parameters, at the cost of many extra normalization factors (the latters are responsible for the blow-up of the variance with increasing $p$).

\subsection{Conclusions}
In this work, generalizing techniques from \cite{1910.08187}, we evaluated the performance of QAOA in the infinite-size limit for the MAX-CUT problem applied to different families of sparse random graphs. While the analysis only rigorously holds in the $n \to \infty$ and large degree limit, it provides good enough initial variational parameters to optimize small-scale and very sparse instances. The latter finding suggests that more efficient training algorithms could be designed for QAOA or other variational quantum algorithms by gaining a detailed theoretical understanding of toy problems, even with impractical assumptions. In addition, we proposed a numerical algorithm to evaluate the performance of QAOA on an average finite-size instance of the SK model, where the analytic methods exposed in \cite{1910.08187} fail; however, the algorithm remains limited to low depth. It would be desirable to identify more cases where the performance of QAOA can be classically evaluated on large problem instances (beyond full classical simulation). Finally, from a theoretical point of view, a shortcoming of our work is its inability to address the $D$-spin model for $D \geq 3$ and $p > 1$. This case would be of considerably higher fundamental interest (given non-approximability by shallow classical \cite{Chen2019} and quantum \cite{2108.06049} local algorithms), but we expect the analysis to require significantly different techniques from those used in this work.

\subsection*{Acknowledgements}

This project has received funding from the European Research Council (ERC) under the European Union's Horizon 2020 research and innovation programme (grant agreement No.\ 817581). Google Cloud credits were provided by Google via the EPSRC Prosperity Partnership in Quantum Software for Modeling and Simulation (EP/S005021/1). This work was supported by the EPSRC Centre for Doctoral Training in Delivering Quantum Technologies, grant ref. EP/S021582/1. The data used to generate the figures is available at \url{https://github.com/sami-b95/predicting_qaoa_parameters_data}.

% ------------------------------------------------------------------------------

\bibliographystyle{hapalike}
\bibliography{bibliography}

% ------------------------------------------------------------------------------

\appendix

\section{Technical derivations}
\subsection{Preliminaries}
\label{sec:preliminaries}
In this section, we derive a general expression for the moment-generating function of QAOA on an average instance for random optimization problems satisfying a certain permutation invariance condition. The latter will be formalized by the introduction of an explicitly permutation-invariant basis of a subspace of the tensor power of the Hilbert space, called \textit{configuration basis}. The name \textit{configuration basis} is in reference to \cite{1910.08187}, which evaluates the energy of QAOA on the Sherrington-Kirkpatrick model at infinite size; the authors take advantage of the permutation symmetry of the problem by grouping tensor products of computational basis states under what they call ``configurations" and summing over these configurations. This is in essence equivalent to the approach developed here. However, the formalism used in this work differs as it accounts for the permutation invariance at an earlier stage,. The resulting gain in generality facilitates the exposition of different random optimization problems without repetition and gives further insight into the problems addressable by this method.

To derive an expression for the moments of QAOA on an average instance, the first step will be to reorganize the tensor network describing these moments. For that purpose, we will occasionally regard matrices acting on $\mathcal{H}$ as a vector from $\mathcal{H} \otimes \mathcal{H}$, an identification commonly referred to as \textit{vectorization}:
\begin{defn}[Matrix vectorization]
Given a matrix $A$ acting on an $n$-qubit space: $A := \sum_{x, y \in \{0, 1\}^n}A_{xy}\ket{x}\bra{y}$, the vectorization of $A$, denoted by $\Vect(A)$, as:
\begin{align}
    \Vect(A) & := \sum_{x, y \in \{0, 1\}^n}A_{xy}\ket{x}\ket{y}.
\end{align}
\end{defn}
From this definition, the moment-generating function of QAOA can be rewritten using standard graphical methods for tensor networks \cite{Bridgeman2017}:
\begin{lem}[QAOA moment-generating function, rewritten]
\label{prop:qaoa_tn_reorganized}
\begin{align}
\label{eq:qaoa_tn_reorganized}
    & \braket{\Psi_{\textnormal{QAOA}}(H_C, \bm{\beta}, \bm{\gamma})|e^{i\frac{\gamma'}{2}H_C}|\Psi_{\textnormal{QAOA}}(H_C, \bm{\beta}, \bm{\gamma})}\nonumber\\
    & = \left(\bigotimes_{0 \leq j < \lceil p/2 \rceil}\Vect\left(e^{i\frac{\beta_{2j}}{2}\sum_{0 \leq k < n}X_k}\right)^T \otimes \bigotimes_{0 \leq j < \lfloor p/2 \rfloor}\Vect\left(e^{-i\frac{\beta_{2\lfloor p/2 \rfloor - 1 - 2j}}{2}\sum_{0 \leq k < n}X_k}\right)^T \otimes \bra{+}\right)\nonumber\\
    & \hspace*{30px} \bigotimes_{0 \leq j < p}e^{i\frac{\gamma_j}{2}H_C} \otimes e^{i\frac{\gamma'}{2}H_C} \otimes \bigotimes_{0 \leq j < p}e^{-i\frac{\gamma_{p - 1 - j}}{2}H_C}\nonumber\\
    & \left(\ket{+} \otimes \bigotimes_{0 \leq j < \lfloor p/2\rfloor}\Vect\left(e^{i\frac{\beta_{1 + 2j}}{2}\sum_{0 \leq k < n}X_k}\right) \otimes \bigotimes_{0 \leq j < \lceil p/2 \rceil}\Vect\left(e^{-i\frac{\beta_{2\lceil p/2 \rceil - 2 - 2j}}{2}\sum_{0 \leq k < n}X_k}\right)\right)
\end{align}
\end{lem}
To analyze the latter expression further, we will express the vector and operator in the middle in the configuration basis, defined below.
\begin{defn}[Configuration basis]
Let $n \geq 1$ and $r \geq 1$ be integers. The \textit{configuration basis} on $n$ qubits at level $r$ is spanned by the vectors
\begin{align*}
    \ket{(n_s)_{s \in \{0, 1\}^r}} & := \frac{1}{\sqrt{\binom{n}{\left(n_s\right)_{s \in \{0, 1\}^r}}}}\sum_{\substack{x_{r - 1}, \ldots, x_0 \in \{0, 1\}^n\\\forall s \in \{0, 1\}^r,\,\left|\left\{j \in [n]\,:\,\left((x_{r - 1})_j, \ldots, (x_0)_j\right) = s\right\}\right| = n_s}}\ket{x_{r - 1}} \ldots \ket{x_0},
\end{align*}
where $\left(n_s\right)_{s \in \{0, 1\}^r}$ is a $2^r$-tuple of non-negative integers summing to $n$  and, following the notation from \cite{1910.08187}, $\binom{n}{\left(n_s\right)_s} = \frac{n!}{\prod_{s \in \{0, 1\}^r}n_s!}$ is a multinomial coefficient.
\end{defn}
These states account for the invariance under qubit permutations of random problem instances such as the Sherrington-Kirkpatrick model discussed in \cite{1910.08187}. More precisely, any vector from $\left(\mathbf{C}^{2^n}\right)^{\otimes r}$ that is invariant under all permutations $\sigma^{\otimes r}$, where $\sigma$ is an arbitrary $n$-qubit permutation, can be decomposed in this basis; similarly, any operator commuting with all such permutations is diagonal in this basis. For instance, for $r = 1$, vectors $\left(\ket{n_0, n_1}\right)_{\substack{n_0, n_1 \geq 0\\n_0 + n_1 = n}}$ span the symmetric $n$-qubit subspace; precisely, $\ket{n_0, n_1}$ is the uniform superposition of $n$-bit bitstrings of Hamming weight $n_1$. Here, we used the notational shortcut $\ket{n_0, n_1}$ for $\ket{(n_0, n_1)}$ and will continue to do so in the following. As another example, for $n = 1$, the states $\ket{(n_s)_{s \in \{0, 1\}^r}}$ are the product states $\ket{b_{r - 1}} \otimes \ldots \otimes \ket{b_0}$ where a single bit $b_j$ is $1$. We will now explicitly express the vectors and operator appearing in equation \ref{eq:qaoa_tn_reorganized} in this basis.

We start with $\Vect\left(e^{-i\frac{\beta}{2}\sum_{0 \leq k < n}X_k}\right)$:
\begin{lem}
    \begin{align}
        & \ket{\Vect\left(e^{-i\frac{\beta}{2}\sum_{0 \leq k < n}X_k}\right)}\nonumber\\
        & = \sum_{\substack{n_{00}, n_{01}, n_{10}, n_{11} \geq 0\\n_{00} + n_{01} + n_{10} + n_{11} = n}}\sqrt{\binom{n}{\left(n_s\right)_{s \in \{0, 1\}^2}}}(-1)^{n_{01} + n_{10}}\left(\cos\frac{\beta}{2}\right)^{n_{00} + n_{11}}\left(i\sin\frac{\beta}{2}\right)^{n_{01} + n_{10}}\nonumber\\
        & \hspace*{300px} \ket{n_{00}, n_{01}, n_{10}, n_{11}}
    \end{align}
\begin{proof}
This can be inferred from the coefficients of the Wigner D-matrix, see e.g. \cite[appendix II]{rose_1957}. Alternatively, we give a self-contained proof here.
\begin{align*}
    e^{-i\frac{\beta}{2}\sum_{0 \leq k < n}X_k} & = \sum_{0 \leq j \leq n}e^{-i\frac{\beta}{2}(n - 2j)}\sum_{\substack{z \in \{+, -\}^n\\z\textrm{ has } j -}}\ket{z}\bra{z}\\
    & = \sum_{0 \leq j < n}e^{-i\frac{\beta}{2}(n - 2j)}\sum_{x, y \in \{0, 1\}^n}\sum_{\substack{z \in \{+, -\}^n\\z\textrm{ has } j -}}\braket{x|z}\braket{z|y}\ket{x}\bra{y}
\end{align*}
Now, let us fix $x, y \in \{0, 1\}^n$ and evaluate $\sum_z\braket{x|z}\braket{z|y}$. For $s \in \{0, 1\}^2$, let $n_s := \left\{j \in [n]\,:\,(x_j, y_j) = s\right\}$. By invariance of $\sum_z\ket{z}\bra{z}$ under qubit permutations, one may assume:
\begin{align*}
    x & = 0^{n_{00} + n_{01}}1^{n_{10} + n_{11}}\\
    y & = 0^{n_{00}}1^{n_{01}}0^{n_{10}}1^{n_{11}}
\end{align*}
Then, choosing the $j$ $-$ of bitstring $z$ amounts to choosing $j_{00}$ $-$ among the first $n_{00}$ qubits, $j_{01}$ among the next $n_{01}$ qubits, etc., so that $j_{00} + j_{01} + j_{10} + j_{11} = j$. For such a choice,
\begin{align*}
    \braket{x|z}\braket{z|y} & = \frac{1}{2^n}(-1)^{j_{01} + j_{10}}
\end{align*}
Therefore,
\begin{align*}
    \sum_{\substack{z \in \{+, -\}^n\\z\textrm{ has } j -}}\braket{x|z}\braket{z|y} & = \sum_{\substack{j_{00}, j_{01}, j_{10}, j_{11}\\j_{00} + j_{01} + j_{10} + j_{11} = j}}\binom{n_{00}}{j_{00}}\binom{n_{01}}{j_{01}}\binom{n_{10}}{j_{10}}\binom{n_{11}}{j_{11}}\frac{1}{2^n}(-1)^{j_{01} + j_{10}}
\end{align*}
and
\begin{align*}
    & e^{-i\frac{\beta}{2}\sum_{0 \leq k < n}X_k}\\
    & = \sum_{0 \leq j < n}e^{-i\frac{\beta}{2}(n - 2j)}\sum_{\substack{n_{00}, n_{01}, n_{10}, n_{11} \geq 0\\n_{00} + n_{01} + n_{10} + n_{11} = n}}\sum_{\substack{x, y \in \{0, 1\}^n\\\forall s \in \{0, 1\}^2\,\left|\left\{j \in [n]\,:\,(x_j, y_j) = s\right\}\right| = n_s}}\ket{x}\bra{y}\\
    & \hspace*{30px} \times \sum_{\substack{j_{00}, j_{01}, j_{10}, j_{11}\\j_{00} + j_{01} + j_{10} + j_{11} = j}}\binom{n_{00}}{j_{00}}\binom{n_{01}}{j_{01}}\binom{n_{10}}{j_{10}}\binom{n_{11}}{j_{11}}\frac{1}{2^n}(-1)^{j_{01} + j_{10}}\\
    & = \sum_{\substack{n_{00}, n_{01}, n_{10}, n_{11} \geq 0\\n_{00} + n_{01} + n_{10} + n_{11} = n}}\sum_{\substack{x, y \in \{0, 1\}^n\\\forall s \in \{0, 1\}^2\,\left|\left\{j \in [n]\,:\,(x_j, y_j) = s\right\}\right| = n_s}}\ket{x}\bra{y}\\
    & \hspace*{30px} \times \sum_{\substack{j_{00}, j_{01}, j_{10}, j_{11} \geq 0}}\binom{n_{00}}{j_{00}}\binom{n_{01}}{j_{01}}\binom{n_{10}}{j_{10}}\binom{n_{11}}{j_{11}}\frac{1}{2^n}(-1)^{j_{01} + j_{10}}e^{-i\frac{\beta}{2}(n - 2j_{00} - 2j_{01} - 2j_{10} - 2j_{11})}\\
    & = \sum_{\substack{n_{00}, n_{01}, n_{10}, n_{11} \geq 0\\n_{00} + n_{01} + n_{10} + n_{11} = n}}\sum_{\substack{x, y \in \{0, 1\}^n\\\forall s \in \{0, 1\}^2\,\left|\left\{j \in [n]\,:\,(x_j, y_j) = s\right\}\right| = n_s}}\ket{x}\bra{y}\frac{e^{-i\frac{\beta}{2}n}}{2^n}\left(1 + e^{i\beta}\right)^{n_{00} + n_{11}}\left(1 - e^{i\beta}\right)^{n_{01} + n_{10}}\\
    & = \sum_{\substack{n_{00}, n_{01}, n_{10}, n_{11} \geq 0\\n_{00} + n_{01} + n_{10} + n_{11} = n}}\left(\cos\frac{\beta}{2}\right)^{n_{00} + n_{11}}\left(i\sin\frac{\beta}{2}\right)^{n_{01} + n_{10}}(-1)^{n_{01} + n_{10}}\sum_{\substack{x, y \in \{0, 1\}^n\\\forall s \in \{0, 1\}^2\,\left|\left\{j \in [n]\,:\,(x_j, y_j) = s\right\}\right| = n_s}}\ket{x}\bra{y}.
\end{align*}
The result follows.
\end{proof}
\end{lem}
Using the decomposition just established, we will express the tensor product vectors appearing in equation \ref{eq:qaoa_tn_reorganized} in the configuration basis. This is done recursively, eliminating one factor of the tensor product after the other. The general step of the recursion uses the following two lemmas:
\begin{lem}
\label{lemma:extended_number_vector_dot_mixer}
    \begin{align}
        & \bra{\left(n_s\right)_{s \in \{0, 1\}^r}}\left\{\mathbf{1}_2^{\otimes (r - 2)} \otimes \ket{\Vect\left(e^{-i\frac{\beta}{2}\sum_{0 \leq k < n}X_k}\right)}\right\}\nonumber\\
        & = (-1)^{\sum_{s' \in \{0, 1\}^{r - 2}}(n_{s'01} + n_{s'10})}\left(\cos\frac{\beta}{2}\right)^{\sum_{s' \in \{0, 1\}^{r - 2}}(n_{s'00} + n_{s'11})}\left(i\sin\frac{\beta}{2}\right)^{\sum_{s' \in \{0, 1\}^{r - 2}}(n_{s'01} + n_{s'10})}\nonumber\\
        & \hspace*{20px} \sqrt{\prod_{s' \in \{0, 1\}^{r - 2}}\binom{\sum_{s \in \{0, 1\}^2}n_{s's}}{\left(n_{s's}\right)_{s \in \{0, 1\}^2}}}\left\langle\left(\sum_{s \in \{0, 1\}^2}n_{s's}\right)_{s' \in \{0, 1\}^{r - 2}}\right|
        \label{eq:extended_number_vector_dot_right_mixer}
    \end{align}
    Similarly,
    \begin{align}
        & \bra{\left(n_s\right)_{s \in \{0, 1\}^r}}\left\{\ket{\Vect\left(e^{-i\frac{\beta}{2}\sum_{0 \leq k < n}X_k}\right)} \otimes \mathbf{1}_2^{\otimes (r - 2)}\right\}\nonumber\\
        & = (-1)^{\sum_{s' \in \{0, 1\}^{r - 2}}(n_{01s'} + n_{10s'})}\left(\cos\frac{\beta}{2}\right)^{\sum_{s' \in \{0, 1\}^{r - 2}}(n_{00s'} + n_{11s'})}\left(i\sin\frac{\beta}{2}\right)^{\sum_{s' \in \{0, 1\}^{r - 2}}(n_{01s'} + n_{10s'})}\nonumber\\
        & \hspace*{20px} \sqrt{\prod_{s' \in \{0, 1\}^{r - 2}}\binom{\sum_{s \in \{0, 1\}^2}n_{ss'}}{\left(n_{ss'}\right)_{s \in \{0, 1\}^2}}}\left\langle\left(\sum_{s \in \{0, 1\}^2}n_{ss'}\right)_{s' \in \{0, 1\}^{r - 2}}\right|
        \label{eq:extended_number_vector_dot_left_mixer}
    \end{align}
    \begin{proof}
        \begin{align}
            & \ket{\Vect\left(e^{-i\frac{\beta}{2}\sum_{0 \leq k < n}X_k}\right)}\nonumber\\
            & = \sum_{\substack{n'_{00}, n'_{01}, n'_{10}, n'_{11} \geq 0\\n'_{00} + n'_{01} + n'_{10} + n'_{11} = n}}\sqrt{\binom{n}{\left(n'_s\right)_{s \in \{0, 1\}^2}}}(-1)^{n'_{01} + n'_{10}}\left(\cos\frac{\beta}{2}\right)^{n'_{00} + n'_{11}}\left(i\sin\frac{\beta}{2}\right)^{n'_{01} + n'_{10}}\ket{\left(n'_s\right)_{s \in \{0, 1\}^2}}\nonumber\\
            & = \sum_{\substack{n'_{00}, n'_{01}, n'_{10}, n'_{11} \geq 0\\y_0, y_1 \in \{0, 1\}^n\\n'_{00} + n'_{01} + n'_{10} + n'_{11} = n\\\forall s \in \{0, 1\}^2,\,\left|\{k \in [n]\,:\,(y_0)_k = s_0, (y_1)_k = s_1\}\right| = n'_s}}(-1)^{n'_{01} + n'_{10}}\left(\cos\frac{\beta}{2}\right)^{n'_{00} + n'_{11}}\left(i\sin\frac{\beta}{2}\right)^{n'_{01} + n'_{10}}\ket{y_1}\ket{y_0}\label{eq:vec_mixer_comp_basis}
        \end{align}
        Fix $x_0, \ldots, x_{r - 1} \in \{0, 1\}^n$ such that $\forall s \in \{0, 1\}^r$, $\left|\{k \in [n]\,:\,\forall 0 \leq j < r, (x_j)_k = s_j\}\right| = n_s$. One wants to compute
        \begin{align}
            \left(\bra{x_{r - 1}} \otimes \ldots \otimes \bra{x_1} \otimes \bra{x_0}\right)\left(\mathbf{1}^{\otimes (r - 2)} \otimes \ket{\Vect\left(e^{-i\frac{\beta}{2}\sum_{0 \leq k < n}X_k}\right)}\right),
        \end{align}
        which reduces to computing
        \begin{align}
            \left(\bra{x_1} \otimes \bra{x_0}\right)\ket{\Vect\left(e^{-i\frac{\beta}{2}\sum_{0 \leq k < n}X_k}\right)}
        \end{align}
        The terms from the sum \ref{eq:vec_mixer_comp_basis} expressing $\ket{\Vect\left(e^{-i\frac{\beta}{2}\sum_{0 \leq k < n}X_k}\right)}$ that contribute are the following:
        \begin{align}
            y_0 & = x_0\\
            y_1 & = x_1\\
            n'_s & = \sum_{s' \in \{0, 1\}^{r - 2}}n_{s's} \qquad \forall s \in \{0, 1\}^2
        \end{align}
        Therefore,
        \begin{align}
            & \left(\bra{x_{r - 1}} \otimes \ldots \otimes \bra{x_0}\right)\left(\mathbf{1}^{\otimes (r - 2)} \otimes \ket{\Vect\left(e^{-i\frac{\beta}{2}\sum_{0 \leq k < n}X_k}\right)}\right)\nonumber\\
            & = (-1)^{\sum_{s' \in \{0, 1\}^{r - 2}}(n_{s'01} + n_{s'10})}\left(\cos\frac{\beta}{2}\right)^{\sum_{s' \in \{0, 1\}^{r - 2}}(n_{s'00} + n_{s'11})}\left(i\sin\frac{\beta}{2}\right)^{\sum_{s' \in \{0, 1\}^{r - 2}}(n_{s'01} + n_{s'10})}\nonumber\\
            & \hspace*{280px} \times \bra{x_{r - 1}} \otimes \ldots \otimes \bra{x_2}
        \end{align}
        (Note that the right-hand side does not explicitly depend on $x_1, x_0$.) Equation \ref{eq:extended_number_vector_dot_right_mixer} then follows from simple algebra and counting after recalling the definition of the generalized number states.

        Equation \ref{eq:extended_number_vector_dot_left_mixer} is proved very similarly.
    \end{proof}
\end{lem}

\begin{lem}
\label{lemma:extended_number_vector_plus}
\begin{align}
    \braket{+|n_0, n_1} & = \frac{1}{\sqrt{2^n}}\sqrt{\binom{n}{n_0, n_1}} = \frac{1}{\sqrt{2^n}}\sqrt{\binom{n}{n_0}} = \frac{1}{\sqrt{2^n}}\sqrt{\binom{n}{n_1}}
\end{align}
\begin{proof}
Simply follows from grouping bitstrings in $\ket{+} = \frac{1}{\sqrt{2^n}}\sum_{k \in \{0, 1\}^n}\ket{k}$ by Hamming weights.
\end{proof}
\end{lem}

In the first equality above, we denoted the binomial coefficient without parentheses: $\binom{n}{n_0, n_1} = \binom{n}{\left(n_0, n_1\right)}$; we will often use this notational shortcut for bi- and multinomial coefficients in the following.

Applying the latter two lemmas recursively gives the following decomposition of the vectors in equation \ref{eq:qaoa_tn_reorganized} in the configuration basis:
\begin{lem}
\label{lemma:tensor_product_extended_number_basis}
\begin{align}
    & \bra{\left(n_s\right)_{s \in \{0, 1\}^{2p + 1}}}\left(\ket{+} \otimes \bigotimes_{0 \leq j < \lfloor p/2\rfloor}\Vect\left(e^{i\frac{\beta_{1 + 2j}}{2}\sum_{0 \leq k < n}X_k}\right) \otimes \bigotimes_{0 \leq j < \lceil p/2 \rceil}\Vect\left(e^{-i\frac{\beta_{2\lceil p/2 \rceil - 2 - 2j}}{2}\sum_{0 \leq k < n}X_k}\right)\right)\nonumber\\
    & = \frac{1}{\sqrt{2^n}}\sqrt{\binom{n}{\left(n_s\right)_{s \in \{0, 1\}^{2p + 1}}}}\nonumber\\
    & \hspace*{10px} \times \prod_{0 \leq j < \lfloor p/2 \rfloor}\left(\cos\frac{\beta_{1 + 2j}}{2}\right)^{\sum_{\substack{s \in \{0, 1\}^{2p + 1}\\s_{2p - 1 - 2j} = s_{2p - 2 - 2j}}}n_s}\left(i\sin\frac{\beta_{1 + 2j}}{2}\right)^{\sum_{\substack{s \in \{0, 1\}^{2p + 1}\\s_{2p - 1 - 2j} \neq s_{2p - 2 - 2j}}}n_s}\nonumber\\
    & \hspace*{10px} \times \prod_{0 \leq j < \lceil p/2 \rceil}\left(\cos\frac{\beta_{2\lceil p/2 \rceil - 2 - 2j}}{2}\right)^{\sum_{\substack{s \in \{0, 1\}^{2p + 1}\\s_{2\lceil p/2 \rceil - 1 - 2j} = s_{2\lceil p/2 \rceil - 2 - 2j}}}n_s}\left(i\sin\frac{\beta_{2\lceil p/2 \rceil - 2 - 2j}}{2}\right)^{\sum_{\substack{s \in \{0, 1\}^{2p + 1}\\s_{2\lceil p/2 \rceil - 1 - 2j} \neq s_{2\lceil p/2 \rceil - 2 - 2j}}}n_s}\nonumber\\
    & \hspace*{10px} \times (-1)^{\sum_{0 \leq j < \lceil p/2 \rceil}\sum_{\substack{s \in \{0, 1\}^{2p + 1}\\s_{2\lceil p/2 \rceil - 1 - 2j} \neq s_{2\lceil p/2 \rceil - 2 - 2j}}}n_s}
\end{align}
Similarly:
\begin{align}
    & \left(\bigotimes_{0 \leq j < \lceil p/2 \rceil}\Vect\left(e^{i\frac{\beta_{2j}}{2}\sum_{0 \leq k < n}X_k}\right) \otimes \bigotimes_{0 \leq j < \lfloor p/2 \rfloor}\Vect\left(e^{-i\frac{\beta_{2\lfloor p/2 \rfloor - 1 - 2j}}{2}\sum_{0 \leq k < n}X_k}\right) \otimes \bra{+}\right)\ket{\left(n_s\right)_{s \in \{0, 1\}^{2p + 1}}}\nonumber\\
    & = \frac{1}{\sqrt{2^n}}\sqrt{\binom{n}{\left(n_s\right)_{s \in \{0, 1\}^{2p + 1}}}}\nonumber\\
    & \hspace*{10px} \times \prod_{0 \leq j < \lceil p/2 \rceil}\left(\cos\frac{\beta_{2j}}{2}\right)^{\sum_{\substack{s \in \{0, 1\}^{2p + 1}\\s_{2p - 2j} = s_{2p - 2j - 1}}}n_s}\left(i\sin\frac{\beta_{2j}}{2}\right)^{\sum_{\substack{s \in \{0, 1\}^{2p + 1}\\s_{2p - 2j} \neq s_{2p - 2j - 1}}}n_s}\nonumber\\
    & \hspace*{10px} \times \prod_{0 \leq j < \lfloor p/2 \rfloor}\left(\cos\frac{\beta_{2\lfloor p/2 \rfloor - 1 - 2j}}{2}\right)^{\sum_{\substack{s \in \{0, 1\}^{2p + 1}\\s_{2\lfloor p/2 \rfloor - 2j} = s_{2\lfloor p/2 \rfloor - 1 - 2j}}}n_s}\left(i\sin\frac{\beta_{2\lfloor p/2 \rfloor - 1 - 2j}}{2}\right)^{\sum_{\substack{s \in \{0, 1\}^{2p + 1}\\s_{2\lfloor p/2 \rfloor - 2j} \neq s_{2\lfloor p/2 \rfloor - 1 - 2j}}}n_s}\nonumber\\
    & \hspace*{10px} \times (-1)^{\sum_{0 \leq j < \lfloor p/2 \rfloor}\sum_{\substack{s \in \{0, 1\}^{2p + 1}\\s_{2\lfloor p/2 \rfloor - 2j} \neq s_{2\lfloor p/2 \rfloor - 1 - 2j}}}n_s}
\end{align}
\begin{proof}
Both identities result from repeatedly applying lemma \ref{lemma:extended_number_vector_dot_mixer} (to eliminate vectorized QAOA mixers one by one) and lemma \ref{lemma:extended_number_vector_plus} (to eliminate the $\ket{+}$ state).
\end{proof}
\end{lem}
Combining the latter two identities gives the simpler result:
\begin{prop}
\label{prop:mixers_generalized_number_basis}
    Using the notations $\mathcal{L}$ and $B_{\bm{\beta}, s}$ introduced in definitions \ref{def:odd_bitstrings} and \ref{def:B_phi}:
    \begin{align}
        \mathcal{L} & := \left\{s \in \{0, 1\}^{2p + 1}\,:\,s_0 \neq s_p\right\}\\
         B_{\bm{\beta}, s} & := \prod_{0 \leq j < p}\left(\cos\beta_j\right)^{(s_j = s_{j + 1}) + (s_{2p - j} = s_{2p - j - 1})}\left(i\sin\beta_j\right)^{(s_j \neq s_{j + 1}) + (s_{2p - j} \neq s_{2p - j - 1})}
    \end{align}
    the following identity holds:
    \begin{align}
        & \left(\bigotimes_{0 \leq j < \lceil p/2 \rceil}\Vect\left(e^{i\frac{\beta_{2j}}{2}\sum_{0 \leq k < n}X_k}\right)^T \otimes \bigotimes_{0 \leq j < \lfloor p/2 \rfloor}\Vect\left(e^{-i\frac{\beta_{2\lfloor p/2 \rfloor - 1 - 2j}}{2}\sum_{0 \leq k < n}X_k}\right)^T \otimes \bra{+}\right)\nonumber\\
        & \hspace*{50px}\ket{\left(n_s\right)_{s \in \{0, 1\}^{2p + 1}}}\bra{\left(n_s\right)_{s \in \{0, 1\}^{2p + 1}}}\nonumber\\
        & \left(\ket{+} \otimes \bigotimes_{0 \leq j < \lfloor p/2\rfloor}\Vect\left(e^{i\frac{\beta_{1 + 2j}}{2}\sum_{0 \leq k < n}X_k}\right) \otimes \bigotimes_{0 \leq j < \lceil p/2 \rceil}\Vect\left(e^{-i\frac{\beta_{2\lceil p/2 \rceil - 2 - 2j}}{2}\sum_{0 \leq k < n}X_k}\right)\right)\nonumber\\
        & = \frac{1}{2^n}\binom{n}{\left(n_s\right)_{s \in \{0, 1\}^{2p + 1}}}(-1)^{\sum_{\substack{0 \leq l \leq p\\s \in \mathcal{L}_l}}n_s}\nonumber\\
        & \hspace*{10px} \times \prod_{0 \leq j < p}\left(\cos\frac{\beta_j}{2}\right)^{\sum_{\substack{s \in \{0, 1\}^{2p + 1}\\s_j = s_{j + 1}}}n_s + \sum_{\substack{s \in \{0, 1\}^{2p + 1}\\s_{2p - j} = s_{2p - j - 1}}}n_s}\left(i\sin\frac{\beta_j}{2}\right)^{\sum_{\substack{s \in \{0, 1\}^{2p + 1}\\s_j \neq s_{j + 1}}}n_s + \sum_{\substack{s \in \{0, 1\}^{2p + 1}\\s_{2p - j} \neq s_{2p - j - 1}}}n_s}\\
        & = \frac{1}{2^n}\binom{n}{\left(n_s\right)_{s \in \{0, 1\}^{2p + 1}}}(-1)^{\sum_{\substack{s \in \mathcal{L}}}n_s}\prod_{s \in \{0, 1\}^{2p + 1}}B_{\bm{\beta}, s}^{n_s}.
    \end{align}
\end{prop}

This completes the decomposition of the vectors from equation \ref{eq:qaoa_tn_reorganized} in the configuration basis. It remains to decompose $\bigotimes_{0 \leq j < p}e^{i\frac{\gamma_j}{2}H_C} \otimes e^{i\frac{\gamma'}{2}H_C} \otimes \bigotimes_{0 \leq j < p}e^{-i\frac{\gamma_{p - 1 - j}}{2}H_C}$. In general, this operator does not even stabilize $\Span\{\ket{n_s}\}_{s \in \{0, 1\}^{2p + 1}}$; however, as will be detailed in the next sections for several examples, averaging it over problem instances makes it diagonal in this basis. (Informally, this is because the random graph models considered here are invariant under vertex permutations.) For the moment, let us assume the latter and denote by $\braket{(n_s)_s|\mathbf{E}_{H_C}\left[\bigotimes_{0 \leq j < p}e^{i\frac{\gamma_j}{2}H_C} \otimes e^{i\frac{\gamma'}{2}H_C} \otimes \bigotimes_{0 \leq j < p}e^{-i\frac{\gamma_j}{2}H_C}\right]|(n_s)_s}$ the (diagonal) matrix coefficients of the averaged Hamiltonian tensor. Combining proposition \ref{prop:mixers_generalized_number_basis} and lemma \ref{prop:qaoa_tn_reorganized} then gives the following expression for the moment-generating function of QAOA on an average instance:

\begin{prop}
\label{prop:expectation_permutation_invariant_average_hamiltonian}
Assume $\mathbf{E}_{H_C}\left[\bigotimes_{0 \leq j < p}e^{i\frac{\gamma_j}{2}H_C} \otimes e^{i\frac{\gamma'}{2}H_C} \otimes \bigotimes_{0 \leq j < p}e^{-i\frac{\gamma_j}{2}H_C}\right]$ is diagonal in the configuration basis $\left(\ket{(n_s)_{s \in \{0, 1\}^{2p + 1}}}\right)_{\substack{n_s \geq 0\\\sum_sn_s = n}}$. Then the following holds:
\begin{align}
    & \mathbf{E}_{H_C}\braket{\Psi_{\textnormal{QAOA}}(H_C, \bm{\beta}, \bm{\gamma})|e^{i\frac{\gamma'}{2}H_C}|\Psi_{\textnormal{QAOA}}(H_C, \bm{\beta}, \bm{\gamma})}\nonumber\\
    & = \sum_{\substack{(n_s)_{s \in \{0, 1\}^{2p + 1}}\\\sum_sn_s = n}}\frac{1}{2^n}\binom{n}{\left(n_s\right)_s}(-1)^{\sum_{s \in \mathcal{L}}n_s}\prod_sB_{\bm\beta, s}^{n_s}\braket{(n_s)_s|\mathbf{E}_{H_C}\left[\bigotimes_{0 \leq j < p}e^{i\frac{\gamma_j}{2}H_C} \otimes e^{i\frac{\gamma'}{2}H_C} \otimes \bigotimes_{0 \leq j < p}e^{-i\frac{\gamma_j}{2}H_C}\right]|(n_s)_s}
\end{align}
\end{prop}

In the next section, we will then need to show that the averaged Hamiltonian tensor is diagonal in the configuration basis. For that purpose, we will use the following technical proposition \ref{prop:average_hamiltonian_tensor} and its simpler corollary \ref{cor:average_hamiltonian_tensor}. Following the statement of these results, we immediately give a simple application of the corollary. Proposition \ref{prop:average_hamiltonian_tensor} requires to apply a complex-valued function of several complex variables $f\left(z_1, \ldots, z_k\right)$ to matrix arguments $M_1, \ldots, M_k$. This operation is well-defined if for instance the matrices $M_1, \ldots, M_k$ are Hermitian and pairwise commute. In this case, given a common eigenvector $\ket{\psi_{j_1, \ldots, j_k}}$ of $M_1, \ldots, M_k$ with respective eigenvalues $\lambda^{(1)}_{j_1}, \ldots, \lambda^{(k)}_{j_k}$, the operator $f\left(M_1, \ldots, M_k\right)$ is defined to act on $\ket{\psi_{j_1, \ldots, j_k}}$ as the scalar $f\left(\lambda^{(1)}_{j_1}, \ldots, \lambda^{(k)}_{j_k}\right)$.

\begin{prop}
\label{prop:average_hamiltonian_tensor}
Let $p \geq 1$, $D \geq 2$, $n \geq D$ and $N_L \geq 1$ be integers. Let $\left(q_l\right)_{1 \leq l \leq N_L}$ be a probability distribution on $\{1, \ldots, N_L\}$. Let $\left(f_{(D_1, D_2, \ldots, D_{N_L})}\right)_{\substack{D_1, \ldots, D_{N_L}\geq 0\\\sum_{1 \leq l \leq N_L}D_l = D}}$ be a family of functions $\{-1, 1\}^{2p + 1} \longrightarrow \mathbf{C}$ and indexed by $N_L$-tuples of nonnegative integers summing to $D$. The following identity holds:
\begin{align}
    & \sum_{\substack{(I_l)_{1 \leq l \leq N_L}\\I_l \subset [n]\\\sqcup_{1 \leq l \leq N_L}I_l = [n]}}\binom{n}{\left(|I_l|\right)_{1 \leq l \leq N_L}}\prod_{1 \leq l \leq N_L}q_l^{|I_l|}\nonumber\\
    & \hspace*{50px} \times \prod_{\{k_1, \ldots, k_D\} \subset [n]}f_{\left(\left|\{k_1, \ldots, k_D\} \cap I_l\right|\right)_{1 \leq l \leq N_L}}\left(\left(Z_{k_D} \ldots Z_{k_1}\right)_{2p}, \ldots, \left(Z_{k_D} \ldots Z_{k_1}\right)_0\right)\ket{\left(n_s\right)_s}\nonumber\\
    & = \sum_{\substack{\left(n_{s, l}\right)_{\substack{s \in \{0, 1\}^{2p + 1}, 1 \leq l \leq N_L}}\\\sum_{l}n_{s, l} = n_s}}\prod_s\binom{n_s}{\left(n_{s, l}\right)_l}\prod_lq_l^{n_{s, l}}\nonumber\\
    & \hspace*{20px} \times \prod_{\substack{\left(D_{s, l}\right)_{s \in \{0, 1\}^{2p + 1}, 1 \leq l \leq N_L}\\\sum_{s, l}D_{s, l} = D}}f_{\left(\sum_sD_{s, l}\right)_{1 \leq l \leq N_L}}\left((-1)^{\sum_{s, l}D_{s, l}s_{2p}}, \ldots, (-1)^{\sum_{s, l}D_{s, l}s_0}\right)^{\prod_{s, l}\binom{n_{s, l}}{D_{s, l}}}\ket{(n_s)_s}
\end{align}
In the equation above, $\left(Z_{k_D}\ldots Z_{k_1}\right)_j$ precisely means $\left(\mathbf{1}_{2^n}\right)^{\otimes(2p - j)} \otimes \left(Z_{k_D}\ldots Z_{k_1}\right) \otimes \left(\mathbf{1}_{2^n}\right)^{\otimes j}$, where $\left(Z_{k_D}\ldots Z_{k_1}\right)$ acts on an $n$-qubit space, as do the other $2p$ tensor product factors. The integers $1, \ldots, N_L$ should be thought of as labels for the vertices $0, \ldots, n - 1$, $q_l$ as the probability of labelling a vertex $l$ and $\sum_sD_{s, l}$ as the number of vertices labelled $l$ among a $D$-subset of vertices $\{k_1, \ldots, k_D\} \subset [n]$.
\begin{proof}
Recall the definition of $\ket{\left(n_s\right)_s}$:
\begin{align}
\label{eq:ns_recall_def}
    \ket{(n_s)_s} & = \frac{1}{\sqrt{\binom{n}{\left(n_s\right)_s}}}\sum_{\substack{x_0, \ldots, x_{2p} \in \{0, 1\}^{2p + 1}\\\left|\left\{j \in [n]\,:\,\left((x_{2p})_j, \ldots, (x_0)_j\right) = s\right\}\right| = n_s}}\ket{x_{2p}} \ldots \ket{x_0}.
\end{align}
In fact, we will show that the operator
\begin{align*}
    & \sum_{\substack{(I_l)_{1 \leq l \leq N_L}\\I_l \subset [n]\\\sqcup_{1 \leq l \leq N_L}I_l = [n]}}\binom{n}{\left(|I_l|\right)_l}\prod_{1 \leq l \leq N_L}q_l^{|I_l|}\prod_{\{k_1, \ldots, k_D\} \subset [n]}f_{\left(\left|\{k_1, \ldots, k_D\} \cap I_l\right|\right)_l}\left(\left(Z_{k_D} \ldots Z_{k_1}\right)_{2p}, \ldots, \left(Z_{k_D} \ldots Z_{k_1}\right)_0\right)
\end{align*}
acts as multiplication by the same scalar
\begin{align*}
    & \sum_{\substack{\left(n_{s, l}\right)_{\substack{s \in \{0, 1\}^{2p + 1}, 1 \leq l \leq N_L}}\\\sum_{l}n_{s, l} = n_s}}\prod_s\binom{n_s}{\left(n_{s, l}\right)_l}\prod_lq_l^{n_{s, l}}\\
    & \hspace*{20px} \times \prod_{\substack{\left(D_{s, l}\right)_{s \in \{0, 1\}^{2p + 1}, 1 \leq l \leq N_L}\\\sum_{s, l}D_{s, l} = D}}f_{\left(\sum_sD_{s, l}\right)_{1 \leq l \leq N_L}}\left((-1)^{\sum_{s, l}D_{s, l}s_{2p}}, \ldots, (-1)^{\sum_{s, l}D_{s, l}s_0}\right)^{\prod_{s, l}\binom{n_{s, l}}{D_{s, l}}}
\end{align*}
on each vector of sum \ref{eq:ns_recall_def}. Let us then fix bitstrings $x_{2p}, \ldots, x_0$ satisfying the constraints in the sum and compute the action of
\begin{align*}
    \sum_{\left(I_l\right)_l}\binom{n}{\left(I_l\right)_l}\prod_lq_l^{|I_l|}\prod_{\{k_1, \ldots, k_D\} \subset [n]}f_{\left(\left|\{k_1, \ldots, k_D\} \cap I_l\right|\right)_l}\left(\left(Z_{k_D} \ldots Z_{k_1}\right)_{2p}, \ldots, \left(Z_{k_D} \ldots Z_{k_1}\right)_0\right)
\end{align*}
on $\ket{x_{2p}} \ldots \ket{x_0}$. To achieve that, let us then partition vertices $j \in [n]$ according to the values of the $j$-th bits of $x_{2p}, \ldots, x_0$ by defining:
\begin{align}
    J_s & := \left\{j \in [n]\,:\,\left((x_{2p})_j, \ldots, (x_0)_j\right) = s\right\}
\end{align}
By definition of $\ket{\left(n_s\right)_s}$, $|J_s| = n_s$. Now, observe that $\left(\binom{n}{\left(I_l\right)_l}\prod_lq_l^{|I_l|}\right)_{\substack{\left(I_l\right)_l\\\sqcup_lI_l = [n]}}$ is a probability distribution on the partition of $[n]$ into subsets of vertices labelled $1, \ldots, N_L$, whereby each vertex is independently labelled $l$ with probability $q_l$. Such a labelling can be equivalently obtained by labelling vertices from each $J_s$ one $s$ after another. The probability that each $J_s$ contains $n_{s, l}$ vertices labelled $l$ is given by $\prod_s\binom{n_s}{\left(n_{s, l}\right)_l}\prod_lq_l^{n_{s, l}}$, where the $(n_{s, l})$ must satisfy $\sum_ln_{s, l} = n_s$. Given such a labelling, choosing a subset $\{k_1, \ldots, k_D\} \subset [n]$ is equivalent to choosing $D_{s, l}$ elements among the vertices of $J_s$ labelled $l$ for all $s$ and $l$, such that $\sum_{s, l}D_{s, l} = D$. There are exactly $\prod_{s, l}\binom{n_{s, l}}{D_{s, l}}$ subsets $\{k_1, \ldots, k_D\}$ satisfying these properties. The intersection of $\{k_1, \ldots, k_D\}$ with vertices labelled then has size $\sum_sD_{s, l}$. Finally, still considering the same subset $\{k_1, \ldots, k_D\}$, $\left(Z_{k_D} \ldots Z_{k_1}\right)_m\ket{x_{2p}} \ldots \ket{x_0} = \prod_{1 \leq r \leq D}(-1)^{(x_m)_{k_r}}\ket{x_{2p}}\ldots\ket{x_0} = (-1)^{\sum_{s, l}D_{s, l}s_m}\ket{x_{2p}}\ldots\ket{x_0}$. The result follows.
\end{proof}
\end{prop}

The result above can be simplified in case the vertices are unlabelled, i.e. $N_L = 1$:
\begin{cor}
\label{cor:average_hamiltonian_tensor}
Let $p \geq 1$, $D \geq 2$ and $n \geq D$ be integers and $f$ an arbitrary function $\{-1, 1\}^{2p + 1} \longrightarrow \mathbf{C}$. The following identity holds:
\begin{align}
\label{cor:average_hamiltonian_tensor}
    & \prod_{\{k_1, \ldots, k_D\} \subset [n]}f\left(\left(Z_{k_D}\ldots Z_{k_1}\right)_{2p}, \ldots, \left(Z_{k_D}\ldots Z_{k_1}\right)_0\right)\ket{\left(n_s\right)_{s \in \{0, 1\}^{2p + 1}}}\nonumber\\
    & = \prod_{\substack{\left(D_s\right)_{s \in \{0, 1\}^{2p + 1}}\\\sum_sD_s = D}}f\left((-1)^{\sum_sD_ss_{2p}}, \ldots, (-1)^{\sum_sD_ss_0}\right)^{\prod_s\binom{n_s}{D_s}}\ket{\left(n_s\right)_{s \in \{0, 1\}^{2p + 1}}}
\end{align}
\end{cor}

As a simple example, consider MaxCut on a random Erdos-Renyi graph of edge probability $q$. Given a random $n$-vertices graph $G = (V, E)$, the corresponding Ising Hamiltonian is:
\begin{align}
    H_C & = \sum_{\{k, l\} \subset [n]}J_{\{k, l\}}Z_kZ_l\\
    J_{\{k, l\}} & = \left\{\begin{array}{ll}
        1 & \textrm{if } \{k, l\} \in E\\
        0 & \textrm{otherwise} 
    \end{array}\right.
\end{align}
For $G \sim \er(n, q)$, the $J_{\{k, l\}}$ are i.i.d $\Bernoulli(q)$. Averaging  $\bigotimes_{0 \leq j < p}e^{i\frac{\gamma_j}{2}H_C} \otimes e^{i\frac{\gamma'}{2}H_C} \otimes \bigotimes_{0 \leq j < p}e^{-i\frac{\gamma_{p - 1 - j}}{2}H_C}$ over instances then gives:
\begin{align}
    & \mathbf{E}_{G \sim \er(n, q)}\left[\bigotimes_{0 \leq j < p}e^{i\frac{\gamma_j}{2}H_C} \otimes e^{i\frac{\gamma'}{2}H_C} \otimes \bigotimes_{0 \leq j < p}e^{-i\frac{\gamma_{p - 1 - j}}{2}H_C}\right]\nonumber\\
    & = \prod_{\{k, l\} \subset [n]}\left(1 - q + q\bigotimes_{0 \leq j < p}e^{i\frac{\gamma_j}{2}Z_kZ_l} \otimes e^{i\frac{\gamma'}{2}Z_kZ_l} \otimes \bigotimes_{0 \leq j < p}e^{-i\frac{\gamma_{p - 1 - j}}{2}Z_kZ_l}\right)\\
    & = \prod_{\{k, l\} \subset [n]}\left(1 - q + qe^{\frac{i}{2}\sum_{0 \leq j < p}\gamma_j\left(\left(Z_kZ_l\right)_{2p - j} - \left(Z_kZ_l\right)_j\right) + \frac{i\gamma'}{2}\left(Z_kZ_l\right)_p}\right)
\end{align}
One can now apply corollary \ref{cor:average_hamiltonian_tensor} with $D = 2$ and
\begin{align}
    f\left(z_{2p}, \ldots, z_0\right) & := 1 - q + qe^{\frac{i}{2}\sum_{0 \leq j < p}\gamma_j\left(z_{2p - j} - z_j\right) + \frac{i\gamma'}{2}z_p},
\end{align}
giving:
\begin{align}
     & \mathbf{E}_{G \sim \er(n, q)}\left[\bigotimes_{0 \leq j < p}e^{i\frac{\gamma_j}{2}H_C} \otimes e^{i\frac{\gamma'}{2}H_C} \otimes \bigotimes_{0 \leq j < p}e^{-i\frac{\gamma_{p - 1 - j}}{2}H_C}\right]\ket{(n_s)_s}\nonumber\\
     & = \prod_{\{s, t\} \subset \{0, 1\}^{2p + 1}}\left(1 - q + qe^{\frac{i}{2}\sum_{0 \leq j < p}\gamma_j\left((-1)^{s_{2p - j} + t_{2p - j}} - (-1)^{s_j + t_j}\right) + \frac{i\gamma'}{2}(-1)^{s_p + t_p}}\right)^{n_sn_t}\nonumber\\
     & \hspace*{30px} \times \prod_{s \in \{0, 1\}^{2p + 1}}\left(1 - q + qe^{\frac{i\gamma'}{2}}\right)^{\frac{n_s(n_s - 1)}{2}}\ket{(n_s)_s}.
\end{align}

Each product in the latter equation comes from a choice of $\left(D_s\right)_{s \in \{0, 1\}^{2p + 1}}$ such that $\sum_sD_s = D = 2$. The first product results from choosing $D_t = D_u = 1$, for distinct $t, u \in \{0, 1\}^{2p + 1}$ and $D_s = 0$ for $s \neq t, u$. The second product results from choosing $D_t = 2$ for some $t \in \{0, 1\}^{2p + 1}$ and $D_s = 0$ for $s \neq t$.

\subsection{QAOA moments for dense and diluted spin models}
\label{sec:qaoa_moments_dense_diluted}
We now specialize the results established in the previous section to the dense and diluted $D$-spin models, which generalise the case of MaxCut discussed above. This means rewriting the QAOA moment-generating function starting with proposition \ref{prop:qaoa_tn_reorganized} and decomposing the vectors and averaged operator in the generalized number basis thanks to proposition \ref{prop:mixers_generalized_number_basis} and corollary \ref{cor:average_hamiltonian_tensor}. It only remains to compute the average
\begin{align}
    & \mathbf{E}_J\left[\bigotimes_{0 \leq j < p}e^{i\frac{\gamma_j}{2}H[J]} \otimes e^{i\frac{\gamma'}{2}H[J]} \otimes \bigotimes_{0 \leq j < p}e^{-i\frac{\gamma_{p - 1 - j}}{2}H[J]}\right]
\end{align}
over random $D$-spin model Hamiltonians $H_C$ as defined in section \ref{sec:optimization_problems}.
\begin{align*}
    & \mathbf{E}_J\left[\bigotimes_{0 \leq j < p}e^{i\frac{\gamma_j}{2}H[J]} \otimes e^{i\frac{\gamma'}{2}H[J]} \otimes \bigotimes_{0 \leq j < p}e^{-i\frac{\gamma_{p - 1 - j}}{2}H[J]}\right]\\
    & = \mathbf{E}_J\left[\bigotimes_{0 \leq j < p}e^{i\frac{\gamma_j}{2}\sum_{\{k_1, \ldots, k_D\} \subset [n]}J_{\{k_1, \ldots, k_D\}}Z_{k_1}\ldots Z_{k_D}} \otimes e^{i\frac{\gamma'}{2}\sum_{\{k_1, \ldots, k_D\} \subset [n]}J_{\{k_1, \ldots, k_D\}}Z_{k_1}\ldots Z_{k_D}}\right.\\
    & \left. \hspace*{40px} \otimes \bigotimes_{0 \leq j < p}e^{-i\frac{\gamma_{p - 1 - j}}{2}\sum_{\{k_1, \ldots, k_D\} \subset [n]}J_{\{k_1, \ldots, k_D\}}Z_{k_1}\ldots Z_{k_D}}\right]\\
    & = \mathbf{E}_J\left[\prod_{\{k_1, \ldots, k_D\} \subset [n]}\bigotimes_{0 \leq j < p}e^{i\frac{\gamma_j}{2}J_{\{k_1, \ldots, k_D\}}Z_{k_1}\ldots Z_{k_D}} \otimes e^{i\frac{\gamma'}{2}J_{\{k_1, \ldots, k_D\}}Z_{k_1}\ldots Z_{k_D}} \otimes \bigotimes_{0 \leq j < p}e^{-i\frac{\gamma_{p - 1 - j}}{2}J_{\{k_1, \ldots, k_D\}}Z_{k_1}\ldots Z_{k_D}}\right]\\
    & = \prod_{\{k_1, \ldots, k_D\} \subset [n]}\mathbf{E}_J\left[\bigotimes_{0 \leq j < p}e^{i\frac{\gamma_j}{2}J_{\{k_1, \ldots, k_D\}}Z_{k_1}\ldots Z_{k_D}} \otimes e^{i\frac{\gamma'}{2}J_{\{k_1, \ldots, k_D\}}Z_{k_1}\ldots Z_{k_D}} \otimes \bigotimes_{0 \leq j < p}e^{-i\frac{\gamma_{p - 1 - j}}{2}J_{\{k_1, \ldots, k_D\}}Z_{k_1}\ldots Z_{k_D}}\right]
\end{align*}
In the last step, the expectation-product exchange is justified by the fact that the $\left(J_{\{k_1, \ldots, k_D\}}\right)_{\{k_1, \ldots, k_D\} \subset [n]}$ are i.i.d, see definition \ref{def:diluted_d_spin}.
For the diluted $D$-spin model with degree parameter $d$, the expectation in the last line evaluates to
\begin{align}
    & 1 - \frac{nd}{D\binom{n}{D}} + \frac{nd}{D\binom{n}{D}}\bigotimes_{0 \leq j < p}e^{i\frac{\gamma_j}{2}Z_{k_1}\ldots Z_{k_D}} \otimes e^{i\frac{\gamma'}{2}Z_{k_1}\ldots Z_{k_D}} \otimes \bigotimes_{0 \leq j < p}e^{-i\frac{\gamma_{p - 1 - j}}{2}Z_{k_1}\ldots Z_{k_D}}\nonumber\\
    & = 1 - \frac{nd}{D\binom{n}{D}} + \frac{nd}{D\binom{n}{D}}e^{i\sum_{0 \leq j < p}\frac{\gamma_j}{2}\left(\left(Z_{k_1}\ldots Z_{k_D}\right)_{2p - j} - \left(Z_{k_1}\ldots Z_{k_D}\right)_j\right) + i\frac{\gamma'}{2}\left(Z_{k_1}\ldots Z_{k_D}\right)_p}.
\end{align}
For the dense $D$-spin model, it evaluates to
\begin{align}
    & \exp\left(-\frac{n^{1 - D}}{2}\left(\sum_{0 \leq j < p}\frac{\gamma_j}{2}\left(\left(Z_{k_1}\ldots Z_{k_D}\right)_{2p - j} - \left(Z_{k_1}\ldots Z_{k_D}\right)_j\right) + \frac{\gamma'}{2}\left(Z_{k_1}\ldots Z_{k_D}\right)_p\right)^2\right).
\end{align}
In any case, by corollary \ref{cor:average_hamiltonian_tensor}, $\mathbf{E}_J\left[\bigotimes_{0 \leq j < p}e^{i\frac{\gamma_j}{2}H[J]} \otimes e^{i\frac{\gamma'}{2}H[J]} \otimes \bigotimes_{0 \leq j < p}e^{-i\frac{\gamma_{p - 1 - j}}{2}H[J]}\right]$ is diagonal in the $\left(\ket{n_s}\right)_{s \in \{0, 1\}^{2p + 1}}$ basis, with explicitly computable coefficients. Invoking proposition \ref{prop:expectation_permutation_invariant_average_hamiltonian} yields the following results:

\begin{prop}
\label{prop:qaoa_dense_spin_model}
The QAOA moment-generating function for the dense $D$-spin model is given by:
\begin{align}
    & \mathbf{E}_{J \sim \textnormal{Dense}(D, n)}\left[\braket{\Psi_{\textnormal{QAOA}}(J, \bm{\beta}, \bm{\gamma})|e^{i\frac{\gamma'}{2}H[J]}|\Psi_{\textnormal{QAOA}}(J, \bm{\beta}, \bm{\gamma})}\right]\nonumber\\
    & = \frac{1}{2^n}\sum_{\substack{\left(n_s\right)_{s \in \{0, 1\}^{2p + 1}}\\n_s \geq 0}}\binom{n}{\left(n_s\right)_s}(-1)^{\sum_{s \in \mathcal{L}}n_s}\prod_sB_{\bm{\beta}, s}^{n_s}\nonumber\\
    & \hspace*{65px} \times \prod_{\substack{\left(D_s\right)_{s \in \{0, 1\}^{2p + 1}}\\\sum_sD_s = D}}\exp\left(-\frac{n^{1 - D}}{2}\left(\varphi\left(\bm{\gamma}, \oplus_ss^{\oplus D_s}\right) + \frac{\gamma'}{2}(-1)^{\sum_sD_ss_p}\right)^2\prod_s\binom{n_s}{D_s}\right).
\end{align}
The first-order moment is given by:
\begin{align}
    & \mathbf{E}_{J \sim \textnormal{Dense}(D, n)}\left[\braket{\Psi_{\textnormal{QAOA}}(J, \bm{\beta}, \bm{\gamma})|H[J]|\Psi_{\textnormal{QAOA}}(J, \bm{\beta}, \bm{\gamma})}\right]\nonumber\\
    & = \frac{1}{2^n}\sum_{\substack{\left(n_s\right)_{s \in \{0, 1\}^{2p + 1}}\\n_s \geq 0}}\binom{n}{\left(n_s\right)_s}(-1)^{\sum_{s \in \mathcal{L}}n_s}\prod_sB_{\bm{\beta}, s}^{n_s}\prod_{\substack{\left(D_s\right)_{s \in \{0, 1\}^{2p + 1}}\\\sum_sD_s = D}}\exp\left(-\frac{n^{1 - D}}{2}\varphi\left(\bm{\gamma}, \oplus_ss^{\oplus D_s}\right)^2\prod_s\binom{n_s}{D_s}\right)\nonumber\\
    & \hspace*{65px} \times \sum_{\substack{\left(D_s\right)_{s \in \{0, 1\}^{2p + 1}}\\\sum_sD_s = D}}\prod_s\binom{n_s}{D_s}in^{1 - D}(-1)^{\sum_sD_ss_p}\varphi\left(\bm{\gamma}, \oplus_ss^{\oplus D_s}\right)
\end{align}
\end{prop}

\begin{prop}
\label{prop:qaoa_diluted_spin_model}
The QAOA moment-generating function for the diluted $D$-spin model with degree parameter $d$ is given by:
\begin{align}
    & \mathbf{E}_{J \sim \textnormal{Diluted}(D, n, d)}\left[\braket{\Psi_{\textnormal{QAOA}}(J, \bm{\beta}, \bm{\gamma})|e^{i\frac{\gamma'}{2}H[J]}|\Psi_{\textnormal{QAOA}}(J, \bm{\beta}, \bm{\gamma})}\right]\nonumber\\
    & = \frac{1}{2^n}\sum_{\substack{\left(n_s\right)_{s \in \{0, 1\}^{2p + 1}}\\n_s \geq 0}}\binom{n}{\left(n_s\right)_s}(-1)^{\sum_{s \in \mathcal{L}}n_s}\prod_sB_{\bm{\beta}, s}^{n_s}\nonumber\\
    & \hspace*{65px} \times \prod_{\substack{\left(D_s\right)_{s \in \{0, 1\}^{2p + 1}}\\\sum_sD_s = D}}\left(1 - \frac{nd}{D\binom{n}{D}} + \frac{nd}{D\binom{n}{D}}e^{i\left(\varphi\left(\bm{\gamma}, \oplus_ss^{\oplus D_s}\right) + \frac{\gamma'}{2}(-1)^{\sum_sD_ss_p}\right)}\right)^{\prod_s\binom{n_s}{D_s}}
\end{align}
The first-order moment is given by:
\begin{align}
    & \mathbf{E}_{J \sim \textnormal{Diluted}(D, n, d)}\left[\braket{\Psi_{\textnormal{QAOA}}(J, \bm{\beta}, \bm{\gamma})|H[J]|\Psi_{\textnormal{QAOA}}(J, \bm{\beta}, \bm{\gamma})}\right]\nonumber\\
    & = \frac{1}{2^n}\sum_{\substack{\left(n_s\right)_{s \in \{0, 1\}^{2p + 1}}\\n_s \geq 0}}\binom{n}{\left(n_s\right)_s}(-1)^{\sum_{s \in \mathcal{L}}n_s}\prod_sB_{\bm{\beta}, s}^{n_s}\prod_{\substack{\left(D_s\right)_{s \in \{0, 1\}^{2p + 1}}\\\sum_sD_s = D}}\left(1 - \frac{nd}{D\binom{n}{D}} + \frac{nd}{D\binom{n}{D}}e^{i\varphi\left(\bm{\gamma}, \oplus_ss^{\oplus D_s}\right)}\right)^{\prod_s\binom{n_s}{D_s}}\nonumber\\
    & \hspace*{65px} \times \sum_{\substack{\left(D_s\right)_{s \in \{0, 1\}^{2p + 1}}\\\sum_sD_s = D}}\prod_s\binom{n_s}{D_s}\frac{\frac{nd}{D\binom{n}{D}}e^{i\varphi\left(\bm{\gamma}, \oplus_ss^{\oplus D_s}\right)}(-1)^{\sum_sD_ss_p}}{1 - \frac{nd}{D\binom{n}{D}} + \frac{nd}{D\binom{n}{D}}e^{i\varphi\left(\bm{\gamma}, \oplus_ss^{\oplus D_s}\right)}}
\end{align}
\end{prop}
Proposition \ref{prop:qaoa_dense_spin_model} generalizes results derived in \cite{1910.08187} and \cite{2102.12043}. Unfortunately, we do not know of any way of analyzing these formulae further for general $D$ and $p$, even in the limit $n \to \infty$. In the following sections, we consider cases where the analysis is tractable.

\subsection{Analysis for MaxCut on sparse graphs of constant average degree}
\label{sec:analysis_maxcut_constant_degree}
In this section, we specialize the results from section \ref{sec:qaoa_moments_dense_diluted} to the MaxCut problem on sparse Erdos-Renyi graphs, introduced in definition \ref{def:maxcut_erdos_renyi}. The latter corresponds to the diluted $2$-spin model; hence, from proposition \ref{prop:qaoa_diluted_spin_model}:
\begin{prop}
The first-order moment of the QAOA for the MaxCut problem on Erdos-Renyi graphs of expected degree $d$ is given by:
\begin{align}
    & \mathbf{E}_{J \sim \er\left(n, \frac{d}{n - 1}\right)}\left[\braket{\Psi_{\textnormal{QAOA}}(J, \bm{\beta}, \bm{\gamma})|H[J]|\Psi_{\textnormal{QAOA}}(J, \bm{\beta}, \bm{\gamma})}\right]\nonumber\\
    & = \frac{1}{2^n}\sum_{\substack{\left(n_s\right)_{s \in \{0, 1\}^{2p + 1}}\\\sum_sn_s = n}}\binom{n}{\left(n_s\right)_s}(-1)^{\sum_{s \in \mathcal{L}}n_s}\prod_sB_{\bm{\beta}, s}^{n_s}\prod_{\substack{s, t \in \{0, 1\}^{2p + 1}\\s < t}}\left(1 - \frac{d}{n - 1} + \frac{d}{n - 1}e^{i\varphi(\bm{\gamma}, s \oplus t)}\right)^{n_sn_t}\nonumber\\
    & \hspace*{65px} \times \left(\sum_{s \in \{0, 1\}^{2p + 1}}\frac{d}{n - 1}\frac{n_s(n_s - 1)}{2} + \sum_{\substack{s, t \in \{0, 1\}^{2p + 1}\\s < t}}\frac{\frac{d}{n - 1}e^{i\varphi(\bm{\gamma}, s \oplus t)}(-1)^{s_p + t_p}}{1 - \frac{d}{n - 1} + \frac{d}{n - 1}e^{i\varphi(\bm{\gamma}, s \oplus t)}}n_sn_t\right)
\end{align}
\end{prop}
In the previous proposition, we refer to the ordering of bitstrings $s \in \{0, 1\}^{2p + 1}$ introduced in definition \ref{def:bitstrings_ordering}. Swapping the sum on $\left(n_s\right)_s$ with the sum on $s$ or $s, t$, the expression above becomes
\begin{align}
    & \mathbf{E}_{J \sim \er\left(n, \frac{d}{n - 1}\right)}\left[\braket{\Psi_{\textnormal{QAOA}}(J, \bm{\beta}, \bm{\gamma})|H[J]|\Psi_{\textnormal{QAOA}}(J, \bm{\beta}, \bm{\gamma})}\right]\nonumber\\
    & = \sum_{u \in \{0, 1\}^{2p + 1}}\frac{d}{n - 1}\frac{1}{2^n}\sum_{\substack{\left(n_s\right)_s\\\sum_sn_s = n}}(-1)^{\sum_{s \in \mathcal{L}}n_s}\prod_sB_{\bm\beta, s}^{n_s}\prod_{\substack{t, s < t}}\left(1 -  \frac{d}{n - 1} + \frac{d}{n - 1}e^{i\varphi\left(\bm\gamma, s \oplus t\right)}\right)^{n_sn_t}\frac{n_u(n_u - 1)}{2}\nonumber\\
    & + \sum_{\substack{u, v \in \{0, 1\}^{2p + 1}\\u < v}}\frac{\frac{d}{n - 1}e^{i\varphi\left(\bm\gamma, u \oplus v\right)}(-1)^{u_p + v_p}}{1 - \frac{d}{n - 1} + \frac{d}{n - 1}e^{i\varphi\left(\bm\gamma, u \oplus v\right)}}\frac{1}{2^n}\sum_{\substack{(n_s)_s\\\sum_sn_s = n}}\binom{n}{(n_s)_s}(-1)^{\sum_{s \in \mathcal{L}}n_s}\prod_sB_{\bm\beta, s}^{n_s}\nonumber\\
    & \hspace*{220px} \times \prod_{t, s < t}\left(1 - \frac{d}{n - 1} + \frac{d}{n - 1}e^{i\varphi\left(\bm\gamma, s \oplus t\right)}\right)^{n_sn_t}n_un_v.
\end{align}
The inner sums on $\left(n_s\right)_s$ constitute the hard part:
\begin{align}
    S_{u, v} & := \frac{1}{2^n}\sum_{\substack{(n_s)_s\\\sum_sn_s = n}}\binom{n}{\left(n_s\right)_s}(-1)^{\sum_{s \in \mathcal{L}}n_s}\prod_sB_{\bm{\beta}, s}^{n_s}\prod_{\substack{t, s < t}}\left(1 - \frac{d}{n - 1} + \frac{d}{n - 1}e^{i\varphi(\bm{\gamma}, s \oplus t)}\right)^{n_sn_t}\frac{n_un_v}{n^2}\label{eq:s_uv},\\
    S_u & := \frac{1}{2^n}\sum_{\substack{(n_s)_s\\\sum_sn_s = n}}\binom{n}{\left(n_s\right)_s}(-1)^{\sum_{s \in \mathcal{L}}n_s}\prod_sB_{\bm{\beta}, s}^{n_s}\prod_{\substack{t, s < t}}\left(1 - \frac{d}{n - 1} + \frac{d}{n - 1}e^{i\varphi(\bm{\gamma}, s \oplus t)}\right)^{n_sn_t}\frac{n_u(n_u - 1)}{n^2}\label{eq:s_u}.
\end{align}
Each of these sums contains $\binom{n + 2^{2p + 1} - 1}{2^{2p + 1} - 1}$ terms, making them intractable to compute exactly even for moderate $n$ and $p$. Fortunately, they simplify drastically in the $n \to \infty$ limit, with a computational time scaling only exponentially in $p$.

More precisely, closely related expressions were shown \cite{1910.08187} to admit limits as $n \to \infty$; the methods in \cite{1910.08187} applied to the Sherrington-Kirkpatrick model but extend readily to the Erdos-Renyi MaxCut problem discussed here. Therefore, we will not reproduce the detailed calculations but merely state the adapted results. However, it will still be useful to detail the first step of the derivation, which will be required in the analysis of the Monte Carlo algorithm for the SK energy in section \ref{sec:analysis_sk_monte_carlo}. The following proposition establishes a formula that allows to take the infinite size limit following the methods of \cite{1910.08187}. The next result: proposition \ref{prop:infinite_size_second_order_moment} states the $n \to \infty$ limits of $S_u$ and $S_{u, v}$ properly speaking.

\begin{prop}
\label{prop:moment_exact_expression}
Define (to simplify the notation):
\begin{align}
    C_{\bm{\gamma}, s} & := \left(1 - \frac{d}{n - 1} + \frac{d}{n - 1}e^{i\varphi(\bm{\gamma}, s)}\right)
\end{align}
The sums $S_{u,v}$ and $S_u$ defined in equations \ref{eq:s_uv} and \ref{eq:s_u} reduce to:
\begin{align}
    S_{u, v} & = (-1)^{\mathbf{1}_{u \in \mathcal{L}} + \mathbf{1}_{v \in \mathcal{L}}}B_{\bm{\beta}, u}B_{\bm{\beta}, v}C_{\bm{\gamma}, u \oplus v}\frac{n - 1}{n}\sum_{\substack{\left(m_{\{s, F(s)\}}\right)_{s \in \mathcal{L} - \mathcal{L}_p}\\0 \leq n' \leq n\\\sum_{s \in \mathcal{L} - \mathcal{L}_p}m_{\{s, F(s)\}} = n'}}\frac{1}{2^n}\binom{n'}{\left(m_{\{s, F(s)\}}\right)_{s \in \mathcal{L} - \mathcal{L}_p}}\binom{n - 2}{n'}\nonumber\\
    & \hspace*{10px} \times \prod_{\substack{t \in \mathcal{L}\\L(t) \leq p - 1}}\left[B_{\bm{\beta}, t}\left(-C_{\bm{\gamma}, u \oplus t}C_{\bm{\gamma}, v \oplus t}\prod_{\substack{s \in \mathcal{L}\\s < t}}C_{\bm{\gamma}, s \oplus t}^{m_{\{s, F(s)\}}} + C_{\bm{\gamma}, u \oplus F(t)}C_{\bm{\gamma}, v \oplus F(t)}\prod_{\substack{s \in \mathcal{L}\\s < t}}C_{\bm{\gamma}, s \oplus F(t)}^{m_{\{s, F(s)\}}}\right)\right]^{m_{\{t, F(t)\}}}\nonumber\\
    & \hspace*{10px} \times \left(-\sum_{t \in \mathcal{L}_p}B_{\bm{\beta}, t}\prod_{\substack{s \in \mathcal{L}\\L(s) < p}}C_{\bm{\gamma}, s \oplus t}^{m_{\{s, F(s)\}}} + \sum_{t \in \mathcal{L}_p'}B_{\bm{\beta}, t}\prod_{\substack{s \in \mathcal{L}\\L(s) < p}}C_{\bm{\gamma}, s \oplus t}^{m_{\{s, F(s)\}}}\right)^{n - n' - 2}\label{eq:s_uv_cancelled},\\
    S_u & = B_{\bm{\beta}, u}^2\frac{n - 1}{n}\sum_{\substack{\left(m_{\{s, F(s)\}}\right)_{s \in \mathcal{L} - \mathcal{L}_p}\\0 \leq n' \leq n\\\sum_{s \in \mathcal{L} - \mathcal{L}_p}m_{\{s, F(s)\}} = n'}}\frac{1}{2^n}\binom{n'}{\left(m_{\{s, F(s)\}}\right)_{s \in \mathcal{L} - \mathcal{L}_p}}\binom{n - 2}{n'}\nonumber\\
    & \hspace*{10px} \times \prod_{\substack{t \in \mathcal{L}\\L(t) \leq p - 1}}\left[B_{\bm{\beta}, t}\left(-C_{\bm{\gamma}, u \oplus t}^2\prod_{\substack{s \in \mathcal{L}\\s < t}}C_{\bm{\gamma}, s \oplus t}^{m_{\{s, F(s)\}}} + C_{\bm{\gamma}, u \oplus F(t)}^2\prod_{\substack{s \in \mathcal{L}\\s < t}}C_{\bm{\gamma}, s \oplus F(t)}^{m_{\{s, F(s)\}}}\right)\right]^{m_{\{t, F(t)\}}}\nonumber\\
    & \hspace*{10px} \times \left(-\sum_{t \in \mathcal{L}_p}B_{\bm{\beta}, t}\prod_{\substack{s \in \mathcal{L}\\L(s) < p}}C_{\bm{\gamma}, s \oplus t}^{m_{\{s, F(s)\}}} + \sum_{t \in \mathcal{L}_p'}B_{\bm{\beta}, t}\prod_{\substack{s \in \mathcal{L}\\L(s) < p}}C_{\bm{\gamma}, s \oplus t}^{m_{\{s, F(s)\}}}\right)^{n - n' - 2}\label{eq:s_u_cancelled}.
\end{align}

\begin{proof}
Let us establish for instance equation \ref{eq:s_uv_cancelled} ---equation \ref{eq:s_u_cancelled} is very similar. Assume for definiteness $u < v$, $u \neq F(v)$ and $L(v) < p$  (hence $L(u) < p$). The other cases are easy variations. One starts by summing over the $n_s$ such that $s \in \mathcal{L}_p \sqcup \mathcal{L}'_p$ (bitstrings that are symmetric about the $p^{\textrm{th}}$ index), under the constraint $\sum_{t \in \mathcal{L}_p \sqcup \mathcal{L}_p'}n_t = n - n'$. For that purpose, it helps to rewrite the product of $C$ as
\begin{align*}
    \prod_{\substack{s, t \in \{0, 1\}^{2p + 1}\\s < t}}C_{\bm{\gamma}, s \oplus t}^{n_sn_t} & = \prod_{\substack{s, t\\L(s) < L(t) = p}}C_{\bm{\gamma}, s \oplus t}^{n_sn_t}\prod_{\substack{s, t\\s < t, L(t) < p}}C_{\bm{\gamma}, s \oplus t}^{n_sn_t}.
\end{align*}
The factors in the summand of $S_{u, v}$ that involve $\left(n_s\right)_{s\,:\,L(s) = p}$ are the multinomial coefficient and
\begin{align*}
    & \prod_{\substack{s, t\\L(s) < L(t) = p}}C_{\bm{\gamma}, s \oplus t}^{n_sn_t}.
\end{align*}
The sum $S_{u, v}$ becomes:
\begin{align*}
    & \sum_{\substack{\left(n_s\right)_{\substack{s\,:\,L(s) < p}}\\0 \leq n' \leq n}}\frac{1}{2^n}\binom{n'}{\left(n_s\right)_{s, L(s) < p}}\binom{n}{n'}(-1)^{\sum_{s \in \mathcal{L} - \mathcal{L}_p}n_s}\prod_{s, L(s) < p}B_{\bm{\beta}, s}^{n_s}\prod_{\substack{s, t\\s < t, L(t) < p}}C_{\bm{\gamma}, s \oplus t}^{n_sn_t}\nonumber\\
    & \hspace*{50px} \times \left(-\sum_{t \in \mathcal{L}_p}B_{\bm{\beta}, t}\prod_{s, L(s) < p}C_{\bm{\gamma}, s \oplus t}^{n_s} + \sum_{t \in \mathcal{L}_p'}B_{\bm{\beta}, t}\prod_{s, L(s) < p}C_{\bm{\gamma}, s \oplus t}^{n_s}\right)^{n - n'}\frac{n_un_v}{n^2}
\end{align*}
We then define, for all $s \in \mathcal{L} - \mathcal{L}_p$, new variables $m_{\{s, F(s)\}}$ by
\begin{align*}
    m_{\{s, F(s)\}} & := n_s + n_{F(s)}
\end{align*}
We rewrite the factor involving $C$ accordingly:
\begin{align*}
    \prod_{\substack{s, t \in \{0, 1\}^{2p + 1}\\s < t, L(t) < p}}C_{\bm{\gamma}, s \oplus t}^{n_sn_t} & = \prod_{\substack{s, t \in \mathcal{L} - \mathcal{L}_p\\s < t}}C_{\bm{\gamma}, s \oplus t}^{n_sn_t}C_{\bm{\gamma}, s \oplus F(t)}^{n_sn_{F(t)}}C_{\bm{\gamma}, F(s) \oplus t}^{n_{F(s)}n_t}C_{\bm{\gamma}, F(s) \oplus F(t)}^{n_{F(s)}n_{F(t)}}\\
    & = \prod_{\substack{s, t \in \mathcal{L} - \mathcal{L}_p\\s < t}}C_{\bm{\gamma}, s \oplus t}^{m_{\{s, F(s)\}}n_t}C_{\bm{\gamma}, s \oplus F(t)}^{m_{\{s, F(s)\}}n_{F(t)}},
\end{align*}
where we used that $F$ is a one-one mapping between even and odd bitstrings (proposition \ref{prop:partial_flip_parity}) in the first line and proposition \ref{prop:beta_phi_invariance_partial_flip} in the last line.
Next:
\begin{align*}
    & \left(-\sum_{t \in \mathcal{L}_p}B_{\bm{\beta}, t}\prod_{s, L(s) < p}C_{\bm{\gamma}, s \oplus t}^{n_s} + \sum_{t \in \mathcal{L}_p'}B_{\bm{\beta}, t}\prod_{s, L(s) < p}C_{\bm{\gamma}, s \oplus t}^{n_s}\right)^{n - n'}\\
    & = \left(-\sum_{t \in \mathcal{L}_p}B_{\bm{\beta}, t}\prod_{\substack{s \in \mathcal{L}\\L(s) < p}}C_{\bm{\gamma}, s \oplus t}^{m_{\{s, F(s)\}}} + \sum_{t \in \mathcal{L}_p'}B_{\bm{\beta}, t}\prod_{\substack{s \in \mathcal{L}\\L(s) < p}}C_{\bm{\gamma}, s \oplus t}^{m_{\{s, F(s)\}}}\right)^{n - n'}
\end{align*}
We now sum over $n_s, n_{F(s)}$ for $s \in \mathcal{L} - \mathcal{L}_p$ with $m_{\{s, F(s)\}}$ fixed, i.e. setting $n_{F(s)} = m_{\{s, F(s)\}} - n_s$. The cases $s \in \{u, F(u), v, F(v)\}$ require special care. The sum becomes:
\begin{align*}
    & \frac{1}{n^2}\sum_{\substack{\left(m_{\{s, F(s)\}}\right)_{s \in \mathcal{L} - \mathcal{L}_p}\\0 \leq n' \leq n\\\sum_{s \in \mathcal{L} - \mathcal{L}_p}m_{\{s, F(s)\}} = n'}}\frac{1}{2^n}\frac{n'!}{(m_{\{u, F(u)\}} - 1)!(m_{\{v, F(v)\}} - 1)!\prod_{s \in \mathcal{L} - \{u, F(u), v, F(v)\}}m_{\{s, F(s)\}}!}\binom{n}{n'}\\
    & \hspace*{20px} \times (-1)^{\mathbf{1}_{u \in \mathcal{L}} + \mathbf{1}_{v \in \mathcal{L}}}B_{\bm{\beta}, u}B_{\bm{\beta}, v}C_{\bm{\gamma}, u \oplus v}\\
    & \hspace*{20px} \times \left[C_{\bm{\gamma}, u \oplus v}B_{\bm{\beta}, u}\left((-1)^{u \in \mathcal{L}}\prod_{\substack{s \in \mathcal{L}\\s < u\\s < F(u)}}C_{\bm{\gamma}, s \oplus u}^{m_{\{s, F(s)\}}} + (-1)^{u \notin \mathcal{L}}\prod_{\substack{s \in \mathcal{L}\\s < u\\ s < F(u)}}C_{\bm{\gamma}, s \oplus F(u)}^{m_{\{s, F(s)\}}}\right)\right]^{m_{\{u, F(u)\}} - 1}\\
    & \hspace*{20px} \times \left[B_{\bm{\beta}, v}\left((-1)^{v \in \mathcal{L}}\prod_{\substack{s \in \mathcal{L}\\s < v\\s < F(v)}}C_{\bm{\gamma}, s \oplus s_1}^{m_{\{s, F(s)\}}} + (-1)^{v \notin \mathcal{L}}\prod_{\substack{s \in \mathcal{L}\\s < v\\s < F(v)}}C_{\bm{\gamma}, s \oplus F(v)}^{m_{\{s, F(s)\}}}\right)\right]^{m_{\{v, F(v)\}} - 1}\\
    & \hspace*{20px} \times \prod_{\substack{t \in \mathcal{L} - \{u, F(u), v, F(v)\}\\L(t) \leq p - 1}}\left[C_{\bm{\gamma}, u \oplus t}^{\mathbf{1}_{t < u, F(u)}}C_{\bm{\gamma}, v \oplus t}^{\mathbf{1}_{t < v, F(v)}}B_{\bm{\beta}, t}\left(-\prod_{\substack{s \in \mathcal{L}\\s < t}}C_{\bm{\gamma}, s \oplus t}^{m_{\{s, F(s)\}}} + \prod_{\substack{s \in \mathcal{L}\\s < t}}C_{\bm{\gamma}, s \oplus F(t)}^{m_{\{s, F(s)\}}}\right)\right]^{m_{\{t, F(t)\}}}\\
    & \hspace*{20px} \times \left(-\sum_{t \in \mathcal{L}_p}B_{\bm{\beta}, t}\prod_{\substack{s \in \mathcal{L}\\L(s) < p}}C_{\bm{\gamma}, s \oplus t}^{m_{\{s, F(s)\}}} + \sum_{t \in \mathcal{L}_p'}B_{\bm{\beta}, t}\prod_{\substack{s \in \mathcal{L}\\L(s) < p}}C_{\bm{\gamma}, s \oplus t}^{m_{\{s, F(s)\}}}\right)^{n - n'}
\end{align*}
Next, we do a change of variables $m_{\{u, F(u)\}} \leftarrow m_{\{u, F(u)\}} + 1, m_{\{v, F(v)\}} \leftarrow m_{\{v, F(v)\}} + 1$, giving
\begin{align*}
    & (-1)^{\mathbf{1}_{u \in \mathcal{L}} + \mathbf{1}_{v \in \mathcal{L}}}B_{\bm{\beta}, u}B_{\bm{\beta}, v}C_{\bm{\gamma}, u \oplus v}\frac{n - 1}{n}\sum_{\substack{\left(m_{\{s, F(s)\}}\right)_{s \in \mathcal{L} - \mathcal{L}_p}\\0 \leq n' \leq n\\\sum_{s \in \mathcal{L} - \mathcal{L}_p}m_{\{s, F(s)\}} = n' - 2}}\frac{1}{2^n}\binom{n' - 2}{\left(m_{\{s, F(s)\}}\right)_{s \in \mathcal{L} - \mathcal{L}_p}}\binom{n - 2}{n' - 2}\\
    & \hspace*{10px} \times \prod_{\substack{t \in \mathcal{L}\\L(s') \leq p - 1}}\left[C_{\bm{\gamma}, u \oplus t}^{\mathbf{1}_{t < u, F(u)}}C_{\bm{\gamma}, v \oplus t}^{\mathbf{1}_{t < v, F(v)}}B_{\bm{\beta}, t}\left(-C_{\bm{\gamma}, u \oplus t}^{\mathbf{1}_{u < t}}C_{\bm{\gamma}, v \oplus t}^{\mathbf{1}_{v < t}}\prod_{\substack{s \in \mathcal{L}\\s < t}}C_{\bm{\gamma}, s \oplus t}^{m_{\{s, F(s)\}}}\right.\right.\\
    & \left.\left. \hspace*{200px} + C_{\bm{\gamma}, u \oplus F(t)}^{\mathbf{1}_{u < F(t)}}C_{\bm{\gamma}, v \oplus F(t)}^{\mathbf{1}_{v < F(t)}}\prod_{\substack{s \in \mathcal{L}\\s < s'}}C_{\bm{\gamma}, s \oplus F(s')}^{m_{\{s, F(s)\}}}\right)\right]^{m_{\{s', F(s')\}}}\\
    & \hspace*{10px} \times \left(-\sum_{t \in \mathcal{L}_p}B_{\bm{\beta}, t}C_{\bm{\gamma}, u \oplus t}^{\mathbf{1}_{L(u) < p}}C_{\bm{\gamma}, v \oplus t}^{\mathbf{1}_{L(v) < p}}\prod_{\substack{s \in \mathcal{L} - \mathcal{L}_p}}C_{\bm{\gamma}, s \oplus t}^{m_{\{s, F(s)\}}}\right.\nonumber\\
    & \left.\hspace*{40px} + \sum_{t \in \mathcal{L}_p'}B_{\bm{\beta}, t}C_{\bm{\gamma}, u \oplus t}^{\mathbf{1}_{L(u) < p}}C_{\bm{\gamma}, v \oplus t}^{\mathbf{1}_{L(v) < p}}\prod_{\substack{s \in \mathcal{L} - \mathcal{L}_p}}C_{\bm{\gamma}, s \oplus t}^{m_{\{s, F(s)\}}}\right)^{n - n'}.
\end{align*}
One can then remove the $C$ with characteristic functions in exponents using proposition \ref{prop:beta_phi_invariance_partial_flip}, yielding equation \ref{eq:s_uv_cancelled}.
\end{proof}
\end{prop}

To the best of our knowledge, these expressions for $S_{u, v}$ and $S_u$ remain intractable for $p > 1$. Fortunately, the analysis is simplified in the limit $n \to \infty$: 
\begin{prop}[Adapted from proof of {\cite[lemma 2]{1910.08187}}]
\label{prop:infinite_size_second_order_moment}
Recalling the notation from proposition \ref{prop:moment_exact_expression},
\begin{align}
    S^{\infty}_{u, v} & := \lim_{n \to \infty}S_{u, v}\nonumber\\
    & = \sum_{\substack{\left(m_{\{s, F(s)\}}\right)_{s \in \mathcal{L} - \mathcal{L}_p}\\m_{\{s, F(s)\}} \geq 0}}\frac{(-1)^{\mathbf{1}_{u \in \mathcal{L}} + \mathbf{1}_{v \in \mathcal{L}}}B_{\bm{\beta}, u}B_{\bm{\beta}, v}}{4\prod_{s \in \mathcal{L} - \mathcal{L}_p}m_{\{s, F(s)\}}!}\nonumber\\
    & \hspace*{20px} \times \prod_{t \in \mathcal{L} - \mathcal{L}_p}\left[\frac{d}{2}B_{\bm{\beta}, t}\Bigg(\left(e^{i\varphi(\bm{\gamma}, u \oplus F(t))} - e^{i\varphi(\bm{\gamma}, u \oplus t)}\right) + \left(e^{i\varphi(\bm{\gamma}, v \oplus F(t))} - e^{i\varphi(\bm{\gamma}, v \oplus t)}\right)\right.\nonumber\\
    & \hspace*{110px} + \sum_{\substack{s \in \mathcal{L}\\s < t}}m_{\{s, F(s)\}}\left(e^{i\varphi(\bm{\gamma}, s \oplus F(t))} - e^{i\varphi(\bm{\gamma}, s \oplus t)}\right)\Bigg)\Bigg]^{m_{\{t, F(t)\}}}\nonumber\\
    & \hspace*{20px} \times \exp\left(-d\left(1 - \frac{1}{2}\sum_{t \in \mathcal{L}_p \sqcup \mathcal{L}_p'}(-1)^{t \in \mathcal{L}_p}B_{\bm{\beta}, t}e^{i\varphi(\bm{\gamma}, u \oplus t)}\right)\right. \nonumber\\
    & \hspace*{70px} -d\left(1 - \frac{1}{2}\sum_{t \in \mathcal{L}_p \sqcup \mathcal{L}_p'}(-1)^{t \in \mathcal{L}_p}B_{\bm{\beta}, t}e^{i\varphi(\bm{\gamma}, v \oplus t)}\right) \nonumber\\
    & \left. \hspace*{70px} -d\sum_{s \in \mathcal{L} - \mathcal{L}_p}m_{\{s, F(s)\}}\left(1 - \frac{1}{2}\sum_{t \in \mathcal{L}_p \sqcup \mathcal{L}_p'}(-1)^{t \in \mathcal{L}_p}B_{\bm{\beta}, t}e^{i\varphi(\bm{\gamma}, s \oplus t)}\right)\right)
    \label{eq:s_uv_infinite_size}
\end{align}
and
\begin{align}
    S^{\infty}_{u} & := \lim_{n \to \infty}S_{u}\nonumber\\
    & = \sum_{\substack{\left(m_{\{s, F(s)\}}\right)_{s \in \mathcal{L} - \mathcal{L}_p}\\m_{\{s, F(s)\}} \geq 0}}\frac{(-1)^{\mathbf{1}_{u \in \mathcal{L}} + \mathbf{1}_{v \in \mathcal{L}}}B_{\bm{\beta}, u}B_{\bm{\beta}, v}}{4\prod_{s \in \mathcal{L} - \mathcal{L}_p}m_{\{s, F(s)\}}!}\nonumber\\
    & \hspace*{20px} \times \prod_{t \in \mathcal{L} - \mathcal{L}_p}\left[\frac{d}{2}B_{\bm{\beta}, t}\Bigg(2\left(e^{i\varphi(\bm{\gamma}, u \oplus F(t))} - e^{i\varphi(\bm{\gamma}, u \oplus t)}\right)\right.\nonumber\\
    & \hspace*{110px} + \sum_{\substack{s \in \mathcal{L}\\s < t}}m_{\{s, F(s)\}}\left(e^{i\varphi(\bm{\gamma}, s \oplus F(t))} - e^{i\varphi(\bm{\gamma}, s \oplus t)}\right)\Bigg)\Bigg]^{m_{\{t, F(t)\}}}\nonumber\\
    & \hspace*{20px} \times \exp\left(-2d\left(1 - \frac{1}{2}\sum_{t \in \mathcal{L}_p \sqcup \mathcal{L}_p'}(-1)^{t \in \mathcal{L}_p}B_{\bm{\beta}, t}e^{i\varphi(\bm{\gamma}, u \oplus t)}\right)\right. \nonumber\\
    & \left. \hspace*{70px} -d\sum_{s \in \mathcal{L} - \mathcal{L}_p}m_{\{s, F(s)\}}\left(1 - \frac{1}{2}\sum_{t \in \mathcal{L}_p \sqcup \mathcal{L}_p'}(-1)^{t \in \mathcal{L}_p}B_{\bm{\beta}, t}e^{i\varphi(\bm{\gamma}, s \oplus t)}\right)\right)
    \label{eq:s_u_infinite_size}
\end{align}
\end{prop}
The exponential in equation \ref{eq:s_uv_infinite_size} comes from taking the limit of
\begin{align}
    \frac{1}{2^n}\left(-\sum_{t \in \mathcal{L}_p}B_{\bm{\beta}, t}\prod_{\substack{s \in \mathcal{L}\\s \in \mathcal{L} - \mathcal{L}_p}}C_{\bm{\gamma}, s \oplus t}^{m_{\{s, F(s)\}}} + \sum_{t \in \mathcal{L}_p'}B_{\bm{\beta}, t}\prod_{\substack{s \in \mathcal{L} - \mathcal{L_p}}}C_{\bm{\gamma}, s \oplus t}^{m_{\{s, F(s)\}}}\right)^{n - n' - 2}
\end{align}
as $n \to \infty$ (with $\left(m_s\right)_{s \in \mathcal{L} - \mathcal{L}_p}$ fixed) in the summand of equation \ref{eq:s_uv_cancelled}:
\begin{align*}
    & \frac{1}{2^n}\left(-\sum_{t \in \mathcal{L}_p}B_{\bm\beta, t}\prod_{\substack{s \in \mathcal{L} - \mathcal{L}_p}}C_{\bm\gamma, s \oplus t}^{m_{\{s, F(s)\}}} + \sum_{t \in \mathcal{L}'_p}B_{\bm\beta, t}\prod_{\substack{s \in \mathcal{L} - \mathcal{L}_p}}C_{\bm\gamma, s \oplus t}^{m_{\{s, F(s)\}}}\right)^{n - n' - 2}\\
    & = \frac{1}{2^n}\left(\sum_{t \in \mathcal{L}_p \sqcup \mathcal{L}'_p}(-1)^{t \in \mathcal{L}}B_{\bm\beta, t}\prod_{\substack{s \in \mathcal{L} - \mathcal{L}_p}}C_{\bm\gamma, s \oplus t}^{m_{\{s, F(s)\}}}\right)^{n - n' - 2}\\
    & = \frac{1}{2^n}\left(\sum_{t \in \mathcal{L}_p \sqcup \mathcal{L}'_p}(-1)^{t \in \mathcal{L}}B_{\bm\beta, t}\prod_{\substack{s \in \mathcal{L} - \mathcal{L}_p}}\left(1 - \frac{d}{n - 1} + \frac{d}{n - 1}e^{i\varphi\left(\bm\gamma, s \oplus t\right)}\right)^{m_{\{s, F(s)\}}}\right)^{n - n' - 2}\\
    & = \frac{1}{2^n}\left(\sum_{t \in \mathcal{L}_p \sqcup \mathcal{L}'_p}(-1)^{t \in \mathcal{L}}B_{\bm\beta, t} - \frac{d}{n - 1}\sum_{t \in \mathcal{L}_p \sqcup \mathcal{L}'_p}(-1)^{t \in \mathcal{L}}B_{\bm\beta, t}\sum_{\substack{s \in \mathcal{L} - \mathcal{L}_p}}\left(1 - e^{i\varphi\left(\bm\gamma, s \oplus t\right)}\right)m_{\{s, F(s)\}} + \mathcal{O}\left(\frac{1}{n^2}\right)\right)^{n - n' - 2}
\end{align*}
The sums in the brackets can be simplified thanks to the following lemma:
\begin{lem}
\label{lemma:sum_betas}
\begin{align}
    \sum_{t \in \mathcal{L}_p \sqcup \mathcal{L}'_p}(-1)^{t \in \mathcal{L}}B_{\bm{\beta}, s} & = 2.
\end{align}
\begin{proof}
\begin{align*}
    & \sum_{t \in \mathcal{L}_p \sqcup \mathcal{L}'_p}(-1)^{t \in \mathcal{L}}B_{\bm{\beta}, s}\\
    & = \sum_{t \in \mathcal{L}_p \sqcup \mathcal{L}'_p}(-1)^{t_0 + t_p}\prod_{0 \leq j < p}\left(\cos\frac{\beta_j}{2}\right)^{\mathbf{1}_{s_j = s_{j + 1}} + \mathbf{1}_{s_{2p - j} = s_{2p - j - 1}}}\left(i\sin\frac{\beta_j}{2}\right)^{\mathbf{1}_{s_j \neq s_{j + 1}} + \mathbf{1}_{s_{2p - j} \neq s_{2p - j - 1}}}\\
    & = \sum_{t \in \mathcal{L}_p \sqcup \mathcal{L}'_p}(-1)^{t_0 + t_p}\prod_{0 \leq j < p}\left(\cos^2\frac{\beta_j}{2}\right)^{\mathbf{1}_{s_j = s_{j + 1}}}\left(-\sin^2\frac{\beta_j}{2}\right)^{\mathbf{1}_{s_j \neq s_{j + 1}}}\\
    & = \begin{pmatrix}
        1\\
        -1
    \end{pmatrix}^T\begin{pmatrix}
        \cos^2\frac{\beta_{p - 1}}{2} & -\sin^2\frac{\beta_{p - 1}}{2}\\
        -\sin^2\frac{\beta_{p - 1}}{2} & \cos^2\frac{\beta_{p - 1}}{2}
    \end{pmatrix}\ldots\begin{pmatrix}
        \cos^2\frac{\beta_0}{2} & -\sin^2\frac{\beta_0}{2}\\
        -\sin^2\frac{\beta_0}{2} & \cos^2\frac{\beta_0}{2}
    \end{pmatrix}\begin{pmatrix}
        1\\
        -1
    \end{pmatrix}\\
    & = 2.
\end{align*}
To go from the second to the third line, we used that $s_{2p - j} = s_j$ for $s \in \mathcal{L}_p \sqcup \mathcal{L}'_p$.
\end{proof}
\end{lem}
Using the lemma,
\begin{align*}
    &\frac{1}{2^n}\left(-\sum_{t \in \mathcal{L}_p}B_{\bm\beta, t}\prod_{\substack{s \in \mathcal{L} - \mathcal{L}_p}}C_{\bm\gamma, s \oplus t}^{m_{\{s, F(s)\}}} + \sum_{t \in \mathcal{L}'_p}B_{\bm\beta, t}\prod_{\substack{s \in \mathcal{L} - \mathcal{L}_p}}C_{\bm\gamma, s \oplus t}^{m_{\{s, F(s)\}}}\right)^{n - n' - 2}\\
    & = \frac{1}{2^{n' + 2}}\left(1 - \frac{d}{n - 1}\sum_{s \in \mathcal{L} - \mathcal{L}_p}m_{\{s, F(s)\}} + \frac{d}{2(n - 1)}\sum_{t \in \mathcal{L}_p \sqcup \mathcal{L}'_p}(-1)^{t \in \mathcal{L}}B_{\bm\beta, t}\sum_{\substack{s \in \mathcal{L} - \mathcal{L}_p}}e^{i\varphi\left(\bm\gamma, s \oplus t\right)}m_{\{s, F(s)\}}\right.\\
    & \left. \hspace*{60px} + \mathcal{O}\left(\frac{1}{n^2}\right)\right)^{n - n' - 2}\\
    & \xrightarrow[]{n \to \infty}\frac{1}{2^{n' + 2}}\exp\left(-d\sum_{s \in \mathcal{L} - \mathcal{L}_p}m_{\{s, F(s)\}}\left(1 - \frac{1}{2}\sum_{t \in \mathcal{L}_p \sqcup \mathcal{L}'_p}(-1)^{t \in \mathcal{L}}B_{\bm\beta, t}\right)\right).
\end{align*}

Besides, arguments similar to \cite[lemma 2]{1910.08187} show one may extend the summation $\sum\limits_{0 \leq n' \leq n}$ in equation \ref{eq:s_uv_cancelled} to infinity at the same time as taking the limit of the summand.

The quantities $S^{\infty}_{u, v}$ and $S^{\infty}_u$ defined in proposition \ref{prop:infinite_size_second_order_moment} can be evaluated by an algorithm of complexity $\mathcal{O}\left(16^p\right)$, which is a straightforward adaption of the algorithm from \cite{1910.08187}.
\begin{prop}[Adapted from \cite{1910.08187}]
\label{prop:s_uv_eval}
Recalling the notation from proposition \ref{prop:infinite_size_second_order_moment}, $S^{\infty}_{u, v}$ and $S^{\infty}_u$ can be expressed as follows:
\begin{align}
    S^{\infty}_{u, v} & = \frac{(-1)^{u \in \mathcal{L}}B_{\bm{\beta}, u}R_u(-1)^{v \in \mathcal{L}}B_{\bm{\beta}, v}R_v}{4}\\
    S^{\infty}_u & = \frac{B_{\bm{\beta}, u}^2R_u^2}{4}
\end{align}
where the $\left(R_s\right)_{s \in \{0, 1\}^{2p + 1}}$ are defined recursively as follows:
\begin{align}
    R_s & := e^{-d\left(1 - \frac{1}{2}\sum_{t \in \mathcal{L}_p \sqcup \mathcal{L}'_p}(-1)^{t \in \mathcal{L}}B_{\bm{\beta}, t}e^{i\varphi(\bm{\gamma}, s \oplus t)}\right)} \qquad \textrm{for } L(s) \in \{p - 1, p\}\\
    R_s & := e^{-d\left(1 - \frac{1}{2}\sum_{t \in \mathcal{L}_p \sqcup \mathcal{L}'_p}(-1)^{t \in \mathcal{L}}B_{\bm{\beta}, t}e^{i\varphi(\bm{\gamma}, s \oplus t)}\right)}e^{\frac{d}{2}\sum_{t, L(s) < L(t) < p}(-1)^{t \in \mathcal{L}}B_{\bm{\beta}, t}R_te^{i\varphi(\bm{\gamma}, s \oplus t)}}\nonumber\\
    & \hspace*{210px} \textrm{for } L(s) \leq p - 2
\end{align}
\end{prop}
From there, the QAOA energy takes the following form in the infinite size limit:
\begin{prop}
Recalling the $\left(R_s\right)_{s \in \{0, 1\}^{2p + 1}}$ defined in proposition \ref{prop:s_uv_eval},
\label{prop:qaoa_infinite_size_limit_energy}
\begin{align}
    & \lim_{n \to \infty}\mathbf{E}_{J \sim \er\left(n, \frac{d}{n - 1}\right)}\left[\left\langle\Psi_{\textnormal{QAOA}}(J, \bm{\beta}, \bm{\gamma})\bigg|\frac{H[J]}{n}\bigg|\Psi_{\textnormal{QAOA}}(J, \bm{\beta}, \bm{\gamma})\right\rangle\right]\nonumber\\
    & = \frac{d}{4}\left(\frac{1}{2}\sum_{s \in \{0, 1\}^{2p + 1}}R_s^2B_{\bm{\beta}, s}^2 + \sum_{\substack{s, t \in \{0, 1\}^{2p + 1}\\s < t}}(-1)^{s \in \mathcal{L}}(-1)^{s_p}B_{\bm{\beta}, s}R_s(-1)^{t \in \mathcal{L}}(-1)^{t_p}B_{\bm{\beta}, t}R_te^{i\varphi(\bm{\gamma}, s \oplus t)}\right)
\end{align}
\end{prop}
This identity can be compared to the analogous result established in \cite{1910.08187} for the energy of the QAOA applied to the Sherrington-Kirkpatrick model:
\begin{prop}[Energy of SK-QAOA, rephrased from \cite{1910.08187}]
\label{prop:sk_qaoa_energy_eval}
Let $p \geq 1$ an integer and $\bm{\beta}, \bm{\gamma} \in \mathbf{R}^p$. Recursively define $\left(\widetilde{R}_s\right)_{s \in \{0, 1\}^{2p + 1}}$ by:
\begin{align}
    \widetilde{R}_s & := e^{-\frac{1}{4}\sum_{t \in \mathcal{L}_p \sqcup \mathcal{L}'_p}(-1)^{t \in \mathcal{L}}B_{\bm{\beta}, t}\varphi(\bm{\gamma}, s \oplus t)^2} \qquad L(s) \in \{p - 1, p\}\\
    \widetilde{R}_s & := e^{-\frac{1}{4}\sum_{t \in \mathcal{L}_p \sqcup \mathcal{L}'_p}(-1)^{t \in \mathcal{L}}B_{\bm{\beta}, t}\varphi(\bm{\gamma}, s \oplus t)^2}e^{-\frac{1}{4}\sum_{t, L(s) < L(t) < p}(-1)^{t \in \mathcal{L}}B_{\bm{\beta}, t}\varphi(\bm{\gamma}, s \oplus t)^2R_t}
\end{align}
The energy of level-$p$ QAOA applied to the Sherrington-Kirkpatrick model in the infinite size limit is given by:
\begin{align}
    & \lim_{n \to \infty}\mathbf{E}_{J \sim SK(n)}\left[\braket{\Psi_{\textnormal{QAOA}}(J, \bm{\beta}, \bm{\gamma})|\frac{H(J)}{n}|\Psi_{\textnormal{QAOA}}(J, \bm{\beta}, \bm{\gamma})}\right]\nonumber\\
    & = \frac{1}{4}\sum_{s, t \in \{0, 1\}^{2p + 1}}i\varphi\left(\bm{\gamma}, s \oplus t\right)(-1)^{s \in \mathcal{L}}(-1)^{s_p}B_{\bm{\beta}, s}R_s(-1)^{t \in \mathcal{L}}(-1)^{t_p}B_{\bm{\beta}, t}R_t
\end{align}
\end{prop}
The close connection between the expressions for the Sherrington-Kirkpatrick and Erdos-Renyi models allows us to establish the following result:
\begin{repthm}{th:qaoa_sk_qaoa_maxcut}
Let $d \geq 3$ and $p \geq 1$ integers, $\bm{\beta}^{SK}, \bm{\gamma}^{SK} \in \mathbf{R}^{p}$. Consider a random $n$-vertices Erdos-Renyi graph of average degree $d$: $G \sim \er\left(n, \frac{d}{n - 1}\right)$. Then the expected energy of MaxCut-QAOA on $G$ is given as follows as $n \to \infty$:
{
\small
\begin{align}
    \label{eq:qaoa_sk_qaoa_maxcut_relation}
    & \lim_{n \to \infty}\mathbf{E}_{J \sim \er\left(n, \frac{d}{n - 1}\right)}\left\langle\Psi_{\textnormal{QAOA}}\left(J, \frac{\bm{\beta}^{SK}}{\sqrt{d}}, \frac{\bm{\gamma}^{SK}}{\sqrt{d}}\right)\bigg|\frac{H[J]}{n}\bigg|\Psi_{\textnormal{QAOA}}\left(J, \frac{\bm{\beta}^{SK}}{\sqrt{d}}, \frac{\bm{\gamma}^{SK}}{\sqrt{d}}\right)\right\rangle\nonumber\\
    & = \frac{1}{\sqrt{d}}\lim_{n \to \infty}\mathbf{E}_{J \sim \textnormal{SK}(n)}\left\langle\Psi_{\textnormal{QAOA}}\left(J, \bm{\beta}^{SK}, \bm{\gamma}^{SK}\right)\bigg|\frac{H[J]}{n}\bigg|\Psi_{\textnormal{QAOA}}\left(J, \bm{\beta}^{SK}, \bm{\gamma}^{SK}\right)\right\rangle + \mathcal{O}\left(\frac{1}{d}\right)
\end{align}
}
\end{repthm}
This results from the following succession of lemmas:
\begin{lem}
\label{lemma:R_symmetries}
Let $p$, $\bm{\beta}$, $\bm{\gamma}$ and $\left(R_s\right)_{s \in \{0, 1\}^{2p + 1}}$ be as in proposition \ref{prop:s_uv_eval}. The following holds:
\begin{align}
    R_{F(s)} & = R_s\\
    R_{\overline{s}} & = R_s,
\end{align}
where $\overline{s}$ denotes the bitstring obtained by flipping all bits of $s$.
\begin{proof}
These equations can be proved by induction, decreasingly with the level of symmetry $L(s)$. Let us consider the first equation for instance. For $L(s) \in \{p - 1, p\}$,
\begin{align*}
    R_{F(s)} & = e^{-d\left(1 - \frac{1}{2}\sum_{t \in \mathcal{L}_p \sqcup \mathcal{L}'_p}(-1)^{t \in \mathcal{L}}B_{\bm{\beta}, t}e^{i\varphi\left(\bm{\gamma}, F(s) \oplus t\right)}\right)}
\end{align*}
Now, $\varphi\left(\bm{\gamma}, F(s) \oplus t\right) = \varphi\left(\bm{\gamma}, s \oplus t\right)$ using proposition \ref{prop:beta_phi_invariance_partial_flip}, hence $R_{F(s)} = R_s$. Now, assume $L(s) < p - 1$ and $R_t = R_{F(t)}$ for all $t, L(t) > L(s)$. Then
\begin{align*}
    R_{F(s)} & = e^{-d\left(1 - \frac{1}{2}\sum_{t \in \mathcal{L}_p \sqcup \mathcal{L}'_p}(-1)^{t \in \mathcal{L}}B_{\bm{\beta}, t}e^{i\varphi(\bm{\gamma}, F(s) \oplus t)}\right)}e^{\frac{d}{2}\sum_{t, L(s) < L(t) < p}(-1)^{t \in \mathcal{L}}B_{\bm{\beta}, t}R_te^{i\varphi\left(\bm{\gamma}, F(s) \oplus t\right)}}\\
    & = e^{-d\left(1 - \frac{1}{2}\sum_{t \in \mathcal{L}_p \sqcup \mathcal{L}'_p}(-1)^{t \in \mathcal{L}}B_{\bm{\beta}, t}e^{i\varphi(\bm{\gamma}, s \oplus t)}\right)}e^{\frac{d}{2}\sum_{t, L(s) < L(t) < p}(-1)^{t \in \mathcal{L}}B_{\bm{\beta}, t}R_te^{i\varphi\left(\bm{\gamma}, F(s) \oplus t\right)}}\\
    & = e^{-d\left(1 - \frac{1}{2}\sum_{t \in \mathcal{L}_p \sqcup \mathcal{L}'_p}(-1)^{t \in \mathcal{L}}B_{\bm{\beta}, t}e^{i\varphi(\bm{\gamma}, s \oplus t)}\right)}e^{\frac{d}{2}\sum_{t, L(s) < L(t) < p}(-1)^{t \in \mathcal{L}}B_{\bm{\beta}, t}R_te^{i\varphi\left(\bm{\gamma}, F(s) \oplus F(t)\right)}}\\
    & = e^{-d\left(1 - \frac{1}{2}\sum_{t \in \mathcal{L}_p \sqcup \mathcal{L}'_p}(-1)^{t \in \mathcal{L}}B_{\bm{\beta}, t}e^{i\varphi(\bm{\gamma}, s \oplus t)}\right)}e^{\frac{d}{2}\sum_{t, L(s) < L(t) < p}(-1)^{t \in \mathcal{L}}B_{\bm{\beta}, t}R_te^{i\varphi\left(\bm{\gamma}, s \oplus t\right)}},
\end{align*}
where the second line follows from the same argument as in the case $s \in \{p - 1, p\}$ and the third line from induction.
\end{proof}
\end{lem}

\begin{lem}
Let $p$, $\bm{\beta}$, $\bm{\gamma}$ and $\left(R_s\right)_{s \in \{0, 1\}^{2p + 1}}$ be as in proposition \ref{prop:s_uv_eval}. The recursion defining $R_s$ can be rewritten as:
\begin{align*}
    R_s & := e^{-d\left(1 - \frac{1}{2}\sum_{t \in \mathcal{L}_p \sqcup \mathcal{L}'_p}(-1)^{t \in \mathcal{L}}B_{\bm{\beta}, t}\cos\varphi(\bm{\gamma}, s \oplus t)\right)} \qquad \textrm{for } L(s) \in \{p - 1, p\}\\
    R_s & := e^{-d\left(1 - \frac{1}{2}\sum_{t \in \mathcal{L}_p \sqcup \mathcal{L}'_p}(-1)^{t \in \mathcal{L}}B_{\bm{\beta}, t}\cos\varphi(\bm{\gamma}, s \oplus t)\right)}e^{\frac{d}{2}\sum_{t, L(s) < L(t) < p}(-1)^{t \in \mathcal{L}}B_{\bm{\beta}, t}R_t\cos\varphi(\bm{\gamma}, s \oplus t)}\nonumber\\
    & \hspace*{210px} \textrm{for } L(s) \leq p - 2
\end{align*}
\begin{proof}
For $L(s) \in \{p - 1, p\}$,
\begin{align*}
    R_s & = e^{-d\left(1 - \frac{1}{2}\sum_{t \in \mathcal{L}_p \sqcup \mathcal{L}'_p}(-1)^{t \in \mathcal{L}}B_{\bm{\beta}, t}e^{i\varphi(\bm{\gamma}, s \oplus t)}\right)}
\end{align*}
Now,
\begin{align*}
    \sum_{t \in \mathcal{L}_p \sqcup \mathcal{L}'_p}(-1)^{t \in \mathcal{L}}B_{\bm{\beta}, t}e^{i\varphi(\bm{\gamma}, s \oplus t)} & = \sum_{t \in \mathcal{L}_p \sqcup \mathcal{L}'_p}(-1)^{\overline{t} \in \mathcal{L}}B_{\bm{\beta}, \overline{t}}e^{i\varphi(\bm{\gamma}, s \oplus \overline{t})}\\
    & = \sum_{t \in \mathcal{L}_p \sqcup \mathcal{L}'_p}(-1)^{t \in \mathcal{L}}B_{\bm{\beta}, t}e^{-i\varphi(\bm{\gamma}, s \oplus t)}
\end{align*}
Averaging the l.h.s and r.h.s gives the result for $s \in \{p - 1, p\}$. Next, assume $L(s) < p - 1$. Then
\begin{align*}
    R_s & = e^{-d\left(1 - \frac{1}{2}\sum_{t \in \mathcal{L}_p \sqcup \mathcal{L}'_p}(-1)^{t \in \mathcal{L}}B_{\bm{\beta}, t}e^{i\varphi(\bm{\gamma}, s \oplus t)}\right)}e^{\frac{d}{2}\sum_{t, L(s) < L(t) < p}(-1)^{t \in \mathcal{L}}B_{\bm{\beta}, t}R_te^{i\varphi(\bm{\gamma}, s \oplus t)}}\\
    & = e^{-d\left(1 - \frac{1}{2}\sum_{t \in \mathcal{L}_p \sqcup \mathcal{L}'_p}(-1)^{t \in \mathcal{L}}B_{\bm{\beta}, t}\cos\varphi(\bm{\gamma}, s \oplus t)\right)}e^{\frac{d}{2}\sum_{t, L(s) < L(t) < p}(-1)^{t \in \mathcal{L}}B_{\bm{\beta}, t}R_te^{i\varphi(\bm{\gamma}, s \oplus t)}}\\
    & = e^{-d\left(1 - \frac{1}{2}\sum_{t \in \mathcal{L}_p \sqcup \mathcal{L}'_p}(-1)^{t \in \mathcal{L}}B_{\bm{\beta}, t}\cos\varphi(\bm{\gamma}, s \oplus t)\right)}e^{\frac{d}{2}\sum_{t, L(s) < L(t) < p}(-1)^{t \in \mathcal{L}}B_{\bm{\beta}, t}R_t\cos\varphi(\bm{\gamma}, s \oplus t)},
\end{align*}
where the second line is obtained as in the case $L(s) \in \{p - 1, p\}$ and the third line follows from the same change of variable trick ($t \to \overline{t}$), using additionally that $R_{\overline{t}} = R_t$ (lemma \ref{lemma:R_symmetries}).
\end{proof}
\end{lem}

\begin{lem}
\label{lemma:r_vs_r_tilde}
Let $p \geq 1, d \geq 2$ integers and $\bm{\beta}^{SK}, \bm{\gamma}^{SK} \in \mathbf{R}^p$. Let $\left(R_s\right)_{s \in \{0, 1\}^{2p + 1}}$ be defined as in proposition \ref{prop:s_uv_eval}, letting $\bm{\beta} := \bm{\beta}^{SK}, \bm{\gamma} := \frac{\bm{\gamma}^{SK}}{\sqrt{d}}$ there and $\left(\widetilde{R}_s\right)_{s \in \{0, 1\}^{2p + 1}}$ be defined as in proposition \ref{prop:sk_qaoa_energy_eval}, letting $\bm{\beta} := \bm{\beta}^{SK}, \bm{\gamma} := \bm{\gamma}^{SK}$ there. Then, as $d \to \infty$,
\begin{align}
    R_s & = \widetilde{R}_s + \mathcal{O}\left(\frac{1}{d}\right)
\end{align}
\begin{proof}
This can be proved by induction, decreasingly with $L(s)$. For $L(s) \in \{p - 1, p\}$,
\begin{align*}
    R_s & = e^{-d\left(1 - \frac{1}{2}\sum_{t \in \mathcal{L}_p \sqcup \mathcal{L}'_p}(-1)^{t \in \mathcal{L}}B_{\bm{\beta}^{SK}, t}\cos\varphi\left(\frac{\bm{\gamma}^{SK}}{\sqrt{d}}, s \oplus t\right)\right)}\\
    & = e^{-d\left[1 - \frac{1}{2}\sum_{t \in \mathcal{L}_p \sqcup \mathcal{L}'_p}(-1)^{t \in \mathcal{L}}B_{\bm{\beta}^{SK}, t}\left(1 - \frac{1}{2d}\varphi(\bm{\gamma}^{SK}, s \oplus t)^2 + \mathcal{O}\left(\frac{1}{d^2}\right)\right)\right]}\\
    & = e^{-d\left(1 - \frac{1}{2}\sum_{t \in \mathcal{L}_p \sqcup \mathcal{L}'_p}(-1)^{t \in \mathcal{L}}B_{\bm{\beta}^{SK}, s}\right)}e^{-\frac{1}{4}\sum_{t \in \mathcal{L}_p \sqcup \mathcal{L}'_p}(-1)^{t \in \mathcal{L}}B_{\bm{\beta}^{SK}, t}\varphi(\bm{\gamma}^{SK}, s \oplus t)^2} + \mathcal{O}\left(\frac{1}{d}\right)
\end{align*}
It remains to show that the first exponential is $1$. This follows from lemma \ref{lemma:sum_betas}.

This proves the result for $L(s) \in \{p - 1, p\}$. Next, for $L(s) < p - 1$,
\begin{align*}
    R_s & = e^{-d\left(1 - \frac{1}{2}\sum_{t \in \mathcal{L}_p \sqcup \mathcal{L}'_p}(-1)^{t \in \mathcal{L}}B_{\bm{\beta}^{SK}, t}\cos\varphi\left(\frac{\bm{\gamma}^{SK}}{\sqrt{d}}, s \oplus t\right)\right)}e^{\frac{d}{2}\sum_{t, L(s) < L(t) < p}(-1)^{t \in \mathcal{L}}B_{\bm{\beta}^{SK}, t}R_t\cos\varphi\left(\frac{\bm{\gamma}^{SK}}{\sqrt{d}}, s \oplus t\right)}
\end{align*}
The first factor is treated the same way as before. For the second factor,
\begin{align*}
    & \frac{d}{2}\sum_{t, L(s) < L(t) < p}(-1)^{t \in \mathcal{L}}B_{\bm{\beta}^{SK}, t}R_t\cos\varphi\left(\frac{\bm{\gamma}^{SK}}{\sqrt{d}}, s \oplus t\right)\\
    & = \frac{d}{2}\sum_{t, L(s) < L(t) < p}(-1)^{t \in \mathcal{L}}B_{\bm{\beta}^{SK}, t}R_t - \frac{1}{4}\sum_{t, L(s) < L(t) < p}(-1)^{t \in \mathcal{L}}B_{\bm{\beta}^{SK}, t}R_t\varphi\left(\bm{\gamma}^{SK}, s \oplus t\right)^2 + \mathcal{O}\left(\frac{1}{d}\right)
\end{align*}
Since the first term cancels:
\begin{align*}
    \sum_{t, L(s) < L(t) < p}(-1)^{t \in \mathcal{L}}B_{\bm{\beta}^{SK}, t}R_t & = \sum_{t, L(s) < L(t) < p}(-1)^{F(t) \in \mathcal{L}}B_{\bm{\beta}^{SK}, F(t)}R_{F(t)}\\
    & = -\sum_{t, L(s) < L(t) < p}(-1)^{t \in \mathcal{L}}B_{\bm{\beta}^{SK}, t}R_t,
\end{align*}
this completes the proof.
\end{proof}
\end{lem}
Using lemma \ref{lemma:r_vs_r_tilde}, theorem \ref{th:qaoa_sk_qaoa_maxcut} follows from straightforward calculations. Indeed, from proposition \ref{prop:qaoa_infinite_size_limit_energy},
\begin{align*}
    & \lim_{n \to \infty}\mathbf{E}_{J \sim \er\left(n, \frac{d}{n - 1}\right)}\left[\left\langle\Psi_{\textnormal{QAOA}}\left(J, \bm{\beta}^{SK}, \frac{\bm{\gamma}^{SK}}{\sqrt{d}}\right)\bigg|\frac{H(J)}{nd/2}\bigg|\Psi_{\textnormal{QAOA}}\left(J, \bm{\beta}^{SK}, \frac{\bm{\gamma}^{SK}}{\sqrt{d}}\right)\right\rangle\right]\nonumber\\
    & = \frac{1}{2}\left(\frac{1}{2}\sum_{s \in \{0, 1\}^{2p + 1}}R_s^2B_{\bm{\beta}^{SK}, s}^2 + \sum_{\substack{s, t \in \{0, 1\}^{2p + 1}\\s < t}}(-1)^{s \in \mathcal{L}}(-1)^{s_p}B_{\bm{\beta}^{SK}, s}(-1)^{t \in \mathcal{L}}(-1)^{t_p}B_{\bm{\beta}^{SK}, t}e^{i\varphi\left(\frac{\bm{\gamma}^{SK}}{\sqrt{d}}, s \oplus t\right)}\right)\\
    & = \frac{1}{4}\left(\sum_s(-1)^{s \in \mathcal{L}}(-1)^{s_p}B_{\bm{\beta}^{SK}, s}R_s\right)^2\\
    & \hspace*{20px} + \frac{1}{4\sqrt{d}}\sum_{s, t \in \{0, 1\}^{2p + 1}}(-1)^{s \in \mathcal{L}}(-1)^{s_p}B_{\bm{\beta}^{SK}, s}R_s(-1)^{t \in \mathcal{L}}(-1)^{t_p}B_{\bm{\beta}^{SK}, t}R_ti\varphi\left(\bm{\gamma}^{SK}, s \oplus t\right) + \mathcal{O}\left(\frac{1}{d}\right).
\end{align*}
The first term vanishes:
\begin{align*}
    \sum_s(-1)^{s \in \mathcal{L}}(-1)^{s_p}B_{\bm{\beta}^{SK}, s}R_s & = \sum_s(-1)^{\overline{s} \in \mathcal{L}}(-1)^{\overline{s}_p}B_{\bm{\beta}^{SK}, \overline{s}}R_{\overline{s}}\\
    & = -\sum_s(-1)^{s \in \mathcal{L}}(-1)^{s_p}B_{\bm{\beta}^{SK}, s}R_s, \qquad (R_s = R_{\overline{s}}\textrm{ by lemma }\ref{lemma:R_symmetries}\textrm{)}
\end{align*}
which yields theorem \ref{th:qaoa_sk_qaoa_maxcut} after applying lemma \ref{lemma:r_vs_r_tilde} and recalling proposition \ref{prop:sk_qaoa_energy_eval}.

\subsection{Analysis for MaxCut on sparse graphs with a degree distribution}
\label{sec:qaoa_maxcut_degree_distribution}
In this section, we derive an expression for the QAOA energy of an average MaxCut instance, where the underlying graph is generated from the pseudo Chung-Lu model (definition \ref{def:pseudo_chung_lu_model}). We will frequently use letters $l, l'$ to refer to the label of a vertex in this model (and not to the level of symmetry of a bitstring). Similar to the derivation of the QAOA energy for the $D$-spin model in section \ref{sec:qaoa_moments_dense_diluted}, we start from the results in section \ref{sec:preliminaries}, notably proposition \ref{prop:qaoa_tn_reorganized}.

It remains to calculate the average of Hamiltonian contributions
\begin{align}
    \mathbf{E}_{J \sim \textrm{PseudoChungLu}\left(n, (q_l)_{1 \leq l \leq N_L}, (d_l)_{1 \leq l \leq N_L}\right)}\left[\bigotimes_{0 \leq j < p}e^{i\frac{\gamma_j}{2}H[J]} \otimes e^{i\frac{\gamma'}{2}H[J]} \otimes \bigotimes_{0 \leq j < p}e^{-i\frac{\gamma_{p - 1 - j}}{2}H[J]}\right].
\end{align}
Recall that a Chung-Lu random graph described by an adjacency matrix $J$ is obtained by choosing a label $l_v \in \{1, \ldots, N_L\}$ independently for each vertex $v \in [n]$, where label $l$ is chosen with probability $q_l$. Then, for each pair of vertices labelled $l, l'$, an edge is created between them with probability $q_{ll'} = \frac{1}{n - 1}\frac{d_ld_{l'}}{\sum_{l''}q_{l''}d_{l''}}$. Given a labelling $\left(l_v\right)_{v \in [n]}$ of the vertices, we denote $I_l := \left\{v \in [n]\,:\,l_v = l\right\}$. Next,
\begin{align*}
    & \mathbf{E}_J\left[\bigotimes_{0 \leq j < p}e^{i\frac{\gamma_j}{2}H[J]} \otimes e^{i\frac{\gamma'}{2}H[J]} \otimes \bigotimes_{0 \leq j < p}e^{-i\frac{\gamma_{p - 1 - j}}{2}H[J]}\right]\\
    & = \mathbf{E}_J\left[\mathbf{E}_J\left[\bigotimes_{0 \leq j < p}e^{i\frac{\gamma_j}{2}H[J]} \otimes e^{i\frac{\gamma'}{2}H[J]} \otimes \bigotimes_{0 \leq j < p}e^{-i\frac{\gamma_{p - 1 - j}}{2}H[J]}\Bigg|(l_v)_{v \in [n]}\right]\right]
\end{align*}
The conditional expectation can be evaluated:
\begin{align*}
    & \mathbf{E}_J\left[\bigotimes_{0 \leq j < p}e^{i\frac{\gamma_j}{2}H[J]} \otimes e^{i\frac{\gamma'}{2}H[J]} \otimes \bigotimes_{0 \leq j < p}e^{-i\frac{\gamma_{p - 1 - j}}{2}H[J]}\Bigg|(l_v)_{v \in [n]}\right]\\
    & = \mathbf{E}_J\left[\prod_{\{j, k\} \subset [n]}\bigotimes_{0 \leq j < p}e^{i\frac{\gamma_j}{2}J_{\{j, k\}}Z_jZ_k} \otimes e^{i\frac{\gamma'}{2}J_{\{j, k\}}Z_kZ_l} \otimes \bigotimes_{0 \leq j < p}e^{-i\frac{\gamma_{p - 1 - j}}{2}J_{\{j, k\}}Z_kZ_l}\Bigg|(l_v)_{v \in [n]}\right]\\
    & = \prod_{\{j, k\} \subset [n]}\mathbf{E}_J\left[\bigotimes_{0 \leq j < p}e^{i\frac{\gamma_j}{2}J_{\{j, k\}}Z_jZ_k} \otimes e^{i\frac{\gamma'}{2}J_{\{j, k\}}Z_kZ_l} \otimes \bigotimes_{0 \leq j < p}e^{-i\frac{\gamma_{p - 1 - j}}{2}J_{\{j, k\}}Z_kZ_l}\Bigg|(l_v)_{v \in [n]}\right]\\
    & = \prod_{\{j, k\} \subset [n]}\left(1 - q_{l_jl_k} + q_{l_jl_k}\bigotimes_{0 \leq j < p}e^{i\frac{\gamma_j}{2}Z_jZ_k} \otimes e^{i\frac{\gamma'}{2}Z_jZ_k} \otimes \bigotimes_{0 \leq j < p}e^{-i\frac{\gamma_{p - 1 - j}}{2}Z_jZ_k}\right).
\end{align*}
Finally, the probability of a labelling, characterized by the labels $\left(l_j\right)_{j \in [n]}$ or the sets $\left(I_l\right)_{1 \leq l \leq N_L}$ defined above, is
\begin{align}
    \binom{n}{\left(|I_l|\right)_{1 \leq l \leq N_L}}\prod_{1 \leq l \leq N_L}q_l^{|I_l|}
\end{align}
This yields:
\begin{align*}
    & \mathbf{E}_J\left[\bigotimes_{0 \leq j < p}e^{i\frac{\gamma_j}{2}H[J]} \otimes e^{i\frac{\gamma'}{2}H[J]} \otimes \bigotimes_{0 \leq j < p}e^{-i\frac{\gamma_{p - 1 - j}}{2}H[J]}\right]\\
    & = \sum_{\substack{(I_l)_{1 \leq l \leq N_L}\\I_l \subset [n]\\\sqcup_{1 \leq l \leq N_L}I_l = [n]}}\binom{n}{\left(|I_l|\right)_l}q_l^{|I_l|}\prod_{\{j, k\} \subset [n]}\left(1 - q_{l_jl_k} + q_{l_jl_k}\bigotimes_{0 \leq j < p}e^{i\frac{\gamma_j}{2}Z_jZ_k} \otimes e^{i\frac{\gamma'}{2}Z_jZ_k} \otimes \bigotimes_{0 \leq j < p}e^{-i\frac{\gamma_{p - 1 - j}}{2}Z_jZ_k}\right)
\end{align*}
By proposition \ref{prop:average_hamiltonian_tensor}, the latter is diagonal in the $\left(\ket{n_s}\right)_{s \in \{0, 1\}^{2p + 1}}$ basis and
\begin{align*}
    & \mathbf{E}_J\left[\bigotimes_{0 \leq j < p}e^{i\frac{\gamma_j}{2}H[J]} \otimes e^{i\frac{\gamma'}{2}H[J]} \otimes \bigotimes_{0 \leq j < p}e^{-i\frac{\gamma_{p - 1 - j}}{2}H[J]}\right]\ket{\left(n_s\right)_s}\\
    & = \frac{1}{2^n}\sum_{\substack{(n_{s, l})_{s \in \{0, 1\}^{2p + 1}, 1 \leq l \leq N_L}\\\sum_ln_{s, l} = n_s}}\left[\prod_{s \in \{0, 1\}^{2p + 1}}\binom{n_s}{\left(n_{s, l}\right)_l}\right]\prod_{s, l}q_l^{n_{s, l}}B_{\bm{\beta}, s}^{n_{s, l}}\left(1 - q_{ll} + q_{ll}e^{i\frac{\gamma'}{2}}\right)^{\frac{n_{s, l}(n_{s, l} - 1)}{2}}\\
    & \hspace*{0.1\textwidth} \times \prod_{\substack{s, l, l'\\l < l'}}\left(1 - q_{ll'} + q_{ll'}e^{i\frac{\gamma'}{2}}\right)^{n_{s, l}n_{s, l'}}\prod_{\substack{s, s', l, l'\\s < s'}}\left(1 - q_{ll'} + q_{ll'}e^{i\varphi(\bm{\gamma}, s \oplus s') + i\frac{\gamma'}{2}}\right)^{n_{s, l}n_{s', l'}}\ket{\left(n_s\right)_s}
\end{align*}
The expected energy of the QAOA follows:
\begin{prop}
\label{prop:qaoa_energy_maxcut_degree_distribution}
Let a graph be generated according to the Chung-Lu model as stated above, with labels $l \in \{1, \ldots, N_L\}$ and degrees $\left(d_l\right)_{1 \leq l \leq N_L}$. The expected energy of the MaxCut-QAOA on an average such instance is:
\begin{align}
    & \mathbf{E}_{J \sim \textnormal{PseudoChungLu}\left(n, \left(q_l\right)_{1 \leq l \leq N_L}, \left(d_l\right)_{1 \leq l \leq N_L}\right)}\left[\braket{\Psi_{\textnormal{QAOA}}(J, \bm\beta, \bm\gamma)|H[J]|\Psi_{\textnormal{QAOA}}(J, \bm\beta, \bm\gamma)}\right]\nonumber\\
    & = \sum_{\left(n_{s, l}\right)_{\substack{s \in \{0, 1\}^{2p + 1}\\1 \leq l \leq N_l}}}\frac{1}{2^n}\binom{n}{\left(n_{s, l}\right)_{s, l}}(-1)^{\sum_{\substack{s \in \mathcal{L}\\l}}n_{s, l}}\prod_{\substack{s \in \{0, 1\}^{2p + 1}\\l}}q_l^{n_{s, l}}B_{\bm{\beta}, s}^{n_{s, l}}\prod_{\substack{s, s'\,:\,s < s'\\l, l'}}\left(1 - q_{ll'} + q_{ll'}e^{i\varphi(\bm\gamma, s \oplus s')}\right)^{n_{s', l'}n_{s, l}}\nonumber\\
    & \hspace*{10px} \times \left(\sum_{\substack{s \in \{0, 1\}^{2p + 1}\\l}}q_{ll}\frac{n_{s, l}(n_{s, l} - 1)}{2} + \sum_{\substack{s \in \{0, 1\}^{2p + 1}\\l, l'\,:\,l < l'}}q_{ll'}n_{s, l}n_{s, l'} + \sum_{\substack{s, s'\,:\,s < s'\\l, l'}}q_{ll'}\frac{(-1)^{s_p + s_p'}e^{i\varphi(\bm{\gamma}, s \oplus s')}}{1 - q_{ll'} + q_{ll'}e^{i\varphi(\bm{\gamma}, s \oplus s')}}n_{s', l'}n_{s, l}\right)
\end{align}
where
\begin{align}
    q_{ll'} & := \frac{1}{n - 1}\frac{d_ld_{l'}}{\overline{d}}\\
    \overline{d} & := \sum_{l''}q_{l''}d_{l''}.
\end{align}
\end{prop}
The latter energy can be evaluated in the infinite size limit in a similar fashion as the MaxCut problem on Erdos-Renyi graphs discussed in section \ref{sec:analysis_maxcut_constant_degree}. Precisely, proposition \ref{prop:qaoa_infinite_size_limit_energy} has the following analogue:
\begin{prop}
\label{prop:qaoa_energy_maxcut_degree_distribution_infinite_size}
Let a graph be generated according to the pseudo-Chung-Lu model as stated above, with labels $l \in \{1, \ldots, N_L\}$ and degrees $\left(d_l\right)_{1 \leq l \leq N_L}$. The expected energy of the MaxCut-QAOA on an average such instance in the infinite size limit:
\begin{align}
    & \lim_{n \to \infty}\mathbf{E}_{J \sim \textnormal{PseudoChungLu}\left(n, \left(q_l\right), \left(d_l\right)\right)}\left[\left\langle\Psi_{\textnormal{QAOA}}(J, \bm{\beta}, \bm{\gamma})\bigg|\frac{H[J]}{n}\bigg|\Psi_{\textnormal{QAOA}}(J, \bm{\beta}, \bm{\gamma})\right\rangle\right]\nonumber\\
    & = \frac{1}{4}\left(\sum_{s, l}\frac{d_ld_{l'}}{\overline{d}}q_l^2B_{\bm\beta, s}^2R_{(s, l)}^2 + \sum_{s, l < l'}\frac{d_ld_{l'}}{\overline{d}}q_lq_{l'}B_{\bm\beta, s}^2R_{(s, l)}R_{(s, l')}\right.\\
    & \left.\hspace*{50px} + \sum_{s < s', l, l'}\frac{d_ld_{l'}}{\overline{d}}(-1)^{s \in \mathcal{L}}q_lB_{\bm\beta, s}R_{(s, l)}(-1)^{s' \in \mathcal{L}}q_{l'}B_{\bm\beta, s'}R_{(s', l')}\right)
\end{align}
where $\left(R_{(s, l)}\right)_{s \in \{0, 1\}^{2p + 1}, 1 \leq l \leq N_L}$ is defined recursively as follows:
\begin{align}
    R_{(s, l)} & = 1 & \qquad L(s) = p\\
    R_{\left(s, l\right)} & = \exp\left(-d_l\left(1 - \frac{1}{2}\sum_{s' \in \mathcal{L}_p \sqcup \mathcal{L}'_p}(-1)^{s' \in \mathcal{L}}B_{\bm{\beta}, s'}e^{i\varphi\left(\bm{\gamma}, s \oplus s'\right)}\right)\right) & \qquad L(s) < p\nonumber\\
    & \hspace*{0.05\textwidth} \times \exp\left(\sum_{\substack{s'\,:\,L(s') > s\\l'}}\frac{d_ld_{l'}}{2\overline{d}}q_{l'}(-1)^{s' \in \mathcal{L}}B_{\bm{\beta}, s'}e^{i\varphi\left(\bm{\gamma}, s \oplus s'\right)}R_{\left(s', l'\right)}\right)
\end{align}
\end{prop}

\subsection{Analysis for diluted $D$-spin model at level $p = 1$}
\label{sec:analysis_diluted_d_spin_model}
In this section, we relate the energy achieved by the $p = 1$ QAOA on the diluted $D$-spin model to the energy achieved on the dense $D$-spin model. The results for the dense $D$-spin model were established in \cite{2102.12043}. We then consider, following proposition \ref{prop:qaoa_diluted_spin_model} applied with $p = 1$:
\begin{align}
    & \mathbf{E}_{J \sim \textnormal{Diluted}(D, n, d)}\left[\braket{\Psi_{\textnormal{QAOA}}(J, \beta, \gamma)|H[J]|\Psi_{\textnormal{QAOA}}(J, \beta, \gamma)}\right]\nonumber\\
    & = \frac{1}{2^n}\sum_{\substack{\left(n_s\right)_{s \in \{0, 1\}^3}\\n_s \geq 0}}\binom{n}{(n_s)_s}(-1)^{\sum_{s \in \mathcal{L}}n_s}\prod_sB_{\beta, s}^{n_s}\prod_{\substack{\{D_s\}_{s \in \{0, 1\}^3}\\\sum_sD_s = D}}\left(1 - \frac{nd}{D\binom{n}{D}} + \frac{nd}{D\binom{n}{D}}e^{i\varphi\left(\gamma, \oplus_ss^{\oplus D_s}\right)}\right)^{\prod_s\binom{n_s}{D_s}}\nonumber\\
    & \hspace*{65px} \times \sum_{\substack{\{D_s\}_{s \in \{0, 1\}^3}\\\sum_sD_s = D}}\prod_s\binom{n_s}{D_s}\frac{\frac{nd}{D\binom{n}{D}}e^{i\varphi\left(\gamma, \oplus_ss^{\oplus D_s}\right)}(-1)^{\sum_sD_ss_p}}{1 - \frac{nd}{D\binom{n}{D}} + \frac{nd}{D\binom{n}{D}}e^{i\varphi\left(\gamma, \oplus_ss^{\oplus D_s}\right)}}\label{eq:diluted_spin_model_p_1}
\end{align}
Here, we denoted by $\beta, \gamma$ (without boldface) the unique angles in the $\bm\beta, \bm\gamma$ vectors. For reference, we reproduce below the values of $B_{\bm\beta, s}$ for all $s \in \{0, 1\}^3$:
\begin{align}
    B_{\beta, 000} = \cos^2\frac{\beta}{2} && B_{\beta, 001} = i\cos\frac{\beta}{2}\sin\frac{\beta}{2} && B_{\beta, 010} = -\sin^2\frac{\beta}{2} && B_{\beta, 011} = i\cos\frac{\beta}{2}\sin\frac{\beta}{2}\\
    B_{\beta, 100} = i\cos\frac{\beta}{2}\sin\frac{\beta}{2} && B_{\beta, 101} = -\sin^2\frac{\beta}{2} && B_{\beta, 110} = i\cos\frac{\beta}{2}\sin\frac{\beta}{2} && B_{\beta, 111} = \cos^2\frac{\beta}{2}
\end{align}
Besides, $\mathcal{L} = \left\{001, 010, 101, 110\right\}$. Analyzing \ref{eq:diluted_spin_model_p_1} requires to estimate the sums
\begin{align}
    \frac{1}{2^n}\sum_{(n_s)_s}\binom{n}{(n_s)_s}(-1)^{\sum_{s \in \mathcal{L}}n_s}\prod_sB_{\beta, s}^{n_s}\prod_{\substack{(D_s)\\\sum_sD_s = D}}\left(1 - \frac{nd}{D\binom{n}{D}} + \frac{nd}{D\binom{n}{D}}e^{i\varphi\left(\gamma, \oplus_ss^{\oplus D_s}\right)}\right)^{\prod_s\binom{n_s}{D_s}}\prod_s\binom{n_s}{(\Delta_s)_s}\label{eq:diluted_spin_model_p1_moment}
\end{align}
for all $\left(\Delta_s\right)_{s \in \{0, 1\}^3}$ such that $\sum_s\Delta_s = D$. This is done in the next proposition, which is partly similar to the results in \cite{2102.12043} but applies to the diluted instead of the dense spin model and is proved using slightly different methods. [For instance, one decomposes polynomials in the variables $\left(n_s\right)_s$ on the $\left(\prod\limits_{s \in \{0, 1\}^{2p + 1}}\binom{n_s}{d_s}\right)_{d_s \geq 0}$ basis instead of the canonical basis $\left(\prod\limits_{s \in \{0, 1\}^{2p + 1}}n_s^{d_s}\right)_{d_s \geq 0}$ used in \cite{2102.12043}. This allows to use straightaway the simple identity $\sum\limits_{\left(n_s\right)_s\,:\,\sum_sn_s = n}\prod_sx_s^{n_s}\prod_s\binom{n_s}{d_s} = \left(\sum_sx_s\right)^{n - \sum_sd_s}\prod_sx_s^{d_s}$. Besides, the asymmetry of the probability distribution for hyperedge weights in the diluted $D$-spin model (as opposed to the dense $D$-spin model) requires an extra technical step not present in \cite{2102.12043}.]
\begin{prop}
\label{prop:diluted_spin_model_p_1_moment_infinite_size}
Let $\left(\Delta_s\right)_{s \in \{0, 1\}^3}$ be integers summing to $D$. Then the sum in equation \ref{eq:diluted_spin_model_p1_moment} evaluates to
\begin{align}
    & \binom{n}{D}\left(\frac{1}{2^D}(-1)^{\sum_{s \in \mathcal{L}}\Delta_s}\binom{D}{(\Delta_s)_s}\left(\prod_sB_{\beta, s}^{\Delta_s}\right)e^{-(\Delta_{001} + \Delta_{011} + \Delta_{100} + \Delta_{110})d(1 - \cos\gamma)} + o(1)\right)
\end{align}
as $n \to \infty$.
\begin{proof}
To lighten the notation, let
\begin{align}
\label{eq:diluted_spin_model_p_1_def_q_C}
    q & := \frac{nd}{D\binom{n}{D}}\\
    C_{\gamma, s} & := \left(1 - q + qe^{i\varphi\left(\gamma, \oplus_ss^{\oplus D_s}\right)}\right).
\end{align}
The sum to estimate is 
\begin{align*}
    & \sum_{\{n_s\}}\frac{1}{2^n}\binom{n}{(n_s)_s}(-1)^{\sum_{s \in \mathcal{L}}n_s}\prod_sB_{\beta, s}^{n_s}\prod_{(D_s)}C_{\gamma, \oplus_ss^{\oplus D_s}}^{\prod_s\binom{n_s}{D_s}}\prod_s\binom{n_s}{\Delta_s}.
\end{align*}
By combining the multinomial coefficients $\binom{n}{(n_s)_s}$ and $\prod_s\binom{n_s}{\Delta_s}$ and shifting variables, this becomes
\begin{align}
\label{eq:diluted_spin_model_p_1_variable_shifted}
    & \binom{n}{D}\binom{D}{(\Delta_s)_s}(-1)^{\sum_{s \in \mathcal{L}}\Delta_s}\prod_sB_{\beta, s}^{\Delta_s}\sum_{\substack{(n_s)_s\\\sum_sn_s = n - D}}\frac{1}{2^n}\binom{n - D}{(n_s)_s}(-1)^{\sum_{s \in \mathcal{L}}n_s}\prod_sB_{\beta, s}^{n_s}\prod_{(D_s)}C_{\gamma, \oplus_ss^{\oplus D_s}}^{\prod_s\binom{n_s + \Delta_s}{D_s}}
\end{align}
Now, let us define variables $n'_{s_2s_0}$ and $\Delta'_{s_2s_0}$ as follows:
\begin{align}
    n'_{s_2s_0} & := n_{s_20s_0} + n_{s_21s_0}\\
    \Delta'_{s_2s_0} & := \Delta_{s_20s_0} + \Delta_{s_21s_0}
\end{align}
We now show that $\prod_{(D_s)}C_{\beta, \oplus_ss^{\oplus D_s}}^{\prod_s\binom{n_s + \Delta_s}{D_s}}$ depends only on $\left(n'_{s_2s_0}\right)_{s_0, s_2 \in \{0, 1\}}$ and $\left(\Delta'_{s_2s_0}\right)_{s_0, s_2 \in \{0, 1\}}$ and not $(n_{s_2s_1s_0})_{s_0, s_1, s_2 \in \{0, 1\}}$ and $(\Delta_{s_2s_1s_0})_{s_0, s_1, s_2 \in \{0, 1\}}$ individually. Indeed,
\begin{align*}
    \prod_{(D_s)}C_{\gamma, \oplus_ss^{\oplus D_s}}^{\prod_s\binom{n_s + \Delta_s}{D_s}} & = \prod_{(D_s)}C_{\gamma, \oplus_{s_0, s_2 \in \{0, 1\}}(s_20s_0)^{\oplus\left(D_{s_20s_0} + D_{s_21s_0}\right)}}^{\prod_s\binom{n_s + \Delta_s}{D_s}}\\
    & = \prod_{\substack{\left(D'_{s_2s_0}\right)_{s_0, s_2 \in \{0, 1\}}\\\sum_{s_0, s_2}D'_{s_2s_0} = D}}\prod_{\substack{\left(D_s\right)_{s \in \{0, 1\}^3}\\D_{s_20s_0} + D_{s_21s_0} = D'_{s_2s_0}}}C_{\gamma, \oplus_{s_0, s_2}(s_20s_0)^{\oplus D'_{s_2s_0}}}^{\prod_{s_0, s_2}\binom{n_{s_20s_0} + \Delta_{s_20s_0}}{D_{s_20s_0}}\binom{n_{s_21s_0} + \Delta_{s_21s_0}}{D_{s_21s_0}}}\\
    & = \prod_{\substack{\left(D'_{s_2s_0}\right)_{s_2, s_0}\\\sum_{s_0, s_2}D'_{s_2s_0} = D}}C_{\gamma, \oplus_{s_0, s_2}(s_20s_0)^{\oplus D'_{s_2s_0}}}^{\sum_{\substack{\left(D_s\right)_{s \in \{0, 1\}^3}\\D_{s_20s_0} + D_{s_21s_0} = D'_{s_2s_0}}}\prod_{s_0, s_2}\binom{n_{s_20s_0} + \Delta_{s_20s_0}}{D_{s_20s_0}}\binom{n_{s_21s_0 + \Delta_{s_21s_0}}}{D_{s_21s_0}}}
\end{align*}
Using the identity:
\begin{align}
    \sum_{0 \leq k \leq c}\binom{a}{k}\binom{b}{c - k} & = \binom{a + b}{c},
\end{align}
the expression above becomes:
\begin{align*}
    \prod_{(D_s)}C_{\gamma, \oplus_ss^{\oplus D_s}}^{\prod_s\binom{n_s + \Delta_s}{D_s}} & = \prod_{\substack{\left(D'_{s_2s_0}\right)_{s_0, s_2}\\\sum_{s_0, s_2}D'_{s_2s_0} = D}}C_{\gamma, \oplus_{s_0, s_2}(s_20s_0)^{D'_{s_2s_0}}}^{\prod_{s_0, s_2}\binom{n'_{s_2s_0} + \Delta'_{s_2s_0}}{D'_{s_2s_0}}},
\end{align*}
which depends only on the $n'_{s_2s_0}$ and $\Delta'_{s_2s_0}$ as announced. We now specify the latter expression further depending on the parity of $D$. If $D$ is even,
\begin{align*}
    \prod_{\left(D_s\right)}C_{q, \bm{\gamma}, \oplus_ss^{\oplus D_s}}^{\prod_s\binom{n_s + \Delta_s}{D_s}} & = \left(1 - q + qe^{-i\gamma}\right)^{\sum_{\substack{\{D'_s\}\\\left(D'_{00}, D'_{11}, D'_{01}, D'_{10}\right)\,\textrm{mod}\,2\,\in\\\{(0, 1, 1, 0), (1, 0, 0, 1)\}}}\binom{n'_{00} + \Delta'_{00}}{D'_{00}}\binom{n'_{11} + \Delta'_{11}}{D'_{11}}\binom{n'_{01} + \Delta'_{01}}{D'_{01}}\binom{n'_{10} + \Delta'_{10}}{D'_{10}}}\\
    & \hspace*{0.05\textwidth} \times \left(1 - q + qe^{i\gamma}\right)^{\sum_{\substack{\{D'_s\}\\\left(D'_{00}, D'_{11}, D'_{01}, D'_{10}\right)\,\textrm{mod}\,2\,\in\\\{(0, 1, 0, 1), (1, 0, 1, 0)\}}}\binom{n'_{00} + \Delta'_{00}}{D'_{00}}\binom{n'_{11} + \Delta'_{11}}{D'_{11}}\binom{n'_{01} + \Delta'_{01}}{D'_{01}}\binom{n'_{10} + \Delta'_{10}}{D'_{10}}}
\end{align*}
If $D$ is odd,
\begin{align*}
    \prod_{\{D_s\}}C_{q, \bm{\gamma}, \oplus_ss^{\oplus D_s}}^{\prod_s\binom{n_s + \Delta_s}{D_s}} & = \left(1 - q + qe^{-i\gamma}\right)^{\sum_{\substack{\{D'_s\}\\\left(D'_{00}, D'_{11}, D'_{01}, D'_{10}\right)\,\textrm{mod}\,2\,\in\\\{(0, 0, 0, 1), (1, 1, 1, 0)\}}}\binom{n'_{00} + \Delta'_{00}}{D'_{00}}\binom{n'_{11} + \Delta'_{11}}{D'_{11}}\binom{n'_{01} + \Delta'_{01}}{D'_{01}}\binom{n'_{10} + \Delta'_{10}}{D'_{10}}}\\
    & \hspace*{0.05\textwidth} \times \left(1 - q + qe^{i\gamma}\right)^{\sum_{\substack{\{D'_s\}\\\left(D'_{00}, D'_{11}, D'_{01}, D'_{10}\right)\,\textrm{mod}\,2\,\in\\\{(0, 0, 1, 0), (1, 1, 0, 1)\}}}\binom{n'_{00} + \Delta'_{00}}{D'_{00}}\binom{n'_{11} + \Delta'_{11}}{D'_{11}}\binom{n'_{01} + \Delta'_{01}}{D'_{01}}\binom{n'_{10} + \Delta'_{10}}{D'_{10}}}.
\end{align*}
In the rest of the proof, we will assume $D$ odd for definiteness. We now come back to the sum over $\left(n_s\right)_s$ in equation \ref{eq:diluted_spin_model_p_1_variable_shifted} and eliminate the $n_s$ such that $L(s) = 0$, i.e. $n_{001}, n_{011}, n_{100}, n_{110}$. This is done by remarking that the only factor in the summand that depends explicitly on the individual $n_s$, and not only on the $n'_{s_2 s_0}$, is $\binom{n - D}{(n_s)_s}(-1)^{\sum_{s \in \mathcal{L}}n_s}$. But
\begin{align*}
    & \sum_{\substack{n_{001}, n_{011}\\n_{001} + n_{011} = n'_{01}}}\binom{n - D}{(n_s)_s}(-1)^{\sum_{\substack{s \in \mathcal{L}}}n_s}\\
    & = \sum_{\substack{n_{001}, n_{011}\\n_{001} + n_{011} = n'_{01}}}\frac{(n - D)!}{n_{000}!n_{001}!n_{010}!n_{011}!n_{100}!n_{101}!n_{110}!n_{111}!}(-1)^{n_{001} + n_{010} + n_{101} + n_{110}}\\
    & = \frac{(n - D)!}{n_{000}!n_{010}!n_{100}!n_{101}!n_{110}!n_{111}!}(-1)^{n_{010} + n_{101} + n_{110}}\frac{\left(1 - 1\right)^{n'_{01}}}{n'_{01}!} \qquad \textrm{(binomial formula)}
\end{align*}
Therefore, summing over $n_{001}, n_{011}$ amounts to setting the latter variables to $0$. The same holds for $n_{100}, n_{110}$ and we are left with the following new form of equation \ref{eq:diluted_spin_model_p_1_variable_shifted}:
\begin{align*}
    & \binom{n}{D}\binom{D}{(\Delta_s)_s}(-1)^{\sum_{s \in \mathcal{L}}\Delta_s}\prod_sB_{\beta, s}^{\Delta_s}\sum_{\substack{n_{000}, n_{010}, n_{101}, n_{111}\\n_{000} + n_{010} + n_{101} + n_{111} = n - D}}\binom{n - D}{n_{000}, n_{010}, n_{101}, n_{111}}(-1)^{n_{010} + n_{101}}\nonumber\\
    & \hspace*{20px} \times \left(\cos^2\frac{\beta}{2}\right)^{n_{000} + n_{111}}\left(-\sin^2\frac{\beta}{2}\right)^{n_{010} + n_{101}}\\
    & \hspace*{20px} \times \left(1 - q + qe^{-i\gamma}\right)^{\sum_{\substack{\{D'_s\}\\\left(D'_{00}, D'_{11}, D'_{01}, D'_{10}\right)\,\textrm{mod}\,2\,\in\\\{(0, 0, 0, 1), (1, 1, 1, 0)\}}}\binom{n'_{00} + \Delta'_{00}}{D'_{00}}\binom{n'_{11} + \Delta'_{11}}{D'_{11}}\binom{\Delta'_{01}}{D'_{01}}\binom{\Delta'_{10}}{D'_{10}}}\\
    & \hspace*{20px} \times \left(1 - q + qe^{i\gamma}\right)^{\sum_{\substack{\{D'_s\}\\\left(D'_{00}, D'_{11}, D'_{01}, D'_{10}\right)\,\textrm{mod}\,2\,\in\\\{(0, 0, 1, 0), (1, 1, 0, 1)\}}}\binom{n'_{00} + \Delta'_{00}}{D'_{00}}\binom{n'_{11} + \Delta'_{11}}{D'_{11}}\binom{\Delta'_{01}}{D'_{01}}\binom{\Delta'_{10}}{D'_{10}}}\\
    & = \binom{n}{D}\binom{D}{(\Delta_s)_s}(-1)^{\sum_{s \in \mathcal{L}}\Delta_s}\prod_sB_{\beta, s}^{\Delta_s}\sum_{0 \leq n'_{00} \leq n - D}\binom{n - D}{n'_{00}}\nonumber\\
    & \hspace*{60px} \times \left(1 - q + qe^{-i\gamma}\right)^{\sum_{\substack{\{D'_s\}\\\left(D'_{00}, D'_{11}, D'_{01}, D'_{10}\right)\,\textrm{mod}\,2\,\in\\\{(0, 0, 0, 1), (1, 1, 1, 0)\}}}\binom{n'_{00} + \Delta'_{00}}{D'_{00}}\binom{n - D - n'_{00} + \Delta'_{11}}{D'_{11}}\binom{\Delta'_{01}}{D'_{01}}\binom{\Delta'_{10}}{D'_{10}}}\\
    & \hspace*{60px} \times \left(1 - q + qe^{i\gamma}\right)^{\sum_{\substack{\{D'_s\}\\\left(D'_{00}, D'_{11}, D'_{01}, D'_{10}\right)\,\textrm{mod}\,2\,\in\\\{(0, 0, 1, 0), (1, 1, 0, 1)\}}}\binom{n'_{00} + \Delta'_{00}}{D'_{00}}\binom{n - D - n'_{00} + \Delta'_{11}}{D'_{11}}\binom{\Delta'_{01}}{D'_{01}}\binom{\Delta'_{10}}{D'_{10}}}
\end{align*}
Now, introduce polar coordinates:
\begin{align}
\label{eq:diluted_spin_model_p_1_polar}
    1 - q + qe^{i\gamma} & =: \rho e^{i\theta}
\end{align}
From the definition of $q$ in equation \ref{eq:diluted_spin_model_p_1_def_q_C}, $\theta = \mathcal{O}\left(n^{1 - D}\right)$. Under this parametrization,
\begin{align*}
    & \exp\left\{\log\left(1 - q + qe^{-i\gamma}\right)\hspace*{-0.06\textwidth}\sum\limits_{\substack{\{D'_s\}\\\left(D'_{00}, D'_{11}, D'_{01}, D'_{10}\right)\,\textrm{mod}\,2\,\in\\\{(0, 0, 0, 1), (1, 1, 1, 0)\}}}\hspace*{-0.06\textwidth}\binom{n'_{00} + \Delta'_{00}}{D'_{00}}\binom{n - D - n'_{00} + \Delta'_{11}}{D'_{11}}\binom{\Delta'_{01}}{D'_{01}}\binom{\Delta'_{10}}{D'_{10}}\right\}\\
    & \hspace*{0.05\textwidth} \times \exp\left\{\log\left(1 - q + qe^{i\gamma}\right)\hspace*{-0.06\textwidth}\sum\limits_{\substack{\{D'_s\}\\\left(D'_{00}, D'_{11}, D'_{01}, D'_{10}\right)\,\textrm{mod}\,2\,\in\\\{(0, 0, 1, 0), (1, 1, 0, 1)\}}}\hspace*{-0.06\textwidth}\binom{n'_{00} + \Delta'_{00}}{D'_{00}}\binom{n - D - n'_{00} + \Delta'_{11}}{D'_{11}}\binom{\Delta'_{01}}{D'_{01}}\binom{\Delta'_{10}}{D'_{10}}\right\}\\
    & = \exp\left\{\log\rho\sum\limits_{\substack{D'_{00}, D'_{11}\\D'_{00} + D'_{11}\,\textrm{even}}}\binom{n'_{00} + \Delta'_{00}}{D'_{00}}\binom{n - D - n'_{00} + \Delta'_{11}}{D'_{11}}\hspace*{-0.06\textwidth}\sum\limits_{\substack{D'_{01}, D'_{10}\\D'_{01} + D'_{10}\,\textrm{odd}\\D'_{01} + D'_{10} = D - D'_{00} - D'_{11}}}\hspace*{-0.06\textwidth}\binom{\Delta'_{01}}{D'_{01}}\binom{\Delta'_{10}}{D'_{10}}\right\}\\
    & \hspace*{0.05\textwidth} \times \exp\left\{i\theta\hspace*{-0.02\textwidth}\sum\limits_{\substack{D'_{00}, D'_{11}\\D'_{00} + D'_{11}\,\textrm{even}}}\binom{n'_{00} + \Delta'_{00}}{D'_{00}}\binom{n - D - n'_{00} + \Delta'_{11}}{D'_{11}}\hspace*{-0.06\textwidth}\sum\limits_{\substack{D'_{01}, D'_{10}\\D'_{01} + D'_{10}\,\textrm{odd}\\D'_{01} + D'_{10} = D - D'_{00} - D'_{11}}}\hspace*{-0.06\textwidth}(-1)^{D'_{00} + D'_{01}}\binom{\Delta'_{01}}{D'_{01}}\binom{\Delta'_{10}}{D'_{10}}\right\}.
\end{align*}
In the last equation, we chose to write $a^x$ as $\exp\left(x\log a\right)$ to avoid overburdening the notation with exponents.
The exponent of $\rho$ in this equation can be further calculated, showing it is independent of $n'_{00}$:
\begin{align*}
    & \sum_{\substack{D'_{00}, D'_{11}\\D'_{00} + D'_{11}\,\textrm{even}}}\binom{n'_{00} + \Delta'_{00}}{D'_{00}}\binom{n - D - n'_{00} + \Delta'_{11}}{D'_{11}}\sum_{\substack{D'_{01}, D'_{10}\\D'_{01} + D'_{10} = D - D'_{00} - D'_{11}}}\binom{\Delta'_{01}}{D'_{01}}\binom{\Delta'_{10}}{D'_{10}}\\
    & = \sum_{\substack{D'_{01}, D'_{10}\\D'_{01} + D'_{10}\,\textrm{odd}}}\binom{\Delta'_{01}}{D'_{01}}\binom{\Delta'_{10}}{D'_{10}}\sum_{\substack{D'_{00}, D'_{11}\\D'_{00} + D'_{11} = D - D'_{01} - D'_{10}}}\binom{n'_{00} + \Delta'_{00}}{D'_{00}}\binom{n - D - n'_{00} + \Delta'_{11}}{D'_{11}}\\
    & = \sum_{\substack{D'_{01}, D'_{10}\\D'_{01} + D'_{10}\,\textrm{odd}}}\binom{\Delta'_{01}}{D'_{01}}\binom{\Delta'_{10}}{D'_{10}}\binom{n - D + \Delta'_{00} + \Delta'_{11}}{D - D'_{01} - D'_{10}}.
\end{align*}
The exponent of $e^{i\theta}$ requires a little more analysis. [This step is not required for the dense $D$-spin model.] First, note that
\begin{align*}
    & \sum_{\substack{D'_{00}, D'_{11}\\D'_{00} + D'_{11}\,\textrm{even}}}\binom{n'_{00} + \Delta'_{00}}{D'_{00}}\binom{n - D - n'_{00} + \Delta'_{11}}{D'_{11}}\sum_{\substack{D'_{01}, D'_{10}\\D'_{01} + D'_{10} = D - D'_{00} - D'_{11}}}(-1)^{D'_{00} + D'_{01}}\binom{\Delta'_{01}}{D'_{01}}\binom{\Delta'_{10}}{D'_{10}}\\
    & = \sum_{\substack{D'_{01}, D'_{10}\\D'_{01} + D'_{10}\,\textrm{odd}}}(-1)^{D'_{01}}\binom{\Delta'_{01}}{D'_{01}}\binom{\Delta'_{10}}{D'_{10}}\sum_{\substack{D'_{00}, D'_{11}\\D'_{00} + D'_{11} = D - D'_{01} - D'_{10}}}(-1)^{D'_{00}}\binom{n'_{00} + \Delta'_{00}}{D'_{00}}\binom{n - D - n'_{00} + \Delta'_{11}}{D'_{11}}
\end{align*}
The inner sum corresponds to the coefficient of $z^{D - D'_{01} - D'_{10}}$ in the series expansion of $(1 - z)^{n'_{00} + \Delta'_{00}}(1 + z)^{n - D - n'_{00} + \Delta'_{11}}$. To estimate this coefficient, let us parametrize:
\begin{align}
    n'_{00} & =: \frac{n - \Delta'_{01} - \Delta'_{10} - 2\Delta'_{00}}{2} - t.
\end{align}
This parametrization is approximately centered around $n'_{00} = \frac{n - D}{2}$, where the binomial coefficient $\binom{n - D}{n'_{00}}$ is maximal. In contrast, as soon as $|t| > n^{3/4}$ for instance, the coefficient is exponentially negligible compared to its maximum value. Using this parametrization,
\begin{align*}
    (1 - z)^{n'_{00} + \Delta'_{00}}(1 + z)^{n - D - n'_{00} + \Delta'_{11}} & = (1 - z)^{\frac{n - \Delta'_{01} - \Delta'_{10}}{2} - t}(1 + z)^{\frac{n - \Delta'_{01} - \Delta'_{10}}{2} + t}\\
    & = (1 - z^2)^{\frac{n - \Delta'_{01} - \Delta'_{10}}{2}}\left(\frac{1 + z}{1 - z}\right)^t
\end{align*}
The coefficient of $z^{D - D'_{01} - D'_{10}}$ is then a polynomial in the variables $\frac{n - \Delta'_{01} - \Delta'_{10}}{2}$ and $t$; the degrees $a, b$ of each monomial $\left(\frac{n - \Delta'_{01} - \Delta'_{10}}{2}\right)^at^b$ satisfy $2a + b = D - D'_{01} - D'_{10}$. Hence, if $|t| \leq n^{4/7}$, as $n \to \infty$,
\begin{align*}
    \left|\left(\frac{n - \Delta'_{01} - \Delta'_{10}}{2}\right)^at^b\right| & = \mathcal{O}\left(n^a|t|^b\right)\\
    & = \mathcal{O}\left(n^{a + \frac{4}{7}b}\right)\\
    & = \mathcal{O}\left(n^{-\frac{a}{7} + \frac{4}{7}(D - D'_{01} - D'_{10})}\right).
\end{align*}
Therefore, recalling $\theta = \mathcal{O}\left(n^{1 - D}\right)$, for $|t| \leq n^{4/7}$, $\left|\theta\left(\frac{n - \Delta'_{01} - \Delta'_{10}}{2}\right)^at^b\right| = \mathcal{O}\left(n^{1 - \frac{3}{7}D}\right) = \mathcal{O}\left(n^{-2/7}\right)$ since $D \geq 3$. Putting everything together,
\begin{align*}
    & \sum_{\{n_s\}}\frac{1}{2^n}\binom{n}{(n_s)_s}(-1)^{\sum_{s \in \mathcal{L}}n_s}\prod_sB_{\bm{\beta}, s}^{n_s}\prod_{(D_s)}C_{q, \bm{\gamma}, \oplus_ss^{\oplus D_s}}^{\prod_s\binom{n_s}{D_s}}\sum_{(D_s)_s}\prod_s\binom{n_s}{\Delta_s}\\
    & = \frac{1}{2^D}(-1)^{\sum_{s \in \mathcal{L}}\Delta_s}\prod_sB_{\bm{\beta}, s}^{\Delta_s}\binom{D}{(\Delta_s)_s}\binom{n}{D}\left|1 - q + qe^{i\gamma}\right|^{\sum_{\substack{D'_{01}, D'_{10}\\D'_{01} + D'_{10}\,\textrm{odd}}}\binom{\Delta'_{01}}{D'_{01}}\binom{\Delta'_{10}}{D'_{10}}\binom{n - \Delta'_{01} - \Delta'_{10}}{D - D'_{01} - D'_{10}}}\left(1 + \mathcal{O}\left(n^{-2/7}\right)\right)
\end{align*}
Finally, recalling the definition of $q$ in equation \ref{eq:diluted_spin_model_p_1_def_q_C},
\begin{align*}
    & \left|1 - q + qe^{i\gamma}\right|^{\sum\limits_{\substack{D'_{01}, D'_{10}\\D'_{01} + D'_{10}\,\textrm{odd}}}\binom{\Delta'_{01}}{D'_{01}}\binom{\Delta'_{10}}{D'_{10}}\binom{n - \Delta'_{01} - \Delta'_{10}}{D - D'_{01} - D'_{10}}}\\
    & = \left|1 - \frac{nd}{D\binom{n}{D}} + \frac{nd}{D\binom{n}{D}}e^{i\gamma}\right|^{\sum\limits_{\substack{D'_{01}, D'_{10}\\D'_{01} + D'_{10}\,\textrm{odd}}}\binom{\Delta'_{01}}{D'_{01}}\binom{\Delta'_{10}}{D'_{10}}\binom{n - \Delta'_{01} - \Delta'_{10}}{D - D'_{01} - D'_{10}}}\\
    & = \left|1 - n^{1 - D}(D - 1)!d(1 - e^{i\gamma}) + \mathcal{O}\left(n^{2 - 2D}\right)\right|^{\sum\limits_{\substack{D'_{01}, D'_{10}\\D'_{01} + D'_{10}\,\textrm{odd}}}\binom{\Delta'_{01}}{D'_{01}}\binom{\Delta'_{10}}{D'_{10}}\binom{n - \Delta'_{01} - \Delta'_{10}}{D - D'_{01} - D'_{10}}}\\
    & = \left(1 - n^{1 - D}(D - 1)!d\left(1 - \cos\gamma\right) + \mathcal{O}\left(n^{2 - 2D}\right)\right)^{\sum\limits_{\substack{D'_{01}, D'_{10}\\D'_{01} + D'_{10}\,\textrm{odd}}}\binom{\Delta'_{01}}{D'_{01}}\binom{\Delta'_{10}}{D'_{10}}\binom{n - \Delta'_{01} - \Delta'_{10}}{D - D'_{01} - D'_{10}}}
\end{align*}
Now, since $\binom{n - \Delta'_{01} - \Delta'_{10}}{D - D'_{01} - D'_{10}} = \Theta\left(n^{D - D'_{01} - D'_{10}}\right)$, only the terms of the sum $\sum\limits_{\substack{D'_{01}, D'_{10}\\D'_{01} + D'_{10}\,\textrm{odd}}}\binom{\Delta'_{01}}{D'_{01}}\binom{\Delta'_{10}}{D'_{10}}\binom{n - \Delta'_{01} - \Delta'_{10}}{D - D'_{01} - D'_{10}}$ such that $D'_{10} + D'_{01} = 1$ contribute to the $n \to \infty$ limit of the last equation. These terms are precisely given by $\left(D'_{01}, D'_{10}\right) \in \left\{(1, 0), (0, 1)\right\}$ and sum up to $\left(\Delta'_{01} + \Delta'_{10}\right)\binom{n - \Delta'_{01} - \Delta'_{10}}{D - 1}$. Therefore,
\begin{align*}
    & \left|1 - q + qe^{i\gamma}\right|^{\sum\limits_{\substack{D'_{01}, D'_{10}\\D'_{01} + D'_{10}\,\textrm{odd}}}\binom{\Delta'_{01}}{D'_{01}}\binom{\Delta'_{10}}{D'_{10}}\binom{n - \Delta'_{01} - \Delta'_{10}}{D - D'_{01} - D'_{10}}}\\
    & \underset{n \to \infty}{\sim} \left(1 - n^{1 - D}(D - 1)!d\left(1 - \cos\gamma\right) + \mathcal{O}\left(n^{2 - 2D}\right)\right)^{\left(\Delta'_{01} + \Delta'_{10}\right)\binom{n - \Delta'_{01} - \Delta'_{10}}{D - 1}}\\
    & \underset{n \to \infty}{\sim} \left(1 - n^{1 - D}(D - 1)!d\left(1 - \cos\gamma\right) + \mathcal{O}\left(n^{2 - 2D}\right)\right)^{\frac{\left(\Delta'_{01} + \Delta'_{10}\right)n^{D - 1}}{(D - 1)!}}\\
    & \xrightarrow[n \to \infty]{} \exp\left(-\left(\Delta'_{01} + \Delta'_{10}\right)d\left(1 - \cos\gamma\right)\right).
\end{align*}
This completes the proof.
\end{proof}
\end{prop}
Injecting the result of proposition \ref{prop:diluted_spin_model_p_1_moment_infinite_size} into equation \ref{eq:diluted_spin_model_p_1} gives:
\begin{prop}
\label{prop:diluted_spin_model_p_1_energy_infinite_size}
In the infinite size limit, the energy of the $p = 1$ QAOA on an average instance of the $D$-spin model with degree parameter $d$ is given by:
\begin{align}
    & \lim_{n \to \infty}\mathbf{E}_{J \sim \textnormal{Diluted}(D, n, d)}\left[\left\langle\Psi_{\textnormal{QAOA}}(J, \beta, \gamma)\bigg|\frac{H[J]}{n}\bigg|\Psi_{\textnormal{QAOA}}(J, \beta, \gamma)\right\rangle\right]\nonumber\\
    & = \frac{d}{2^DD}\sum_{\substack{\left(D_s\right)_{s \in \{0, 1\}^3}\\\sum_sD_s = D}}\binom{D}{(D_s)_s}e^{i\varphi\left(\gamma, \oplus_ss^{\oplus D_s}\right)}(-1)^{\sum_sD_ss_p}(-1)^{\sum_{s \in \mathcal{L}}D_s}e^{-(D_{001} + D_{011} + D_{100} + D_{110})d(1 - \cos\gamma)}\prod_sB_{\beta, s}^{D_s}.
\end{align}
\end{prop}
The latter can finally be related to the energy of the dense $D$-spin model derived in \cite{2102.12043}:
\begin{repprop}{prop:qaoa_dense_qaoa_diluted}
For large $d$, the QAOA energy of the diluted $D$-spin model in the infinite size limit is approximated:
\begin{align}
    & \lim_{n \to \infty}\mathbf{E}_{J \sim \textnormal{Diluted}(D, n, d)}\left[\left\langle\Psi_{\textnormal{QAOA}}\left(J, \beta^{\textnormal{Dense}}, \frac{\gamma^{\textnormal{Dense}}}{\sqrt{(D - 1)!d}}\right)\bigg|\frac{H[J]}{nd/D}\bigg|\Psi_{\textnormal{QAOA}}\left(J, \beta^{\textnormal{Dense}}, \frac{\gamma^{\textnormal{Dense}}}{\sqrt{(D - 1)!d}}\right)\right\rangle\right]\nonumber\\
    & = -\frac{i\gamma^{\textnormal{Dense}}}{2\sqrt{(D - 1)!d}}\left[\left(\cos\beta^{\textnormal{Dense}} - i\sin\beta^{\textnormal{Dense}} e^{-\frac{\left(\gamma^{\textnormal{Dense}}\right)^2}{2(D - 1)!}}\right)^D - \left(\cos\beta^{\textnormal{Dense}} + i\sin\beta^{\textnormal{Dense}} e^{-\frac{\left(\gamma^{\textnormal{Dense}}\right)^2}{2(D - 1)!}}\right)^D\right] + \mathcal{O}\left(\frac{1}{d}\right)\nonumber\\
    & = \frac{1}{\sqrt{(D - 1)!d}}\lim_{n \to \infty}\mathbf{E}_{J \sim \textnormal{Dense}(D, n)}\left[\left\langle\Psi_{\textnormal{QAOA}}\left(J, \beta^{\textnormal{Dense}}, \gamma^{\textnormal{Dense}}\right)\bigg|\frac{H[J]}{n}\bigg|\Psi_{\textnormal{QAOA}}\left(J, \beta^{\textnormal{Dense}}, \gamma^{\textnormal{Dense}}\right)\right\rangle\right] + \mathcal{O}\left(\frac{1}{d}\right)
\end{align}
where the leading term is, up to the factor $\frac{1}{\sqrt{(D - 1)!d}}$, the energy achieved on the dense $D$-spin model by level-1 QAOA evaluated at angles $\beta^{\textnormal{Dense}}, \gamma^{\textnormal{Dense}} = O(1)$.
\begin{proof}
Starting from proposition \ref{prop:diluted_spin_model_p_1_energy_infinite_size},
\begin{align*}
    & \lim_{n \to \infty}\mathbf{E}_{J \sim \textnormal{Diluted}(D, n, d)}\left[\left\langle\Psi_{\textnormal{QAOA}}(J, \beta, \gamma)\bigg|\frac{H[J]}{nd/D}\bigg|\Psi_{\textnormal{QAOA}}(J, \beta, \gamma)\right\rangle\right]\nonumber\\
    & = \frac{1}{2^D}\sum_{\substack{\left(D_s\right)_{s}\\\sum_sD_s = D}}\binom{D}{(D_s)_s}e^{i\varphi\left(\gamma, \oplus_ss^{\oplus D_s}\right)}(-1)^{\sum_sD_ss_p}(-1)^{\sum_{s \in \mathcal{L}}D_s}e^{-(D_{001} + D_{011} + D_{100} + D_{110})d(1 - \cos\gamma)}\prod_sB_{\beta, s}^{D_s}\\
    & = \frac{1}{2^D}\sum_{\substack{\left(D_s\right)\\\sum_sD_s = D}}\binom{D}{(D_s)_s}(-1)^{\sum_sD_ss_p}(-1)^{\sum_{s \in \mathcal{L}}D_s}e^{-\left(D_{001} + D_{011} + D_{100} + D_{110}\right)d(1 - \cos\gamma)}\prod_sB_{\beta, s}^{D_s}\nonumber\\
    & \hspace*{20px} + \frac{1}{2^D}\sum_{\substack{\left(D_s\right)\\\sum_sD_s = D}}\binom{D}{(D_s)_s}i\varphi\left(\gamma, \oplus_ss^{\oplus D_s}\right)(-1)^{\sum_sD_ss_p}(-1)^{\sum_{s \in \mathcal{L}}D_s}\prod_sB_{\beta, s}^{D_s}\\
    & \hspace*{160px} \times e^{-\left(D_{001} + D_{011} + D_{100} + D_{110}\right)d(1 - \cos\gamma)}\\
    & \hspace*{20px} + \mathcal{O}\left(\gamma^2\right)
\end{align*}
The first term cancels by change of variables trick ($s \to \overline{s}$, where $\overline{s}$ is obtained by flipping all bits of $s$) used in section \ref{sec:analysis_maxcut_constant_degree}. The second sum can be explicitly evaluated from the multinomial formula, using the explicit identities:
\begin{align*}
    \varphi\left(\gamma, \oplus_ss^{\oplus D_s}\right) & = \frac{\gamma}{2}\left((-1)^{D_{100} + D_{101} + D_{110} + D_{111}} - (-1)^{D_{001} + D_{011} + D_{101} + D_{111}}\right)\\
    (-1)^{\sum_sD_ss_p} & = (-1)^{D_{010} + D_{011} + D_{110} + D_{111}}\\
    (-1)^{\sum_{s \in \mathcal{L}}D_s} & = (-1)^{D_{001} + D_{010} + D_{101} + D_{110}}\\
    \prod_sB_{\beta, s}^{D_s} & = \left(\cos^2\frac{\beta}{2}\right)^{D_{000} + D_{111}}\left(-\sin^2\frac{\beta}{2}\right)^{D_{010} + D_{101}}\left(i\cos\frac{\beta}{2}\sin\frac{\beta}{2}\right)^{D_{001} + D_{011} + D_{100} + D_{110}}
\end{align*}
This gives:
\begin{align*}
    & \frac{1}{2^D}\sum_{\substack{\left(D_s\right)\\\sum_sD_s = D}}\binom{D}{(D_s)_s}i\varphi\left(\gamma, \oplus_ss^{\oplus D_s}\right)(-1)^{\sum_sD_ss_p}(-1)^{\sum_{s \in \mathcal{L}}D_s}\prod_sB_{\beta, s}^{D_s}\\
    & = -\sum_{\substack{0 \leq k \leq D\\k\,\textrm{ odd}}}\binom{D}{k}i\gamma\left(\cos\beta\right)^{D - k}\left(i\sin\beta\right)^ke^{-kd(1 - \cos\gamma)}\\
    & = -\frac{i\gamma}{2}\left[\left(\cos\beta - i\sin\beta e^{-d(1 - \cos\gamma)}\right)^D - \left(\cos\beta + i\sin\beta e^{-d(1 - \cos\gamma)}\right)^D\right].
\end{align*}
The result follows.
\end{proof}
\end{repprop}

\subsection{Analysis of the Monte-Carlo algorithm for the SK model}
\label{sec:analysis_sk_monte_carlo}
In this section, we analyze algorithms \ref{alg:sample_sk_energy} and \ref{alg:upper_bound_variance_sk_energy}. The first algorithm outputs a real number whose expectation is the energy achieved by QAOA on the finite-size SK-model. The second one calculates an upper bound on the variance of this estimator.

We first require an expression for the energy of QAOA on the Sherrington-Kirkpatrick model. This follows from proposition \ref{prop:qaoa_dense_spin_model} applied with $D = 2$:
\begin{prop}[Expected QAOA energy on the Sherrington-Kirkpatrick model, adapted from \cite{1910.08187}]
\begin{align}
\label{eq:qaoa_sk_expected_energy}
    & \mathbf{E}_{G \sim \textnormal{SK}(n)}\left[\braket{\Psi_{\textnormal{QAOA}}(J, \bm{\beta}, \bm{\gamma})|H(J)|\Psi_{\textnormal{QAOA}}(J, \bm{\beta}, \bm{\gamma})}\right]\nonumber\\
    & = \frac{1}{2^n}\sum_{\substack{\left(n_s\right)_{s \in \{0, 1\}^{2p + 1}}\\\sum_sn_s = n}}\binom{n}{\left(n_s\right)_s}(-1)^{\sum_{s \in \mathcal{L}}n_s}\prod_sB_{\bm{\beta}, s}^{n_s}\prod_{\substack{s, t \in \{0, 1\}^{2p + 1}\\s < t}}\exp\left(-\frac{\varphi(\bm\gamma, s \oplus t)^2}{2n}n_sn_t\right)\nonumber\\
    & \hspace*{65px} \times \sum_{\substack{s, t \in \{0, 1\}^{2p + 1}\\s < t}}\frac{i}{n}(-1)^{s_p + t_p}\varphi\left(\bm\gamma, s \oplus t\right)n_sn_t
\end{align}
\end{prop}

Similar to MaxCut discussed in section \ref{sec:analysis_maxcut_constant_degree}, the problem reduces to estimating sums of the form:
\begin{align}
    S_{u, v} & := \frac{1}{2^n}\sum_{(n_s)_s}\binom{n}{\left(n_s\right)_s}(-1)^{\sum_{s \in \mathcal{L}}n_s}\prod_sB_{\bm{\beta}, s}^{n_s}\prod_{\substack{s, t\\s < t}}\exp\left(-\frac{\varphi(\bm{\gamma}, s \oplus t)^2}{2n}n_sn_t\right)\frac{n_un_v}{n^2}\label{eq:qaoa_sk_def_s_uv}\\
    S_u & := \frac{1}{2^n}\sum_{(n_s)_s}\binom{n}{\left(n_s\right)_s}(-1)^{\sum_{s \in \mathcal{L}}n_s}\prod_sB_{\bm{\beta}, s}^{n_s}\prod_{\substack{s, t\\s < t}}\exp\left(-\frac{\varphi(\bm{\gamma}, s \oplus t)^2}{2n}n_sn_t\right)\frac{n_u(n_u - 1)}{n^2}.\label{eq:qaoa_sk_def_s_u}
\end{align}
Explicitly,
\begin{align}
\label{eq:qaoa_sk_energy_s_uv}
    \mathbf{E}_{G \sim \textrm{SK}(n)}\left[\braket{\Psi_{\textnormal{QAOA}}(J, \bm{\beta}, \bm{\gamma})|H(J)|\Psi_{\textnormal{QAOA}}(J, \bm{\beta}, \bm{\gamma})}\right] & = n\sum_{\substack{s, t \in \{0, 1\}^{2p + 1}\\s < t}}i(-1)^{s_p + t_p}\varphi\left(\bm\gamma, s \oplus t\right)S_{s, t}.
\end{align}
The rescaling of $n_un_v$ (resp. $n_u(n_u - 1)$) by $n^2$ in $S_{u, v}$ (resp. $S_u$), following \cite{1910.08187}, is chosen such that $S_{u, v}$ and $S_u$ converge to finite limits as $n \to \infty$. Algorithm \ref{alg:sample_sk_energy} works by estimating each $S_{u, v}$ independently and combining them as in equation \ref{eq:qaoa_sk_energy_s_uv} to obtain an estimate of the expected QAOA energy. For clarity, we then consider the estimation of a single $S_{u, v}$ in section \ref{sec:s_uv_monte_carlo} before deducing theorem \ref{th:sk_finite_size_monte_carlo} in section \ref{sec:proof_sk_finite_size_monte_carlo}.

\subsubsection{Estimating $S_{u, v}$ by Monte-Carlo sampling}
\label{sec:s_uv_monte_carlo}
In equations \ref{eq:qaoa_sk_def_s_uv} and \ref{eq:qaoa_sk_def_s_u} above, it is straightforward to sum over $n_s, L(s) = p$ (as detailed in \cite{1910.08187} and illustrated again in section \ref{sec:analysis_maxcut_constant_degree}). This yields the same results as in proposition \ref{prop:moment_exact_expression}, with $C_{\bm{\gamma}, s}$ now defined as $\exp\left(-\frac{\varphi(\bm{\gamma}, s)^2}{2n}\right)$ instead of $\left(1 - \frac{d}{n - 1} + \frac{d}{n - 1}e^{i\varphi(\bm{\gamma}, s)}\right)$ there. We will now construct an estimator of the following contribution to $S_{u, v}$ (this also applies to $S_{u}$, obtained by substituting $v \to u$ in the expression of $S_{u, v}$):
\begin{align}
    & \sum_{\substack{\left(m_{\{s, F(s)\}}\right)_{s \in \mathcal{L} - \mathcal{L}_p}\\0 \leq n' \leq n\\\sum_{s \in \mathcal{L} - \mathcal{L}_p}m_{\{s, F(s)\}} = n'}}\frac{1}{2^n}\binom{n'}{\left(m_{\{s, F(s)\}}\right)_{s \in \mathcal{L} - \mathcal{L}_p}}\binom{n - 2}{n'}\nonumber\\
    & \hspace*{10px} \times \prod_{\substack{t \in \mathcal{L}\\L(t) \leq p - 1}}\left[B_{\bm{\beta}, t}\left(-C_{\bm{\gamma}, u \oplus t}C_{\bm{\gamma}, v \oplus t}\prod_{\substack{s \in \mathcal{L}\\s < t}}C_{\bm{\gamma}, s \oplus t}^{m_{\{s, F(s)\}}} + C_{\bm{\gamma}, u \oplus F(t)}C_{\bm{\gamma}, v \oplus F(t)}\prod_{\substack{s \in \mathcal{L}\\s < t}}C_{\bm{\gamma}, s \oplus F(t)}^{m_{\{s, F(s)\}}}\right)\right]^{m_{\{t, F(t)\}}}\nonumber\\
    & \hspace*{10px} \times \left(-\sum_{t \in \mathcal{L}_p}B_{\bm{\beta}, t}\prod_{\substack{s \in \mathcal{L}\\L(s) < p}}C_{\bm{\gamma}, s \oplus t}^{m_{\{s, F(s)\}}} + \sum_{t \in \mathcal{L}_p'}B_{\bm{\beta}, t}\prod_{\substack{s \in \mathcal{L}\\L(s) < p}}C_{\bm{\gamma}, s \oplus t}^{m_{\{s, F(s)\}}}\right)^{n - n' - 2}\\
    & = \frac{1}{4}\sum_{\substack{\left(m_{\{s, F(s)\}}\right)_{s \in \mathcal{L} - \mathcal{L}_p}\\\sum_{s \in \mathcal{L} - \mathcal{L}_p}m_{\left\{s, F(s)\right\}}}}\left(\prod_{\substack{t \in \mathcal{L}\\L(t) \leq p - 1}}\frac{\lambda_{\{t, F(t)\}}\left(\{m_{\{s, F(s)\}}\}_{\substack{s \in \mathcal{L}\\s < t}}\right)^{m_{\{t, F(t)\}}}}{m_{\{t, F(t)\}}!}\right)f\left(\{m_s\}_{s \in \mathcal{L}}\right)\label{eq:sk_expectation_rewriting_pseudo_poisson}
\end{align}
where we defined:
\begin{align}
    \lambda_{\{t, F(t)\}}\left(\left(m_{\{s, F(s)\}}\right)_{\substack{s \in \mathcal{L}\\s < t}}\right) & := \frac{nB_{\bm{\beta}, t}}{2}\left(-C_{\bm{\gamma}, u \oplus t}C_{\bm{\gamma}, v \oplus t}\prod_{\substack{s \in \mathcal{L}\\s < t}}C_{\bm{\gamma}, s \oplus t}^{m_{\{s, F(s)\}}} + C_{\bm{\gamma}, u \oplus F(t)}C_{\bm{\gamma}, v \oplus F(t)}\prod_{\substack{s \in \mathcal{L}\\s < t}}C_{\bm{\gamma}, s \oplus F(t)}^{m_{\{s, F(s)\}}}\right)\label{eq:sk_expectation_def_lambda}\\
    f\left(\left(m_s\right)_{s \in \mathcal{L}}\right) & := \frac{(n - 2)!}{n^{\sum_{s \in \mathcal{L} - \mathcal{L}_p}m_{\{s, F(s)\}}}\left(n - \sum_{s \in \mathcal{L} - \mathcal{L}_p}m_{\{s, F(s)\}} - 2\right)!}\nonumber\\
    & \hspace*{30px} \times \left(\frac{1}{2}\sum_{t \in \mathcal{L}_p \sqcup \mathcal{L}'_p}(-1)^{t \in \mathcal{L}}B_{\bm{\beta}, t}\prod_{\substack{s \in \mathcal{L}\\L(s) < p}}C_{\bm\gamma, s \oplus t}^{m_{\{s, F(s)\}}}\right)^{n - \sum_{s \in \mathcal{L} - \mathcal{L}_p}m_{\{s, F(s)\}} - 2}\label{eq:sk_expectation_def_f}
\end{align}
The rescaling in $n$ for $\lambda_{\left\{t, F(t)\right\}}$ and $f\left((m_s)_{s \in \mathcal{L}}\right)$ is chosen so that these quantities converge as $n \to \infty$ and $\left(m_{\{s, F(s)\}}\right)_{s \in \mathcal{L} - \mathcal{L}_p}$ are fixed. For instance,
\begin{align*}
    & \lambda_{\left\{t, F(t)\right\}}\left(\left(m_{\{s, F(s)\}}\right)_{\substack{s \in \mathcal{L}\\s < t}}\right)\\
    & = \frac{nB_{\bm\beta, t}}{2}\left(-C_{\bm{\gamma}, u \oplus t}C_{\bm{\gamma}, v \oplus t}\prod_{\substack{s \in \mathcal{L}\\s < t}}C_{\bm{\gamma}, s \oplus t}^{m_{\{s, F(s)\}}} + C_{\bm{\gamma}, u \oplus F(t)}C_{\bm{\gamma}, v \oplus F(t)}\prod_{\substack{s \in \mathcal{L}\\s < t}}C_{\bm{\gamma}, s \oplus F(t)}^{m_{\{s, F(s)\}}}\right)\\
    & = \frac{nB_{\bm\beta, t}}{2}\left(-e^{-\frac{\varphi(\bm\gamma, u \oplus t)^2}{2n}}e^{-\frac{\varphi(\bm\gamma, v \oplus t)^2}{2n}}\prod_{\substack{s \in \mathcal{L}\\s < t}}e^{-\frac{\varphi(\bm\gamma, s \oplus t)^2}{2n}m_{\{s, F(s)\}}}\right.\\
    & \left. \hspace*{70px} + e^{-\frac{\varphi(\bm\gamma, u \oplus F(t))^2}{2n}}e^{-\frac{\varphi(\bm\gamma, v \oplus F(t))^2}{2n}}\prod_{\substack{s \in \mathcal{L}\\s < t}}e^{-\frac{\varphi(\bm\gamma, s \oplus F(t))^2}{2n}m_{\{s, F(s)\}}}\right)\\
    & = \frac{nB_{\bm\beta, t}}{2}\left(\frac{\varphi\left(\bm\gamma, u \oplus t\right)^2 - \varphi\left(\bm\gamma, u \oplus F(t)\right)^2 + \varphi\left(\bm\gamma, v \oplus t\right)^2 - \varphi\left(\bm\gamma, v \oplus F(t)\right)^2}{n} \right.\\
    & \left.\hspace*{70px} + \frac{1}{n}\sum_{\substack{s \in \mathcal{L}\\s < t}}\left(\varphi\left(\bm\gamma, s \oplus t\right)^2 - \varphi\left(\bm\gamma, s \oplus F(t)\right)^2\right)m_{\{s, F(s)\}} + \mathcal{O}\left(\frac{1}{n^2}\right)\right)\\
    & \xrightarrow[]{n \to \infty} \frac{B_{\bm\beta, t}}{2}\Bigg(\varphi\left(\bm\gamma, u \oplus t\right)^2 - \varphi\left(\bm\gamma, u \oplus F(t)\right)^2 + \varphi\left(\bm\gamma, v \oplus t\right)^2 - \varphi\left(\bm\gamma, v \oplus F(t)\right)^2\\
    & \left. \hspace*{70px} + \sum_{\substack{s \in \mathcal{L}\\s < t}}\left(\varphi\left(\bm\gamma, s \oplus t\right)^2 - \varphi\left(\bm\gamma, s \oplus F(t)\right)^2\right)m_{\{s, F(s)\}}\right).
\end{align*}
Equation \ref{eq:sk_expectation_rewriting_pseudo_poisson} can now be rewritten:
\begin{align}
    & \frac{1}{4}\sum_{\substack{\left(m_{\{s, F(s)\}}\right)_{s \in \mathcal{L} - \mathcal{L}_p}}}\left(\prod_{\substack{t \in \mathcal{L}\\L(t) \leq p - 1}}\frac{\lambda_{\{t, F(t)\}}\left(\{m_{\{s, F(s)\}}\}_{\substack{s \in \mathcal{L}\\s < t}}\right)^{m_{\{t, F(t)\}}}}{m_{\{t, F(t)\}}!}\right)f\left(\{m_s\}_{s \in \mathcal{L}}\right)\nonumber\\
    & = \frac{1}{4}\sum_{\substack{\left(m_{\{s, F(s)\}}\right)_{s \in \mathcal{L} - \mathcal{L}_p}}}\left(\prod_{\substack{t \in \mathcal{L}\\L(t) \leq p - 1}}e^{-\left|\lambda_{\{t, F(t)\}}\right|}\frac{\left|\lambda_{\{t, F(t)\}}\right|^{m_{\{t, F(t)\}}}}{m_{\{t, F(t)\}}!}\right)\nonumber\\
    & \hspace*{80px} \times \left(\prod_{\substack{t \in \mathcal{L}\\L(t) \leq p - 1}}e^{\left|\lambda_{\{t, F(t)\}}\right|}\left(\frac{\lambda_{\{t, F(t)\}}}{\left|\lambda_{\{t, F(t)\}}\right|}\right)^{m_{\{t, F(t)\}}}\right)f\left(\{m_s\}_{s \in \mathcal{L}}\right)\label{eq:sk_expectation_rewriting_poisson},
\end{align}
where we omitted the arguments of $\lambda_{\{t, F(t)\}}$ in the last line to slightly simplify the notation. The first product in the last inequality corresponds to the joint probability distribution of $4^p - 2^p$ Poisson random variables $m_{\{t, F(t)\}}$ of intensity parameters $\left|\lambda_{\{t, F(t)\}}\right|$. For the smallest bitstring $t$ (according to the order on bitstrings introduced in definition \ref{def:bitstrings_ordering}), $\left|\lambda_{\{t, F(t)\}}\right|$ is a constant; for the following bitstrings $t$, $\left|\lambda_{\{t, F(t)\}}\right|$ depends on the values of the $m_{\{s, F(s)\}}$ such that $s < t$. This interpretation suggests to estimate the sum by importance sampling:
\begin{prop}
\label{prop:sk_energy_importance_sampling}
Let $\left\{m_{\{s, F(s)\}}\right\}_{s \in \mathcal{L} - \mathcal{L}_p}$ be sampled from the joint probability distribution
\begin{align}
    \prod_{\substack{t \in \mathcal{L}\\L(t) \leq p - 1}}e^{-\left|\lambda_{\{t, F(t)\}}\right|}\frac{\left|\lambda_{\{t, F(t)\}}\right|^{m_{\{t, F(t)\}}}}{m_{\{t, F(t)\}}!},
\end{align}
where $\lambda_{\{t, F(t)\}}$ was defined in equation \ref{eq:sk_expectation_def_lambda}. Consider the random variable
\begin{align}
    X & := \frac{1}{4}\left(\prod_{\substack{t \in \mathcal{L}\\L(t) \leq p - 1}}e^{\left|\lambda_{\{t, F(t)\}}\right|}\left(\frac{\lambda_{\{t, F(t)\}}}{\left|\lambda_{\{t, F(t)\}}\right|}\right)^{m_{\{t, F(t)\}}}\right)f\left(\{m_s\}_{s \in \mathcal{L}}\right).
\end{align}
The lowest-order moments of $X$ are:
\begin{align}
    \mathbf{E}\left[X\right] & = \frac{1}{4}\sum_{\substack{\left(m_{\{s, F(s)\}}\right)_{s \in \mathcal{L} - \mathcal{L}_p}}}\left(\prod_{\substack{t \in \mathcal{L}\\L(t) \leq p - 1}}\frac{\lambda_{\{t, F(t)\}}^{m_{\{t, F(t)\}}}}{m_{\{t, F(t)\}}!}\right)f\left(\{m_s\}_{s \in \mathcal{L}}\right)\label{eq:sk_expectation_estimator}\\
    \mathbf{E}\left[\left|X\right|^2\right] & = \frac{1}{4}\sum_{\substack{\left(m_{\{s, F(s)\}}\right)_{s \in \mathcal{L} - \mathcal{L}_p}}}\left(\prod_{\substack{t \in \mathcal{L}\\L(t) \leq p - 1}}e^{\left|\lambda_{\{t, F(t)\}}\right|}\frac{\left|\lambda_{\{t, F(t)\}}\right|^{m_{\{t, F(t)\}}}}{m_{\{t, F(t)\}}!}\right)\left|f(\{m_s\}_{s \in \mathcal{L} - \mathcal{L}_p})\right|^2\label{eq:sk_expectation_estimator_squared}
\end{align}
\end{prop}
Equation \ref{eq:sk_expectation_estimator} shows that the expectation of the estimator output by algorithm \ref{alg:sample_sk_energy} is equal to the QAOA energy. The variance of the estimator can be upper-bounded by upper-bounding the second-order moment in equation \ref{eq:sk_expectation_estimator_squared}. To upper-bound the second-order moment, we start by upper-bounding $\lambda_{\{t, F(t)\}}$ in the next lemma:
\begin{lem}
\label{lemma:sk_bound_lambda}
$\lambda_{\{t, F(t)\}}$ and $f\left(\{m_s\}\right)$ defined in \ref{eq:sk_expectation_def_lambda} and \ref{eq:sk_expectation_def_f} can be upper-bounded as follows:
\begin{align}
    \left|\lambda_{\{t, F(t)\}}\right| & \leq \frac{1}{4}\left|B_{\bm{\beta}, t}\right|\Bigg(\left|\varphi(\bm{\gamma}, u \oplus t)^2 - \varphi(\bm{\gamma}, u \oplus F(t))^2\right| + \left|\varphi(\bm{\gamma}, v \oplus t)^2 - \varphi(\bm{\gamma}, v \oplus F(t))^2\right|\nonumber\\
    & \left.\hspace*{70px} + \sum_{\substack{s \in \mathcal{L}\\s < t}}\left|\varphi(\bm\gamma, s \oplus t)^2 - \varphi(\bm\gamma, s \oplus F(t))^2\right|m_{\{s, F(s)\}}\right)\\
    \left|f(\left(m_s\right))\right| & \leq 1
\end{align}
\begin{proof}
The bound on $\lambda_{\{t, F(t)\}}$ follows from the definition of the latter in equation \ref{eq:sk_expectation_def_lambda}, using that
\begin{align}
    \left|x_1x_2\ldots x_k - y_1y_2\ldots y_k\right| \leq \sum_{1 \leq j \leq k}|x_k - y_k|
\end{align} for $|x_1|, \ldots, |x_k|, |y_1|, \ldots, |y_k| \leq 1$ and $\left|e^{-x} - e^{-y}\right| \leq |x - y|$ for $x, y \geq 0$.

Explicitly, recalling $C_{\bm\gamma, s} = e^{-\frac{\varphi(\bm\gamma, s)^2}{2n}} \in ]-\infty, 1]$,
\begin{align*}
    & \left|\lambda_{\{t, F(t)\}}\right|\\
    & = \frac{n|B_{\bm\beta, t}|}{2}\left|-C_{\bm\gamma, u \oplus t}C_{\bm\gamma, v \oplus t}\prod_{\substack{s \in \mathcal{L}\\s < t}}C_{\bm\gamma, s \oplus t}^{m_{\{s, F(s)\}}} + C_{\bm\gamma, u \oplus F(t)}C_{\bm\gamma, v \oplus F(t)}\prod_{\substack{s \in \mathcal{L}\\s < t}}C_{\bm\gamma, s \oplus F(t)}^{m_{\{s, F(s)\}}}\right|\\
    & \leq \frac{n|B_{\bm\beta, t}|}{2}\left(\left|C_{\bm\gamma, u \oplus t} - C_{\bm\gamma, u \oplus F(t)}\right| + \left|C_{\bm\gamma, v \oplus t} - C_{\bm\gamma, v \oplus F(t)}\right| + \sum_{\substack{s \in \mathcal{L}\\s < t}}\left|C_{\bm\gamma, s \oplus t} - C_{\bm\gamma, s \oplus F(t)}\right|m_{\{s, F(s)\}}\right)\\
    & = \frac{n|B_{\bm\beta, t}|}{2}\left(\left|e^{-\frac{\varphi(\bm\gamma, u \oplus t)^2}{2n}} - e^{-\frac{\varphi(\bm\gamma, u \oplus F(t))^2}{2n}}\right| + \left|e^{-\frac{\varphi(\bm\gamma, v \oplus t)^2}{2n}} - e^{-\frac{\varphi(\bm\gamma, v \oplus F(t))^2}{2n}}\right|\right.\\
    & \left.\hspace*{70px} + \sum_{\substack{s \in \mathcal{L}\\s < t}}\left|e^{-\frac{\varphi(\bm\gamma, s \oplus t)^2}{2n}} - e^{-\frac{\varphi(\bm\gamma, s \oplus F(t))^2}{2n}}\right|m_{\{s, F(s)\}}\right)\\
    & \leq \frac{1}{4}\left|B_{\bm{\beta}, t}\right|\Bigg(\left|\varphi(\bm{\gamma}, u \oplus t)^2 - \varphi(\bm{\gamma}, u \oplus F(t))^2\right| + \left|\varphi(\bm{\gamma}, v \oplus t)^2 - \varphi(\bm{\gamma}, v \oplus F(t))^2\right|\nonumber\\
    & \left.\hspace*{70px} + \sum_{\substack{s \in \mathcal{L}\\s < t}}\left|\varphi(\bm\gamma, s \oplus t)^2 - \varphi(\bm\gamma, s \oplus F(t))^2\right|m_{\{s, F(s)\}}\right).
\end{align*}
The bound on $f\left(\{m_s\}\right)$ is obtained by crudely bounding
\begin{align*}
    & \left|\left(\frac{1}{2}\sum_{t \in \mathcal{L}_p \sqcup \mathcal{L}'_p}(-1)^{t \in \mathcal{L}}B_{\bm{\beta}, t}\prod_{\substack{s \in \mathcal{L}\\L(s) < p}}C_{\bm{\gamma}, s \oplus t}^{m_{\{s, F(s)\}}}\right)^{n - \sum_{s \in \mathcal{L} - \mathcal{L}_p}m_{\{s, F(s)\}} - 2}\right|\\
    & \leq \left(\frac{1}{2}\sum_{t \in \mathcal{L}_p \sqcup \mathcal{L}'_p}|B_{\bm{\beta}, t}|\prod_{\substack{s \in \mathcal{L}\\L(s) < p}}\left|C_{\bm{\gamma}, s \oplus t}\right|^{m_{\{s, F(s)\}}}\right)^{n - \sum_{s \in \mathcal{L} - \mathcal{L}_p}m_{\{s, F(s)\}} - 2}\\
    & \leq \left(\frac{1}{2}\sum_{t \in \mathcal{L}_p \sqcup \mathcal{L}'_p}|B_{\bm{\beta}, t}|\right)^{n - \sum_{s \in \mathcal{L} - \mathcal{L}_p}m_{\{s, F(s)\}} - 2}\\
    & = \left(\frac{1}{2}\begin{pmatrix}
    1\\
    1
    \end{pmatrix}^T\begin{pmatrix}
    \cos^2\frac{\beta_{p - 1}}{2} & \sin^2\frac{\beta_{p - 1}}{2}\\
    \sin^2\frac{\beta_{p - 1}}{2} & \cos^2\frac{\beta_{p - 1}}{2}
    \end{pmatrix}\ldots\begin{pmatrix}
    \cos^2\frac{\beta_0}{2} & \sin^2\frac{\beta_0}{2}\\
    \sin^2\frac{\beta_0}{2} & \cos^2\frac{\beta_0}{2}
    \end{pmatrix}\begin{pmatrix}
    1\\
    1
    \end{pmatrix}\right)^{n - \sum_{s \in \mathcal{L} - \mathcal{L}_p}m_{\{s, F(s)\}} - 2}\\
    & \leq 1.
\end{align*}
\end{proof}
\end{lem}
Replacing $\lambda_{\{t, F(t)\}}$ and $f\left(\{m_s\}\right)$ by their upper bounds in equation \ref{eq:sk_expectation_estimator_squared}, the resulting expression becomes exactly computable. The method is the same as for evaluating the QAOA energy in the infinite size limit.
\begin{prop}
\label{prop:qaoa_sk_upper_bound_variance_s_uv}
Let $\left(\hat{R}_s\right)_{\substack{s \in \mathcal{L}\\L(s) \leq p - 1}}$ be recursively (decreasingly with $L(s)$) defined as follows:
\begin{align}
    \hat{R}_s & = e^{\frac{1}{4}\sum_{t \in \mathcal{L}}\left|B_{\bm{\beta}, t}\right|\left|\varphi(\bm{\gamma}, s \oplus t)^2 - \varphi(\bm{\gamma}, s \oplus F(t))^2\right|} \qquad L(s) = p - 1\\
    \hat{R}_s & = e^{\frac{1}{4}\sum_{t \in \mathcal{L}}\left|B_{\bm{\beta}, t}\right|\left|\varphi(\bm{\gamma}, s \oplus t)^2 - \varphi(\bm{\gamma}, s \oplus F(t))^2\right|}e^{\sum_{\substack{t \in \mathcal{L}\\t > s}}\left|B_{\bm{\beta}, t}\right|\left|\varphi(\bm{\gamma}, s \oplus t)^2 - \varphi(\bm{\gamma}, s \oplus F(t))\right|\hat{R}_t} \qquad L(s) \leq p - 2
\end{align}
Note that these definition do not depend upon bitstrings $u, v$. Let $X$ be as in proposition \ref{prop:sk_energy_importance_sampling}. The following bound holds:
\begin{align}
    \mathbf{E}\left[|X|^2\right] & \leq \frac{1}{4}\left|B_{\bm\beta, u}\hat{R}_uB_{\bm\beta, v}\hat{R}_v\right|.
\end{align}
\end{prop}

\subsubsection{Proof of theorem \ref{th:sk_finite_size_monte_carlo}}
\label{sec:proof_sk_finite_size_monte_carlo}

Proposition \ref{prop:qaoa_sk_upper_bound_variance_s_uv} gives an upper bound on the variance of the estimator calculated by algorithm \ref{alg:sample_sk_energy} for a single $S_{u, v}$. An upper bound on the variance of the estimated QAOA energy (theorem \ref{th:sk_finite_size_monte_carlo}) then follows from equation \ref{eq:qaoa_sk_energy_s_uv} by applying proposition \ref{prop:qaoa_sk_upper_bound_variance_s_uv} to all $u, v \in \{0, 1\}^{2p + 1}$.

\begin{repthm}{th:sk_finite_size_monte_carlo}
Let $n \geq 2$, $p \geq 1$ integers and $\bm\beta, \bm\gamma \in \mathbf{R}^p$ QAOA angles. Given these parameters, algorithm \ref{alg:sample_sk_energy} outputs a sample whose expectation equals the expectation QAOA energy on an average size-$n$ instance of the SK model:
\begin{align}
    & \mathbf{E}_{J \sim \textnormal{SK}(n)}\left[\left\langle\Psi_{\textnormal{QAOA}}(J, \bm\beta, \bm\gamma)\bigg|\frac{H[J]}{n}\bigg|\Psi_{\textnormal{QAOA}}(J, \bm\beta, \bm\gamma)\right\rangle\right]
\end{align}
and variance is upper-bounded by the output of algorithm \ref{alg:upper_bound_variance_sk_energy}; given $\left(\hat{R}_s\right)_{s \in \{0, 1\}^{2p + 1}}$ defined in proposition \ref{prop:qaoa_sk_upper_bound_variance_s_uv}, this is
\begin{align}
    \frac{1}{4}\sum_{\substack{s, t \in \{0, 1\}^{2p + 1}\\s < t}}\left|\varphi\left(\bm\gamma, s \oplus t\right)\right|\hat{R}_s\hat{R}_t.
\end{align}
\begin{proof}
According to equation \ref{eq:qaoa_sk_energy_s_uv}, the expected QAOA energy to estimate is:
\begin{align}
    \mathbf{E}_{G \sim \textrm{SK}(n)}\left[\left\langle\Psi_{\textnormal{QAOA}}(J, \bm{\beta}, \bm{\gamma})\bigg|\frac{H(J)}{n}\bigg|\Psi_{\textnormal{QAOA}}(J, \bm{\beta}, \bm{\gamma})\right\rangle\right] & = \sum_{\substack{s, t \in \{0, 1\}^{2p + 1}\\s < t}}i(-1)^{s_p + t_p}\varphi\left(\bm\gamma, s \oplus t\right)S_{s, t}
\end{align}
Algorithm \ref{alg:sample_sk_energy} produces a sample with the expectation above by producing independent samples with expectations $\left(S_{u, v}\right)_{\{u, v\} \subset \{0, 1\}^{2p + 1}}$ and replacing $S_{s, t}$ by these samples in the equation above. By independence, and invoking proposition \ref{prop:qaoa_sk_upper_bound_variance_s_uv}, the variance of the sample for the expected energy is upper-bounded by
\begin{align}
    & \frac{1}{4}\sum_{\substack{s, t \in \{0, 1\}^{2p + 1}\\s < t}}\left|\varphi(\bm\gamma, s \oplus t)\right|^2|B_{\bm\beta, s}|\hat{R}_s|B_{\bm\beta, t}|\hat{R}_t.
\end{align}
\end{proof}
\end{repthm}

\section{Complementary figures}
\label{sec:extra_figures}

This appendix collects several complementary figures illustrating the numerical experiments discussed in section \ref{subsec:numerical_experiments}.

\subsection{MaxCut-QAOA}
\subsubsection{Erdos-Renyi graphs}
\begin{figure}[!htbp]
    \centering
    \begin{subfigure}{0.48\textwidth}
        \includegraphics[width=\textwidth]{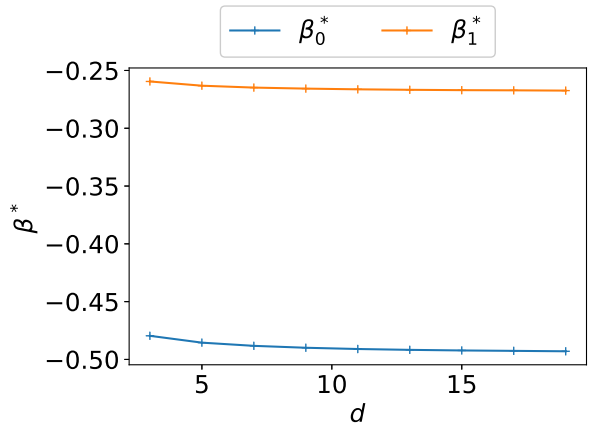}
        \caption{Optimal $\bm\beta$ angles at level $p = 2$.}
    \end{subfigure}
    \hspace*{0.03\textwidth}
    \begin{subfigure}{0.47\textwidth}
        \includegraphics[width=\textwidth]{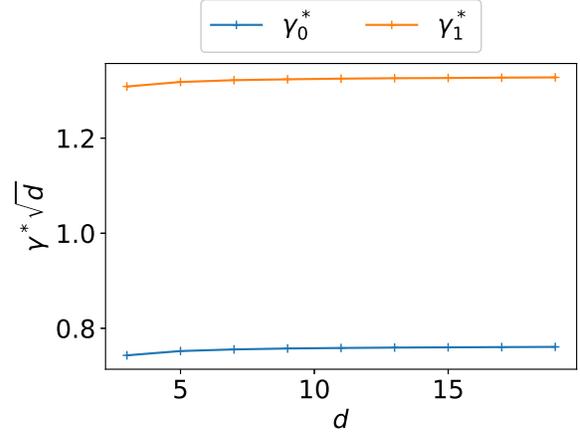}
        \caption{Optimal $\bm\gamma$ angles at level $p = 2$.}
    \end{subfigure}\\
    \vspace*{10px}
    \begin{subfigure}{0.48\textwidth}
        \includegraphics[width=\textwidth]{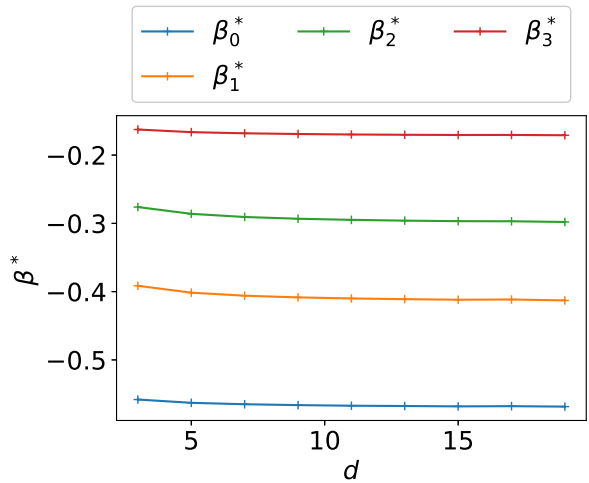}
        \caption{Optimal $\bm\beta$ angles at level $p = 4$.}
    \end{subfigure}
    \hspace*{0.03\textwidth}
    \begin{subfigure}{0.47\textwidth}
        \includegraphics[width=\textwidth]{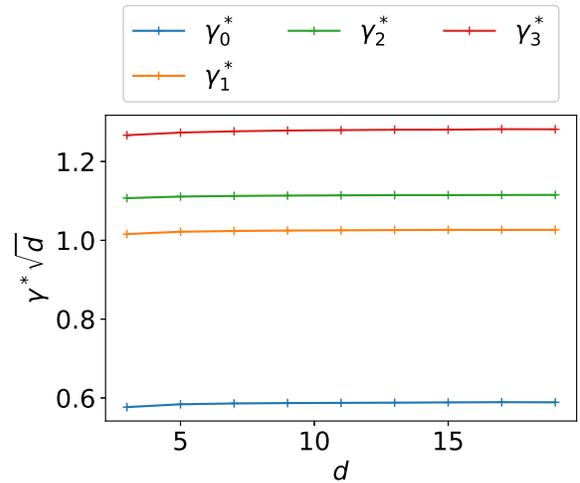}
        \caption{Optimal $\bm\gamma$ angles at level $p = 4$.}
    \end{subfigure}
    \caption{Scaling of optimal MaxCut-QAOA parameters for $p = 5$ and $d \in [3, 19]$. Each point was generated from 1000 optimization trials.}
    \label{fig:maxcut_erdos_renyi_parameters_scaling_complementary}
\end{figure}

The results of optimizing MaxCut-QAOA at levels $p \in \{2, 4\}$ on Erdos-Renyi graphs of varying average degrees (see section \ref{sec:maxcut_erdos_renyi_numerics} for details) are available in figure \ref{fig:maxcut_erdos_renyi_parameters_scaling_complementary}.

\subsubsection{Chung-Lu graphs}

We also optimized angles for QAOA applied to MaxCut on random Chung-Lu graphs; the QAOA energy was computed from proposition \ref{prop:qaoa_energy_maxcut_degree_distribution}. An example is given in figure \ref{fig:maxcut_chung_lu_parameters_scaling}. The random graphs considered there are generated from the pseudo Chung-Lu ensemble defined in section \ref{def:pseudo_chung_lu_model}, with a pair of expected degrees $(d_1, d_2) = (4, 9)$ and corresponding probabilities $(q_1, q_2) = (1 - q, q)$. The optimal $\bm\beta$ and $\bm\gamma$ angles at level $p = 2$ are plotted against $q$; the $\gamma_j^*$ are rescaled by $\sqrt{\overline{d}}$, where $\overline{d} := 4(1 - q) + 9q$ is the average expected degree. With this convention, the rescaled optimal angles are expected to take similar values at $q = 0$ and $q = 1$, which is the case indeed. They depend more noticeably on $q$ for $0 < q < 1$, but still mildly with a variation of a few percents. The sensitivity of $\beta_j^*$ (resp. $\gamma^*_j$) to $q$ decreases (resp. increases) when the QAOA layer level $j$ increases. These results suggest that MaxCut-QAOA requires less mixing (smaller $\bm\beta$ angles in absolute value) for an average graph with an expected degree distribution than for an average graph with a single expected degree.

\begin{figure}[!t]
\centering
\begin{subfigure}{0.47\textwidth}
    \centering
    \includegraphics[width=\textwidth]{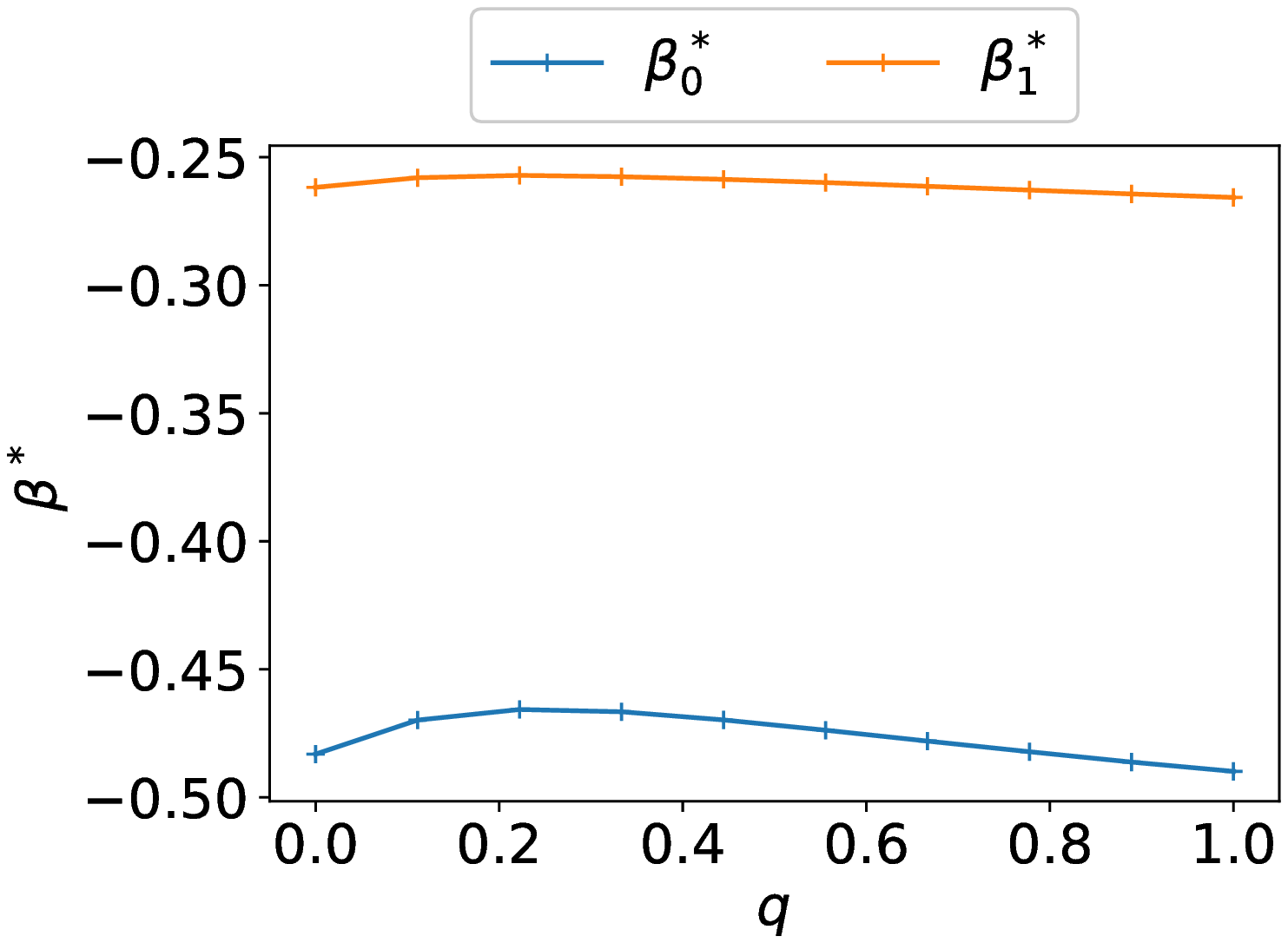}
    \caption{Optimal $\bm\beta$ angles at level $p = 2$. $\beta^*_0$, $\beta^*_1$ respectively vary by $3.6\%$, $1.8\%$ from $q = 0$ to their maximums.}
\end{subfigure}
\hspace*{0.03\textwidth}
\begin{subfigure}{0.44\textwidth}
    \centering
    \includegraphics[width=\textwidth]{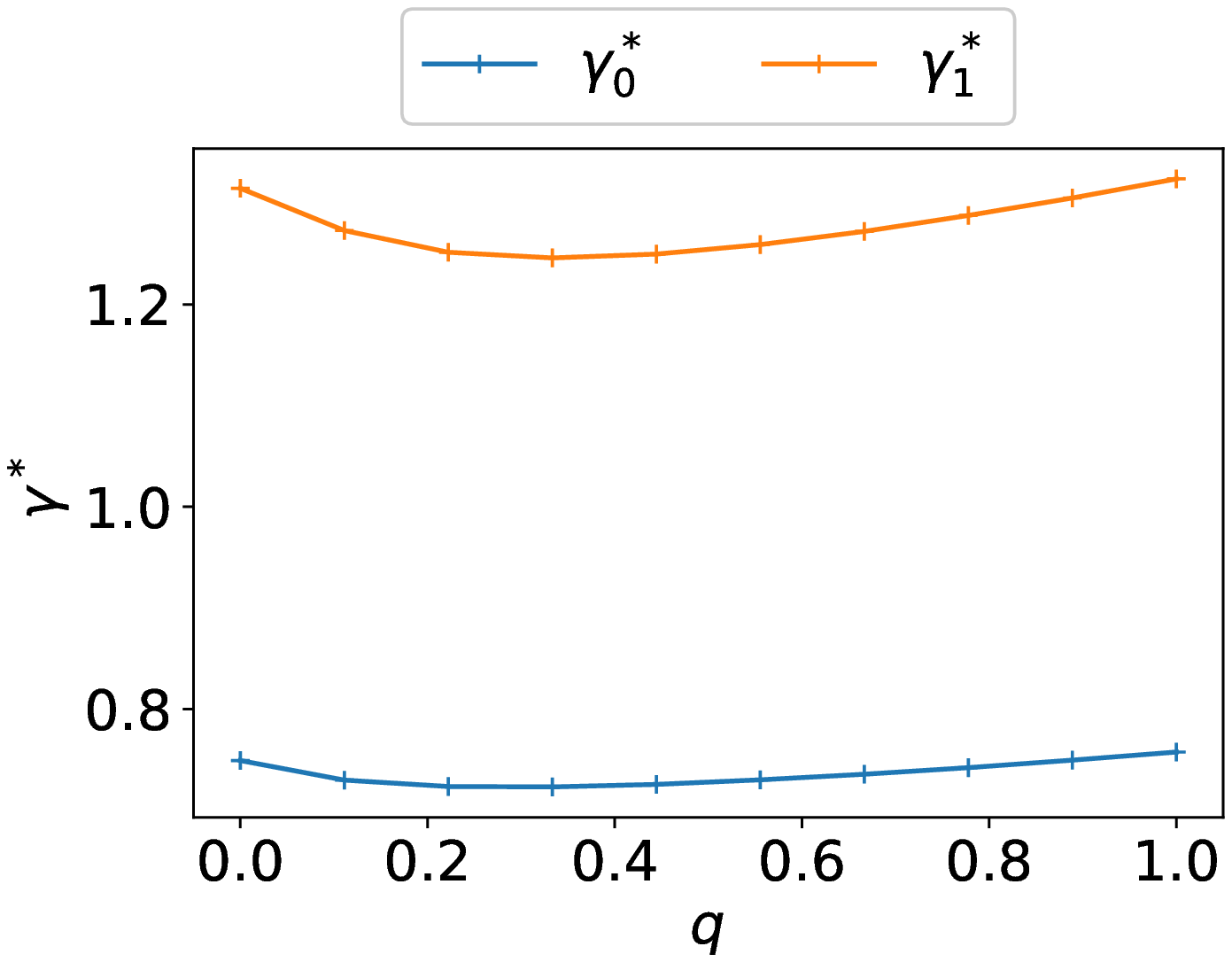}
    \caption{Optimal $\bm\gamma$ angles at level $p = 2$. The angles are rescaled by the average expected degree $\overline{d} := 4q + 9(1 - q)$. $\gamma^*_0\sqrt{\overline{d}}, \gamma^*_1\sqrt{\overline{d}}$ respectively vary by $3.5\%$ and $5.2\%$ between $q = 0$ and their minimums.}
\end{subfigure}
\vspace*{10px}
\begin{subfigure}{0.47\textwidth}
    \centering
    \includegraphics[width=\textwidth]{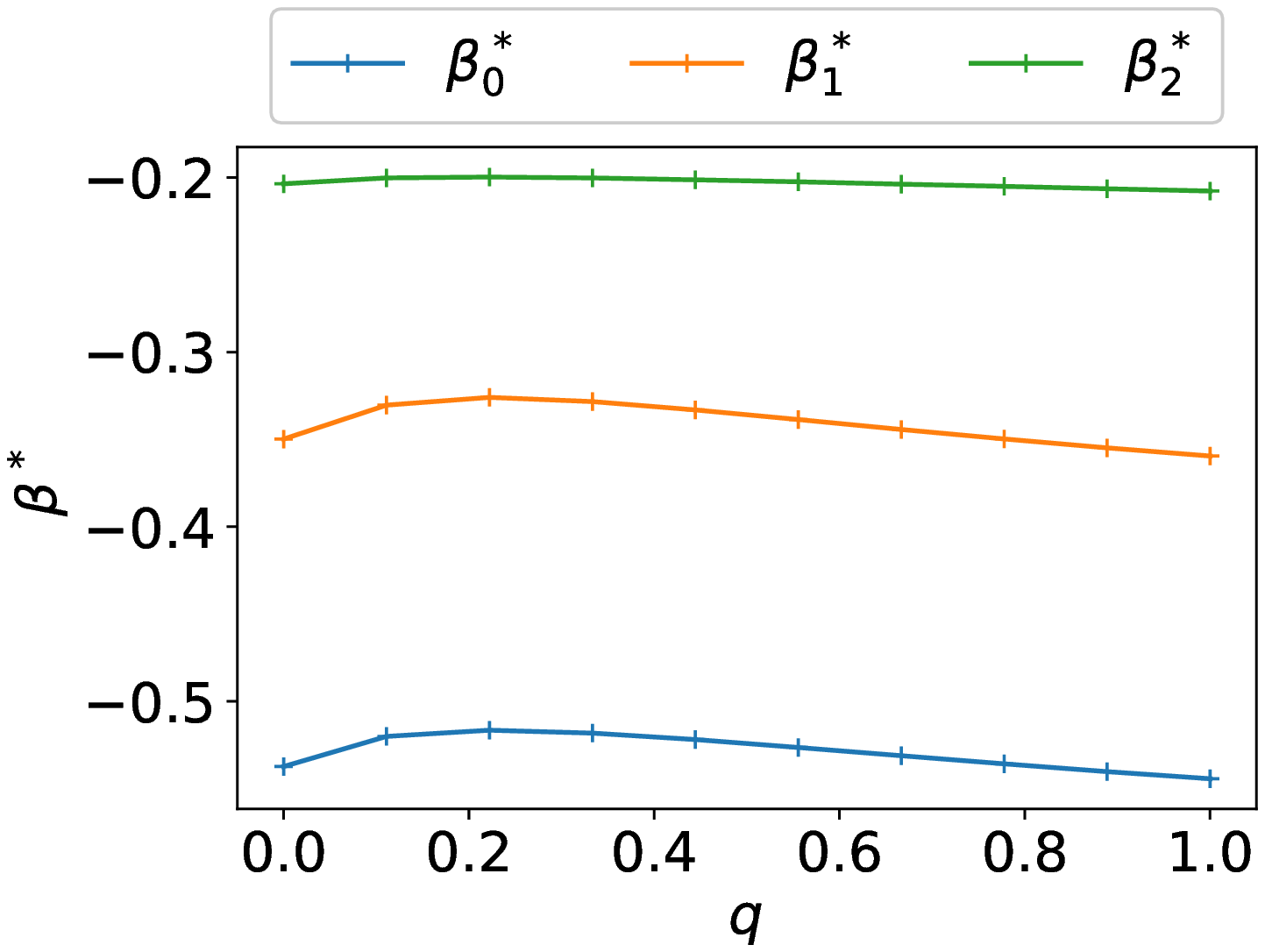}
    \caption{Optimal $\bm\beta$ angles at level $p = 3$.}
\end{subfigure}
\hspace*{0.03\textwidth}
\begin{subfigure}{0.44\textwidth}
    \centering
    \includegraphics[width=\textwidth]{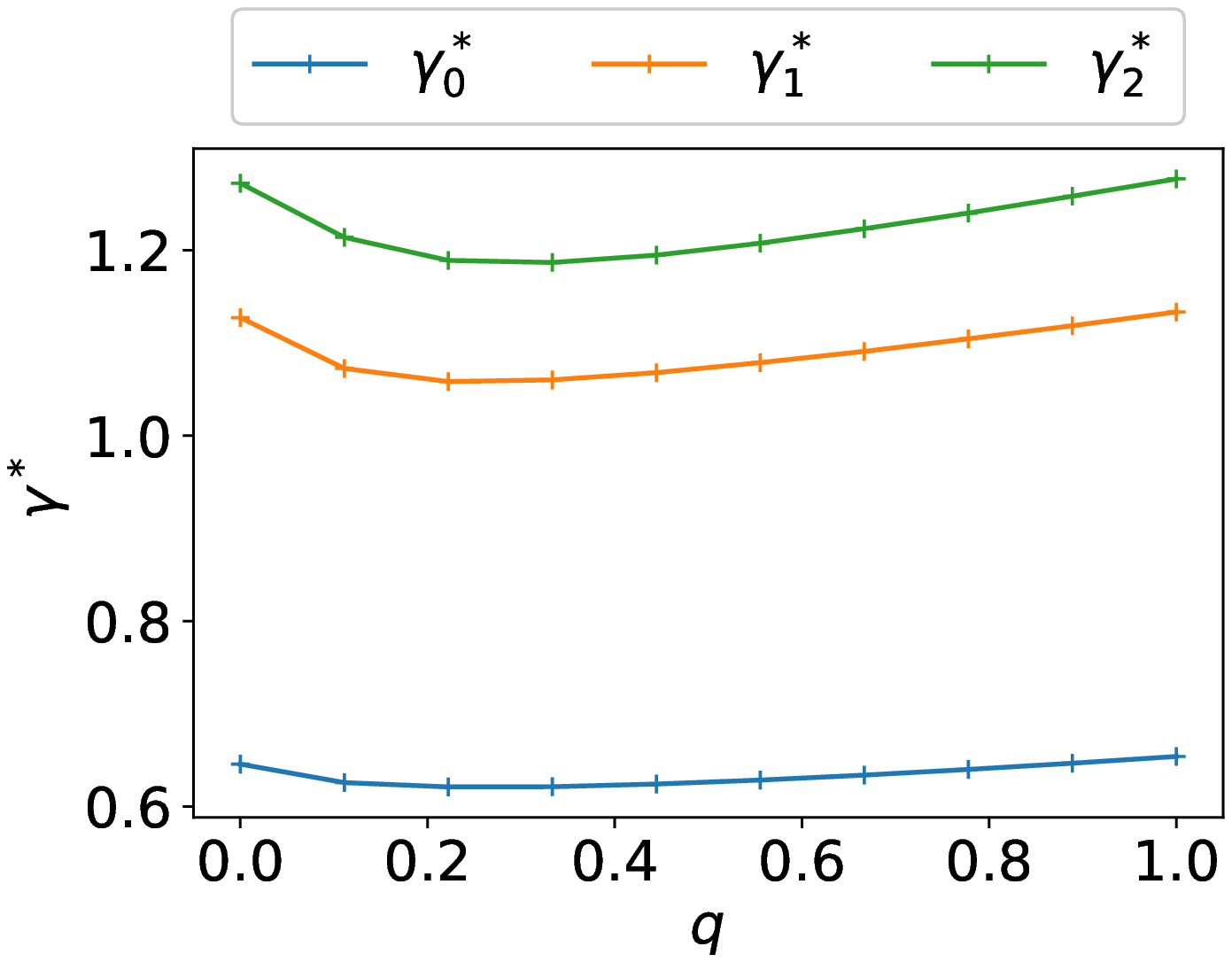}
    \caption{Optimal $\bm\gamma$ angles at level $p = 3$.}
\end{subfigure}
    \caption{Scaling of optimal parameters for MaxCut-QAOA on a random Chung-Lu graph. The graph has expected degrees $4$ and $9$, with respective probabilities $q$ and $1 - q$.}
    \label{fig:maxcut_chung_lu_parameters_scaling}
\end{figure}

\begin{figure}
    \ContinuedFloat
    \centering
    \begin{subfigure}{0.47\textwidth}
        \centering
        \includegraphics[width=\textwidth]{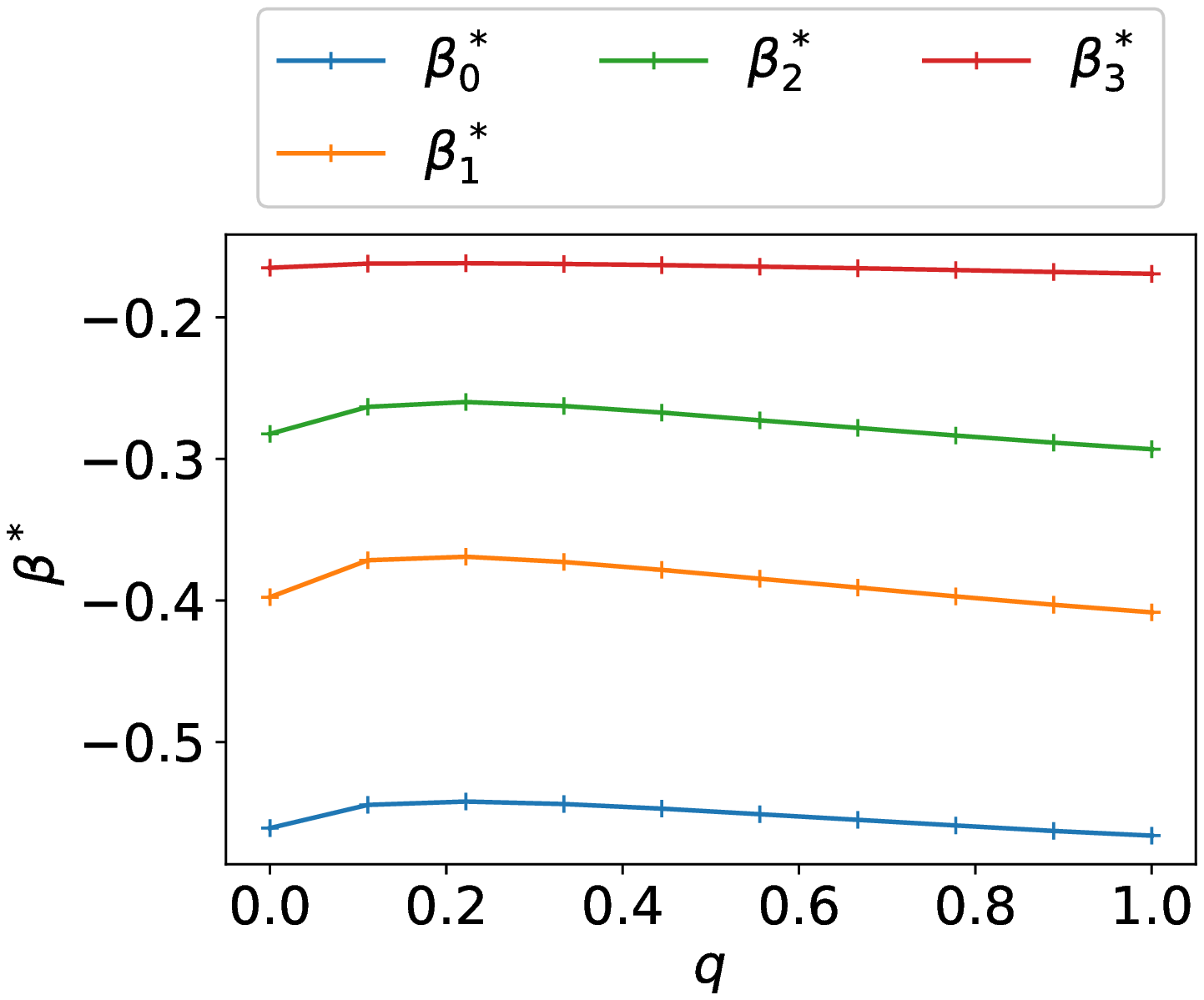}
        \caption{Optimal $\bm\beta$ angles at level $p = 4$.}
    \end{subfigure}
    \hspace*{0.03\textwidth}
    \begin{subfigure}{0.44\textwidth}
        \centering
        \includegraphics[width=\textwidth]{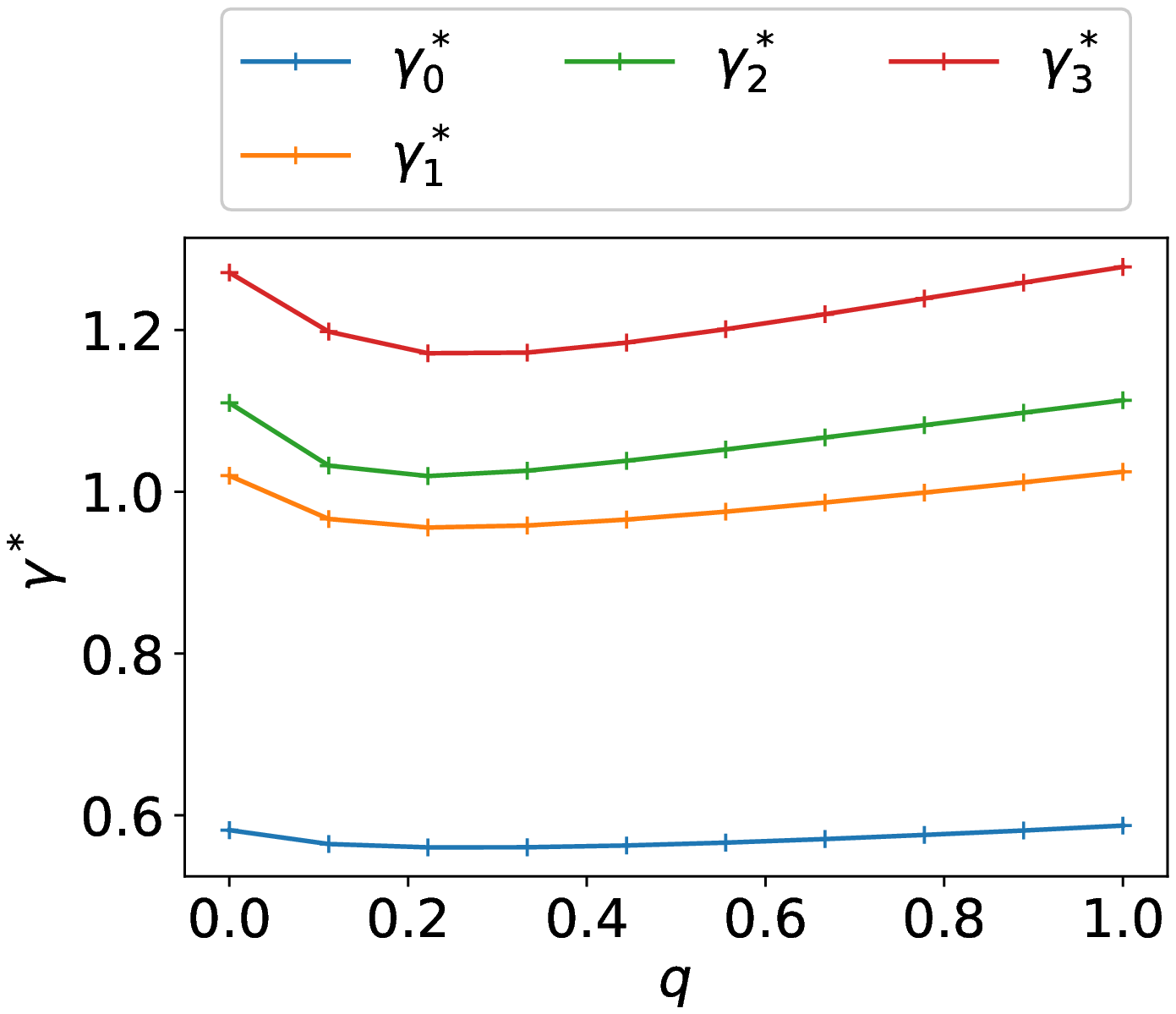}
        \caption{Optimal $\bm\gamma$ angles at level $p = 4$.}
    \end{subfigure}
    \vspace*{10px}
    \begin{subfigure}{0.47\textwidth}
        \centering
        \includegraphics[width=\textwidth]{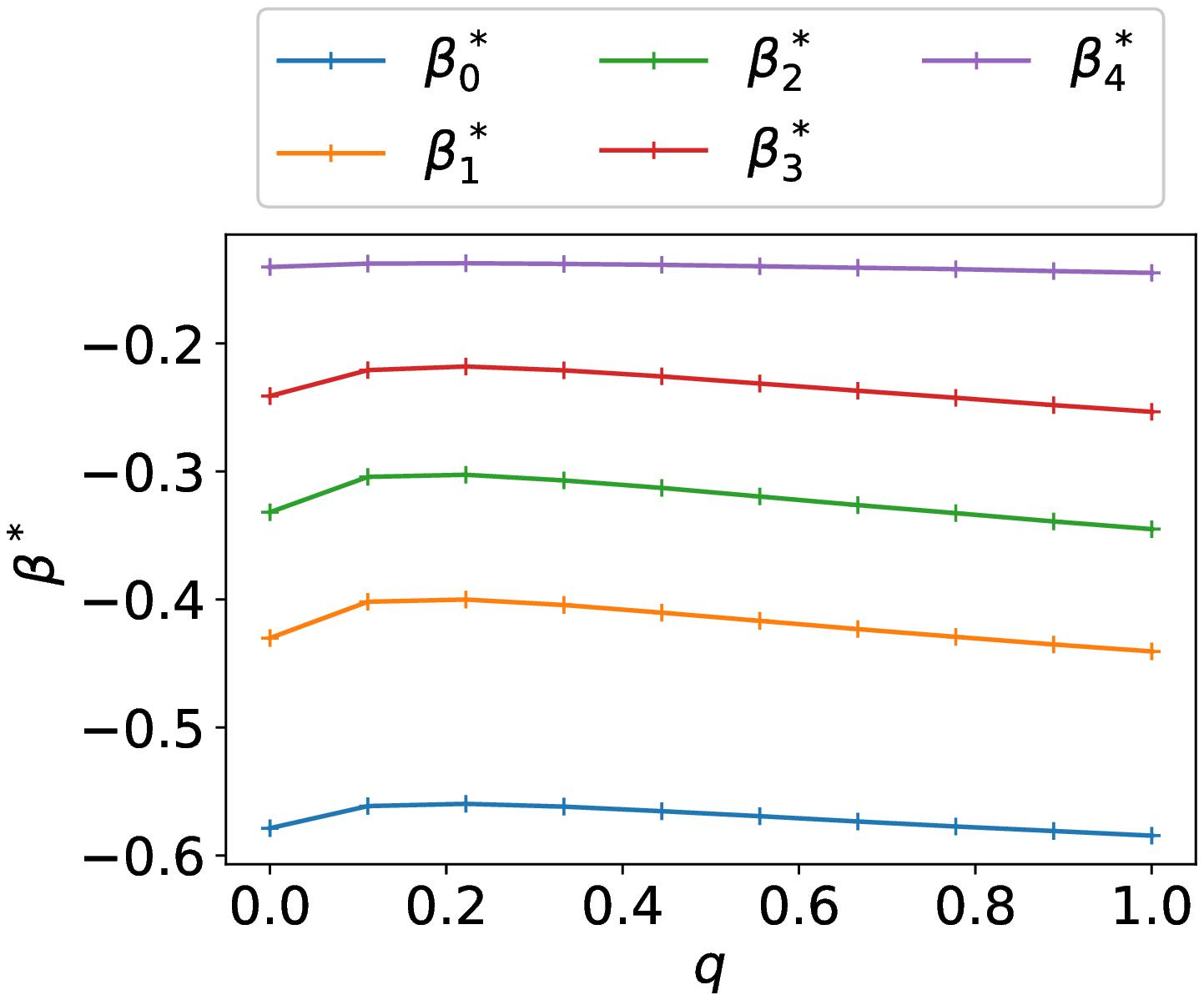}
        \caption{Optimal $\bm\beta$ angles at level $p = 5$.}
    \end{subfigure}
    \hspace*{0.03\textwidth}
    \begin{subfigure}{0.44\textwidth}
        \centering
        \includegraphics[width=\textwidth]{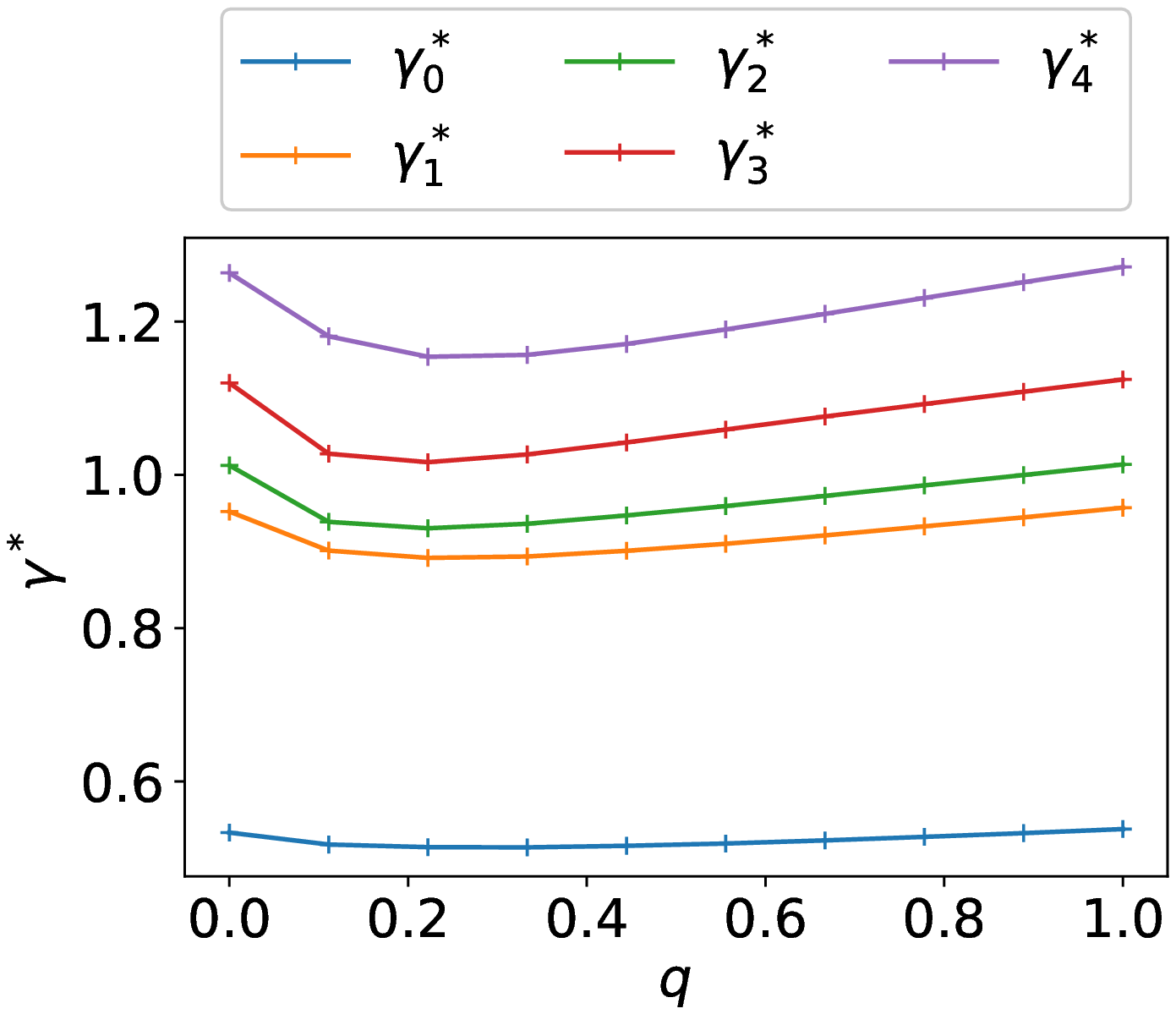}
        \caption{Optimal $\bm\gamma$ angles at level $p = 5$.}
    \end{subfigure}
    \caption{Scaling of optimal parameters for MaxCut-QAOA on a random Chung-Lu graph. The graph has expected degrees $4$ and $9$, with respective probabilities $q$ and $1 - q$.}
    \label{fig:maxcut_chung_lu_parameters_scaling}
\end{figure}

% ------------------------------------------------------------------------------

% ------------------------------------------------------------------------------
\end{document}